\pdfoutput=1
\documentclass[usenatbib]{mn2e}
\usepackage{apjfonts}
\usepackage{amssymb}
\usepackage{amsmath}
\usepackage{ctable}
\usepackage{url}
\usepackage{fixltx2e} 

\newcommand{\paperone}{Paper {\small I}}

\newcommand{\be}{\begin{equation}}
\newcommand{\ee}{\end{equation}}

\newcommand{\etal}{et al.}

\newcommand{\msun}{M_{\sun}}

\newcommand{\fgas}{f_{\rm gas}}

\newcommand{\tauavg}{\tau_{0}}
\newcommand{\tableyes}{\checkmark}
\newcommand{\tableno}{$\times$}
\newcommand{\movieurl}{\url{https://www.cfa.harvard.edu/~phopkins/Site/Research.html}}

\newcommand{\scaleup}{}

\newcommand\plotone[1]
 {\centering \leavevmode \includegraphics[width={0.99\columnwidth}]{#1}}
\newcommand\plotonesmall[1]
 {\centering \leavevmode \includegraphics[width={0.85\columnwidth}]{#1}}

\newcommand{\plotside}[1]
 {\centering \leavevmode \includegraphics[width={0.95\textwidth}]{#1}}
\newcommand{\plotsidesmall}[1]
 {\centering \leavevmode \includegraphics[width={0.90\textwidth}]{#1}}

\newcommand{\acknowledgments}{\begin{small}\section*{Acknowledgments}\end{small}}
\newcommand\altaffilmark[1]{$^{#1}$}
\newcommand\altaffiltext[1]{$^{#1}$}
\title[Stellar Feedback and the ISM]{
The Structure of the Interstellar Medium of Star Forming Galaxies}

\author[Hopkins \etal]{
\parbox[t]{\textwidth}{ 
Philip F.~Hopkins\thanks{E-mail:phopkins@astro.berkeley.edu}\altaffilmark{1},
Eliot Quataert\altaffilmark{1}, \&
Norman Murray\altaffilmark{2,3} 
}
\vspace*{6pt} \\
\altaffiltext{1}{Department of Astronomy and Theoretical Astrophysics Center, University of California Berkeley, Berkeley, CA 94720} \\
\altaffiltext{2}{Canadian Institute for Theoretical Astrophysics, 
60 St.\ George Street, University of Toronto, ON M5S 3H8, Canada} \\
\altaffiltext{3}{Canada Research Chair in Astrophysics} 
}

\date{Submitted to MNRAS, October 14, 2011}
\begin{document}
\maketitle
\label{firstpage}

\begin{abstract}

We develop and implement numerical methods for including stellar feedback in galaxy-scale numerical simulations.   Our models include simplified treatments of heating by Type-I and Type-II supernovae, gas recycling from young stars and AGB winds, heating from the shocked stellar winds, HII photoionization heating, and radiation pressure from stellar photons.  The energetics and time-dependence associated with the feedback are
taken directly from stellar evolution models.    We implement these stellar feedback models in smoothed particle hydrodynamic simulations with pc-scale resolution, modeling galaxies from 
SMC-like dwarfs and Milky-way analogues to massive $z \sim 2$ star forming disks. 
Absent stellar feedback, gas cools rapidly and collapses without limit into dense sub-units, inconsistent with observations.  By contrast, in all cases with feedback, the interstellar medium (ISM) quickly approaches a statistical steady state in which giant molecular clouds (GMCs) continuously form, disperse, and re-form, leading to a multi-phase ISM.   In this paper, we quantify the  properties of the ISM and GMCs in this self-regulated state. In a companion paper we study the galactic winds driven by stellar feedback.

Our primary results on the structure of the ISM in star forming galaxies include:

1)  Star forming galaxies generically self-regulate so that the cool, dense gas maintains Toomre's $Q$  $\sim 1$.   Most of the volume is occupied by relatively diffuse hot gas, while most of the mass is in dense GMC complexes created by self-gravity.   The phase structure of the gas and  the gas mass fraction at high densities are much more sensitive probes of the physics of stellar feedback than integrated quantities such as the Toomre $Q$ or gas velocity dispersion.

2)   Different stellar feedback mechanisms act on different spatial (and density) scales.    Radiation pressure and HII gas pressure are critical for preventing runaway collapse of dense gas in GMCs.   Shocked supernovae ejecta and stellar winds dominate the dynamics of the volume-filling hot gas. However, this gas primarily vents out of the star-forming disk and contributes only modestly to the mid-plane ISM pressure.   

3) The galaxy-averaged star formation rate is determined by feedback, with different mechanisms dominating in different galaxy types. For a given feedback efficiency, restricting star formation to molecular gas or modifying the cooling function has little effect on the star formation rate in the galaxies we model (including an SMC-mass dwarf).  By contrast, changing the feedback mechanisms or assumed  feedback efficiencies directly translates to shifts off of the observed Kennicutt-Schmidt relation.

4) Self-gravity leads to GMCs with an approximately self-similar mass function $\propto M^{-2}$, with a high mass cutoff determined by the characteristic Jeans/Toomre mass of the system.   In all of our galaxy models, GMCs live for a few dynamical times before they are disrupted by stellar feedback.  The net star formation efficiency in GMCs ranges from $\sim 1 \%$ in dwarfs and MW-like spirals to nearly $\sim 10 \%$ in gas-rich rapidly star forming galaxies.   GMCs are approximately virialized, but there is a large dispersion in the virial parameter for a given GMC mass and lower mass GMCs tend to be preferentially unbound.

\end{abstract}


\begin{footnotesize}
\ctable[
  caption={{\normalsize Galaxy Models}\label{tbl:sim.ics}},center,star
  ]{lccccccccccccccc}{
\tnote[ ]{Parameters describing our galaxy models, used as the initial conditions for all of the simulations: \\ 
{\bf (1)} Model name: shorthand for models of a high-redshift massive 
starburst ({\bf HiZ}); local gas-rich galaxy ({\bf Sbc}); MW-analogue ({\bf MW}); and isolated SMC-mass dwarf ({\bf SMC}). 
{\bf (2)} $\epsilon_{g}$: gravitational force softening in our highest-resolution simulations.
{\bf (3)} $m_{i}$: gas particle mass in our highest-resolution simulations. New star particles formed have mass $=0.5\,m_{i}$, disk/bulge particles $\approx m_{i}$, and dark matter halo particles $\approx5\,m_{i}$. 
{\bf (4)} $M_{\rm halo}$: halo mass. 
{\bf (5)} $c$: halo concentration. Values lie on the halo mass-concentration 
relation at the appropriate redshift ($z=0$ for SMC, Sbc, and MW; $z=2$ for HiZ). 
{\bf (6)} $V_{\rm max}$: halo maximum circular velocity. 
{\bf (7)} $M_{\rm bary}$: total baryonic mass.
{\bf (8)} $M_{\rm b}$: bulge mass.
{\bf (9)} $a$: \citet{hernquist:profile} profile scale-length for bulge. 
{\bf (10)} $M_{d}$: stellar disk mass.
{\bf (11)} $r_{d}$: stellar disk scale length. 
{\bf (12)} $h$: stellar disk scale-height.
{\bf (13)} $M_{g}$: gas disk mass.
{\bf (14)} $r_{g}$: gas disk scale length (gas scale-height determined so that $Q=1$).
{\bf (15)} $f_{\rm gas}$: average gas fraction of the disk 
inside of the stellar $R_{e}$ ($M_{\rm g}[<R_{e}]/(M_{\rm g}[<R_{e}]+M_{\rm d}[<R_{e}])$).
The total gas fraction, including the extended disk, is $\sim50\%$ larger.
{\bf (15)} $t_{\rm dyn}$: gas disk dynamical time at the half-gas mass radius. 
}
}{
\hline\hline
\multicolumn{1}{c}{Model} &
\multicolumn{1}{c}{$\epsilon_{\rm g}$} &
\multicolumn{1}{c}{$m_{i}$} &
\multicolumn{1}{c}{$M_{\rm halo}$} & 
\multicolumn{1}{c}{$c$} & 
\multicolumn{1}{c}{$V_{\rm max}$} & 
\multicolumn{1}{c}{$M_{\rm bary}$} & 
\multicolumn{1}{c}{$M_{\rm b}$} & 
\multicolumn{1}{c}{$a$} & 
\multicolumn{1}{c}{$M_{\rm d}$} & 
\multicolumn{1}{c}{$r_{d}$} & 
\multicolumn{1}{c}{$h$} & 
\multicolumn{1}{c}{$M_{\rm g}$} & 
\multicolumn{1}{c}{$r_{g}$} &
\multicolumn{1}{c}{$f_{\rm gas}$} &
\multicolumn{1}{c}{$t_{\rm dyn}$} \\
\multicolumn{1}{c}{\,} &
\multicolumn{1}{c}{[pc]} &
\multicolumn{1}{c}{[$\msun$]} & 
\multicolumn{1}{c}{[$\msun$]} & 
\multicolumn{1}{c}{\,} & 
\multicolumn{1}{c}{[${\rm km\,s^{-1}}$]} & 
\multicolumn{1}{c}{[$\msun$]} & 
\multicolumn{1}{c}{[$\msun$]} & 
\multicolumn{1}{c}{[kpc]} & 
\multicolumn{1}{c}{[$\msun$]} & 
\multicolumn{1}{c}{[kpc]} & 
\multicolumn{1}{c}{[pc]} & 
\multicolumn{1}{c}{[$\msun$]} & 
\multicolumn{1}{c}{[kpc]} &
\multicolumn{1}{c}{\,} &
\multicolumn{1}{c}{[Myr]} \\
\hline
{\bf HiZ} & 3.5 & 1700 & 1.4e12 & 3.5 & 230 & 1.07e11 & 7e9 & 1.2 & 3e10 & 1.6 & 130 & 7e10 & 3.2 & 0.49 & 12 \\ 
{\bf Sbc} & 2.5 & 130 & 1.5e11 & 11 & 86 & 1.05e10 & 1e9 & 0.35 & 4e9 & 1.3 & 320 & 5.5e9 & 2.6 & 0.36 & 22 \\ 
{\bf MW} & 2.5 & 220 & 1.6e12 & 12 & 190 & 7.13e10 & 1.5e10 & 1.0 & 4.73e10 & 3.0 & 300 & 0.9e10 & 6.0 & 0.09 & 31 \\ 
{\bf SMC} & 0.7 & 25 & 2.0e10 & 15 & 46 & 8.9e8 & 1e7 & 0.25 & 1.3e8 & 0.7 & 140 & 7.5e8 & 2.1 & 0.56 & 45 \\ 
\hline\hline\\
}
\end{footnotesize}

\vspace{-0.24in}
\section{Introduction}
\label{sec:intro}


The Kennicutt-Schmidt (KS) law for star formation in galaxies implies a gas consumption time of $\sim50$ dynamical times \citep{kennicutt98}; comparably long gas consumption times are inferred even on the scale of giant molecular clouds (GMCs) and other dense star-forming regions \citep[e.g.][]{krumholz:sf.eff.in.clouds}. 
Moreover, the total fraction of the gas turned into stars in GMCs over their lifetime is only a few percent  \citep{zuckerman:1974.gmc.constraints,williams:1997.gmc.prop,evans:1999.sf.gmc.review,evans:2009.sf.efficiencies.lifetimes}.   By contrast, absent stellar feedback, GMCs experience runaway collapse to densities much higher than observed, and eventually turn a near-unity fraction of their gas into stars 
\citep{hopkins:rad.pressure.sf.fb,tasker:2011.photoion.heating.gmc.evol,
bournaud:2010.grav.turbulence.lmc,dobbs:2011.why.gmcs.unbound,
krumholz:2011.rhd.starcluster.sim,harper-clark:2011.gmc.sims}. 
An analogous problem occurs in cosmological models of galaxy evolution without strong stellar feedback:   gas cools efficiently, collapsing into  clumps 
that runaway to high densities on a few dynamical times and produce an order of magnitude more stars than is observed \citep[e.g.][and references therein]{katz:treesph,somerville99:sam,cole:durham.sam.initial}. 

Neither thermal pressure nor turbulence in and of itself can stave off collapse in the ISM:
 cooling is rapid, so that thermal support is quickly lost, and turbulent support  dissipates on a single crossing time (e.g., \citealt{ostriker:2001.gmc.column.dist}).  Some mechanism must therefore continuously inject energy and momentum into the gas on both GMC and galactic scales, in order to prevent runaway collapse to arbitrarily high densities. Moreover, it is not enough to simply slow down the collapse of gas; if that were the case, the gas would eventually turn into stars on long timescales. Instead, mass must actually be expelled from both GMCs and galaxies as a whole.

A number of physical mechanisms have been proposed as sources of  random motions in the ISM and GMCs:   these include photo-ionization, stellar winds, radiation pressure from both UV and IR photons, proto-stellar jets, cosmic-rays created in supernova shocks, 
and supernovae themselves (this is not an exhaustive list!).  An alternative class of models suggests that the turbulence is not explicitly produced by star formation, but comes from gravitational cascades from larger scales or instabilities that can tap into the differential rotation in galaxies.
\citep[see e.g.][and references therein, for reviews of ISM turbulence and stellar feedback]{mac-low:2004.turb.sf.review,mckee:2007.sf.theory.review}.

Making a realistic comparison between  models of the ISM and observations ultimately requires numerical simulations, since the systems of interest are highly non-linear, chaotic, and time-dependent, with a large number of competing physical processes 
and a huge dynamic range in space and time.   However, most numerical 
experiments  have traditionally been restricted to idealized simulations of single star-forming regions or sub-regions of single GMCs. There are good reasons for doing so, since these calculations are critical for studying the physics of star formation.  They cannot, however, address a large number of observational constraints such as the balance of phases in the ISM; 
 the origin of the {\em global} Schmidt-Kennicutt relation; and observations of GMC/galaxy scaling relations, such as mass functions and the linewidth-size relation.
Likewise, the lifetimes of GMCs and their internal structure cannot be addressed in detail without a model that self-consistently incorporates continuous 
GMC formation, dissolution, and re-formation.   In short, addressing these 
questions requires {\em galaxy-scale} simulations that also self-consistently follow  feedback from sub-GMC through super-galactic scales. 

There are at least two major challenges that numerical simulations must confront to model feedback and GMC-level phase 
structure explicitly. First, simulations must actually resolve the formation of 
GMCs and the phase structure of the ISM 
(spatial resolution $\sim1\,$pc; although in massive gas-rich galaxies GMCs can be much larger, $\sim$kpc in size). 
Numerical simulations of isolated galaxies, galaxy mergers, and cosmological ``zoom-in'' simulations of individual 
halos can now reach this resolution, but it remains prohibitive in full cosmological simulations. 

The second challenge facing simulations is that a wide range of processes appear to be important sources of feedback and turbulence in the ISM of galaxies: e.g., protostellar jets, HII regions, stellar winds, radiation pressure from young stars, and supernovae. It is therefore unlikely that the full problem, 
which spans many orders of magnitude in spatial, mass, and time scales, can be treated with just a single 
simple prescription, without including multiple physical processes.  In particular the physics of the dense ISM is clearly very different from that of the diffuse ISM, and different feedback processes likely dominate in each case. Thus far, however, many calculations that model stellar feedback 
explicitly have invoked {\em only} thermal heating via supernovae
\citep[e.g.][]{thackercouchman00,governato:disk.formation,ceverino:cosmo.sne.fb,
teyssier:2010.clumpy.sb.in.mergers,colin:2010.subgrid.sf.cosmosims,avila-reese:2011.ssfr.cosmosims,pilkington:2011.gas.dwarf.zoomsims}. The thermal energy, however, tends to be rapidly radiated away, especially 
in dense sub-regions of GMCs.   Perhaps even more importantly, the actual time required for SNe to explode (a few Myr) is  {\em longer} than the observed ``lifetime'' of 
low-mass GMCs and so supernovae cannot be the dominant mechanism unbinding GMCs in local galaxies 
\citep{evans:2009.sf.efficiencies.lifetimes}.

In \citet{hopkins:rad.pressure.sf.fb} (hereafter \paperone) we developed a new numerical model 
for the effects of the  momentum injection from massive stars 
on small scales in star-forming regions. 
Using galaxy-scale simulations with pc-scale resolution to resolve the formation of individual 
GMCs, we accounted (approximately) for the momentum deposited as gas and dust in the vicinity of young stars absorbs the stellar radiation.    We showed that this feedback mechanism alone can dramatically suppress runaway star formation, disrupting collapsing 
GMCs once a few percent of their mass is turned into 
stars, and naturally producing galaxies on the observed Schmidt-Kennicutt relation. 
This is, however, just a first step towards a more complete model of the ISM.

In this paper, we therefore extend the models in \paperone\ to include 
HII photo-ionization heating and the momentum injection, gas heating 
(energy injection) and mass recycling from supernovae (Types I and II) and
stellar winds (both early fast winds and late, AGB winds).\footnote{\label{foot:url}Movies of these
  simulations are available at \movieurl}  We also extend the 
radiation pressure model to include not just the small-scale radiation 
pressure force from multiply-scattered IR photons in GMCs, but also the long-range 
radiation pressure force produced by the diffuse interstellar UV through IR radiation field.   Our models of these feedback processes are physically motivated and tied to stellar evolution calculations; they are, however, also necessarily simplified and approximate given the range of spatial scales and processes we consider.   In addition to an improved treatment of stellar feedback, we also extend our cooling 
and star formation models to include explicit models for 
molecular chemistry, allowing for the suppression of star formation 
 via photo-dissociation and inefficient molecule formation.  These are all processes that have been discussed analytically in the literature, and in some cases explored in idealized simulations that focus on an individual physical effect.   However, by including all of these processes together, our study allows us to assess the relative importance of different feedback mechanisms and the interplay between the different feedback and star formation processes in a given galaxy model.    Towards this end, we have implemented this star formation and feedback physics in a range of idealized, observationally-motivated galaxy models 
that span low-redshift dwarfs, Milky Way-like systems, and 
nearby starbursts through to extremely massive, 
high-redshift starburst galaxies.



The remainder of this paper is organized as follows.   In \S \ref{sec:sims} we describe 
the galaxy models and our method of implementing feedback from the various mechanisms outlined above. 
The Appendix contains tests of the numerical methods.
In \S~\ref{sec:morph} we discuss how each feedback mechanism singly and in combination 
effects the observed galaxy morphologies in both gas and stars, for a range of different galaxy properties.
In \S~\ref{sec:sfh} we consider the star formation histories of each model and show how 
they are influenced by each feedback mechanism separately and in concert. 
We investigate the role of molecular chemistry in \S~\ref{sec:appendix.chemistry} and 
show that it has little effect on galaxy properties or global star formation (for the galaxy models we consider).
We examine the global properties of the ISM in \S~\ref{sec:dispersion}, 
including gas disk velocity dispersions, scale heights, and Toomre parameters, 
and we illustrate how different feedback mechanisms contribute to supporting turbulence 
and maintaining marginal disk stability. 
We consider the phase structure of the ISM in \S~\ref{sec:ism.structure}, and show 
how the presence of these feedback mechanisms together naturally gives rise to a multi-phase ISM 
(with cold molecular, warm ionized, and hot diffuse components). In \S~\ref{sec:clumps} we examine the properties of the cold dense gas in the simulated 
GMC population: GMC mass functions, densities, velocity dispersions and scaling laws, 
virial parameters, lifetimes, and star formation efficiencies. We compare each of these 
properties to observations and show how they are affected by the presence or absence 
of different feedback mechanisms. 
Finally, in \S \ref{sec:discussion} we summarize our results and discuss their implications.  

These results presented in this paper provide an explanation for disruption of GMCs and the observed slow rate of star formation in galaxies, but they do not address the question of why galaxies have a baryon fraction well below that of the universe as a whole. In a companion paper we show that the combined effects of radiation pressure and supernovae drive powerful galaxy scale winds, ejecting baryons from star forming galaxies.

\begin{figure}
    \centering
    \plotone{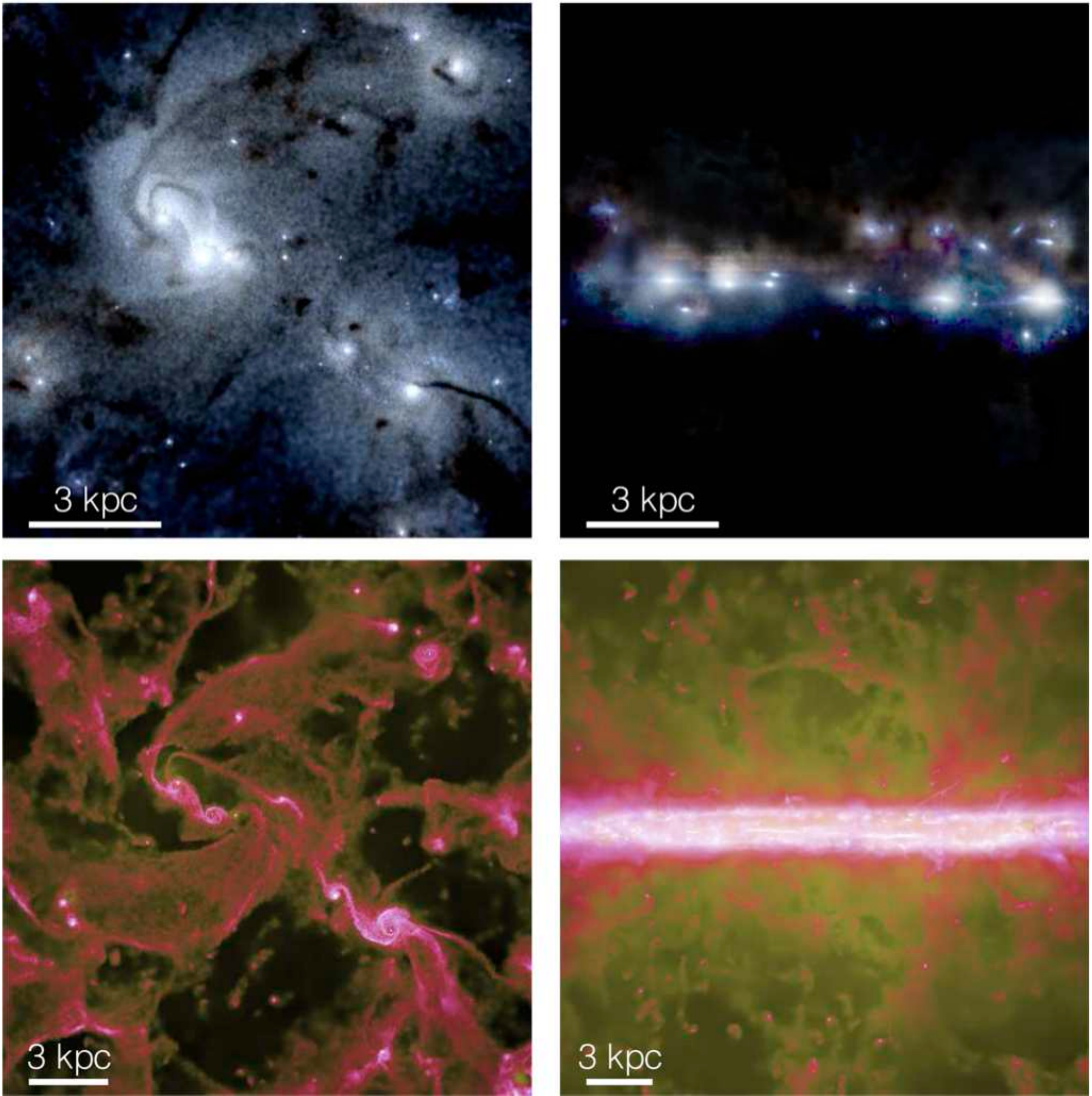}
    \caption{Morphology of the gas and stars in a standard simulation 
    of a HiZ disk model, a massive, 
    $z\sim2-4$ starburst disk with $\dot{M}_{\ast}>100\,\msun\,{\rm yr^{-1}}$, with all feedback mechanisms enabled.  The images are shown
    at $\sim2$ orbital times after the beginning of the simulation when the disk is 
    in a feedback-regulated steady-state. We show  
    face-on ({\em left}) and edge-on {\em right} projections.
    {\em Top:} Stars. The image is a mock $ugr$ (SDSS-band) composite, 
    with the spectrum of all stars calculated from their known age and metallicity, 
    and dust extinction/reddening accounted for from the line-of-sight dust mass. 
    The brightness follows a logarithmic scale with a stretch of $\approx2\,$dex.
    {\em Bottom:} Gas. Brightness encodes projected gas density (logarithmically 
    scaled with a $\approx4\,$dex stretch); color encodes gas temperature 
    with blue being $T\lesssim1000\,$K molecular gas, pink $\sim10^{4}-10^{5}$\,K 
    warm ionized gas, and yellow $\gtrsim10^{6}\,$K hot gas. 
    Gravitational collapse forms massive kpc-scale star cluster complexes
    that give rise to the clumpy morphology (edge on, similar to 
    ``clump-chain'' galaxies) that is further enhanced in the optical by 
    patchy extinction. Violent outflows are present, emerging both from the complexes and 
    the disk as a whole, driven by the massive starburst. These outflows maintain a 
    thick gas disk and disrupt many of the clumps, which continuously re-form. 
    \label{fig:morph.hiz}}
\end{figure}


\begin{figure}
    \centering
    \plotone{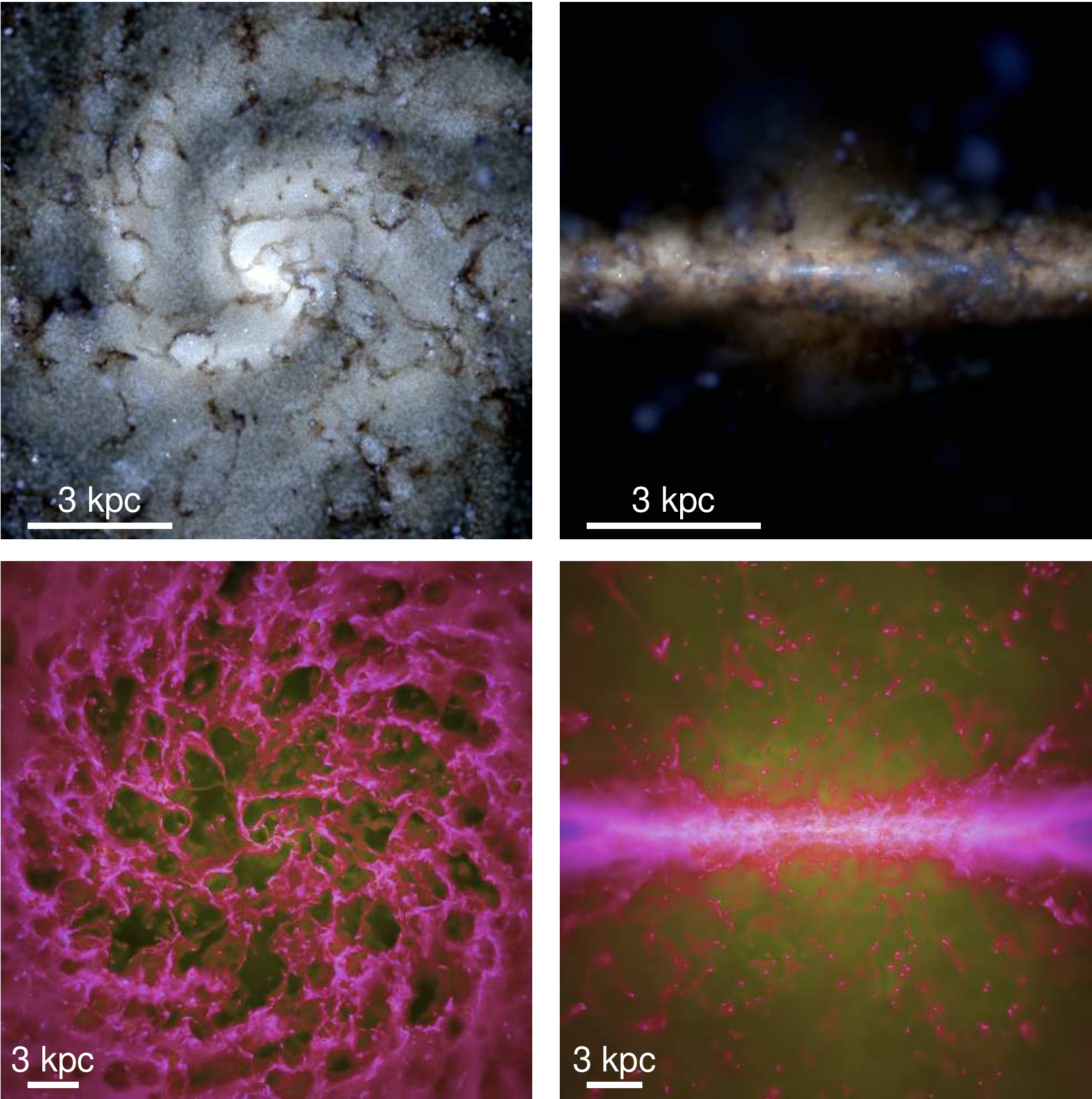}
    \caption{Stellar and gas images as in Figure~\ref{fig:morph.hiz}, but for the Sbc 
    model,  a dwarf starburst galaxy. The disk is again clumpy, but the clumps are 
    relatively much smaller than those in the HiZ model, and do not gravitationally dominate -- the global disk structure more clearly traces spiral 
    structure albeit with a  disturbed, filamentary and irregular pattern in the gas. 
    There is a stellar bar in the central few kpc and a bright starburst, but the 
    disk is quite dusty, especially edge-on.
    Strong outflows arise from the 
    central few kpc, producing a clumpy, multi-phase wind.
    \label{fig:morph.sbcn}}
\end{figure}

\begin{figure}
    \centering
    \plotone{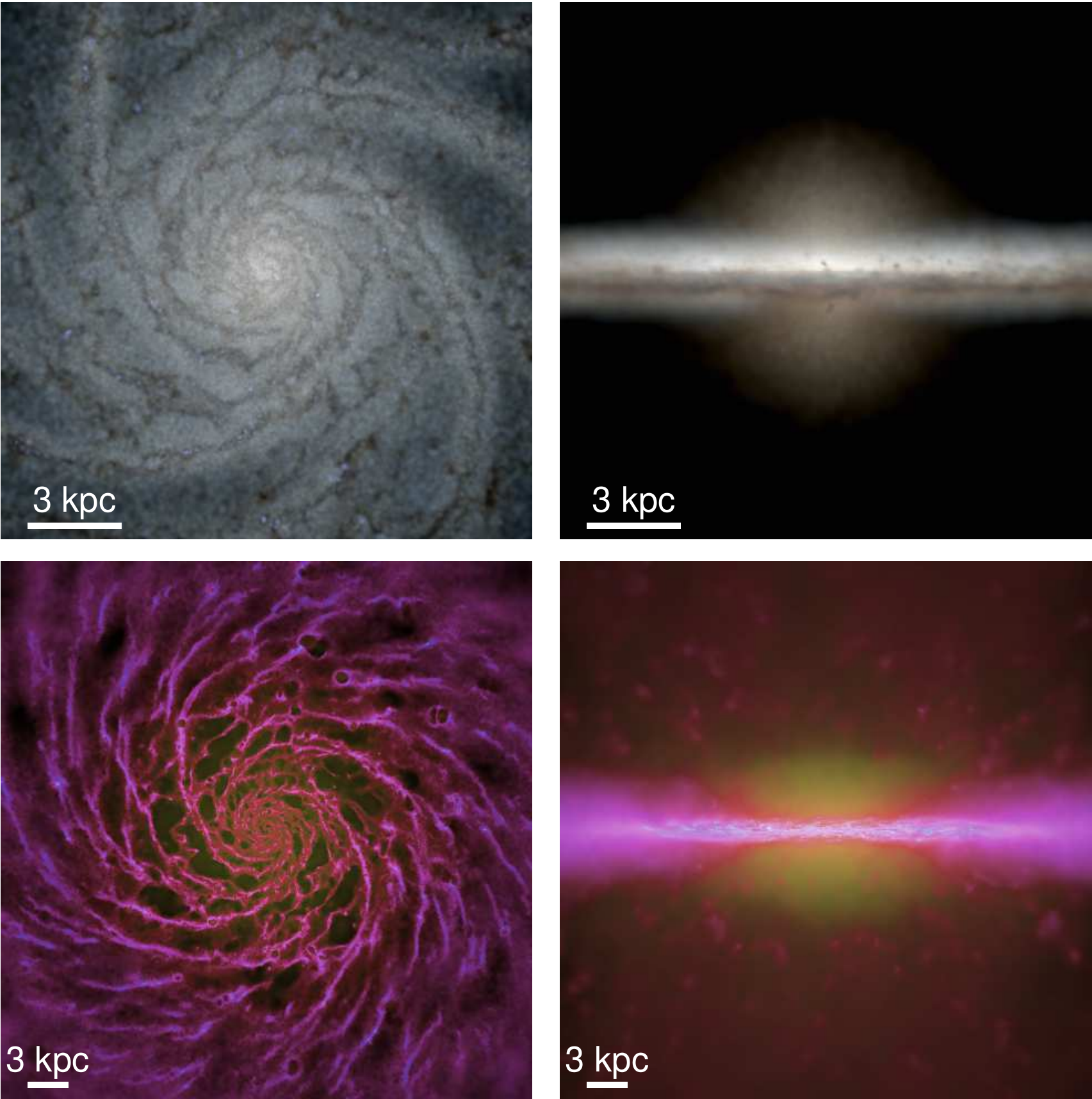}
    \caption{Stellar and gas images as in Figure~\ref{fig:morph.hiz}, but for the Milky Way model.   These images are for a model with a slightly higher dark matter fraction than our standard MW model, such that a bar does not form. The resulting morphology in gas and stars is a canonical spiral 
    (compare e.g.\ M101), with dust lanes along the spiral arms, flocculant morphology 
    and features such as spurs/feathers in the outer arms. 
    The star formation is still concentrated in star clusters 
    (visible as small blue knots, largely along the arms) -- but these are 
    relatively much smaller than the giant clumps in 
    Figure~\ref{fig:morph.hiz}, owing to the much smaller Jeans mass. 
    SNe-driven bubbles/holes are clear in the outer gas disk. 
     \label{fig:morph.mw.spiral}}
\end{figure}

\begin{figure}
    \centering
    \plotone{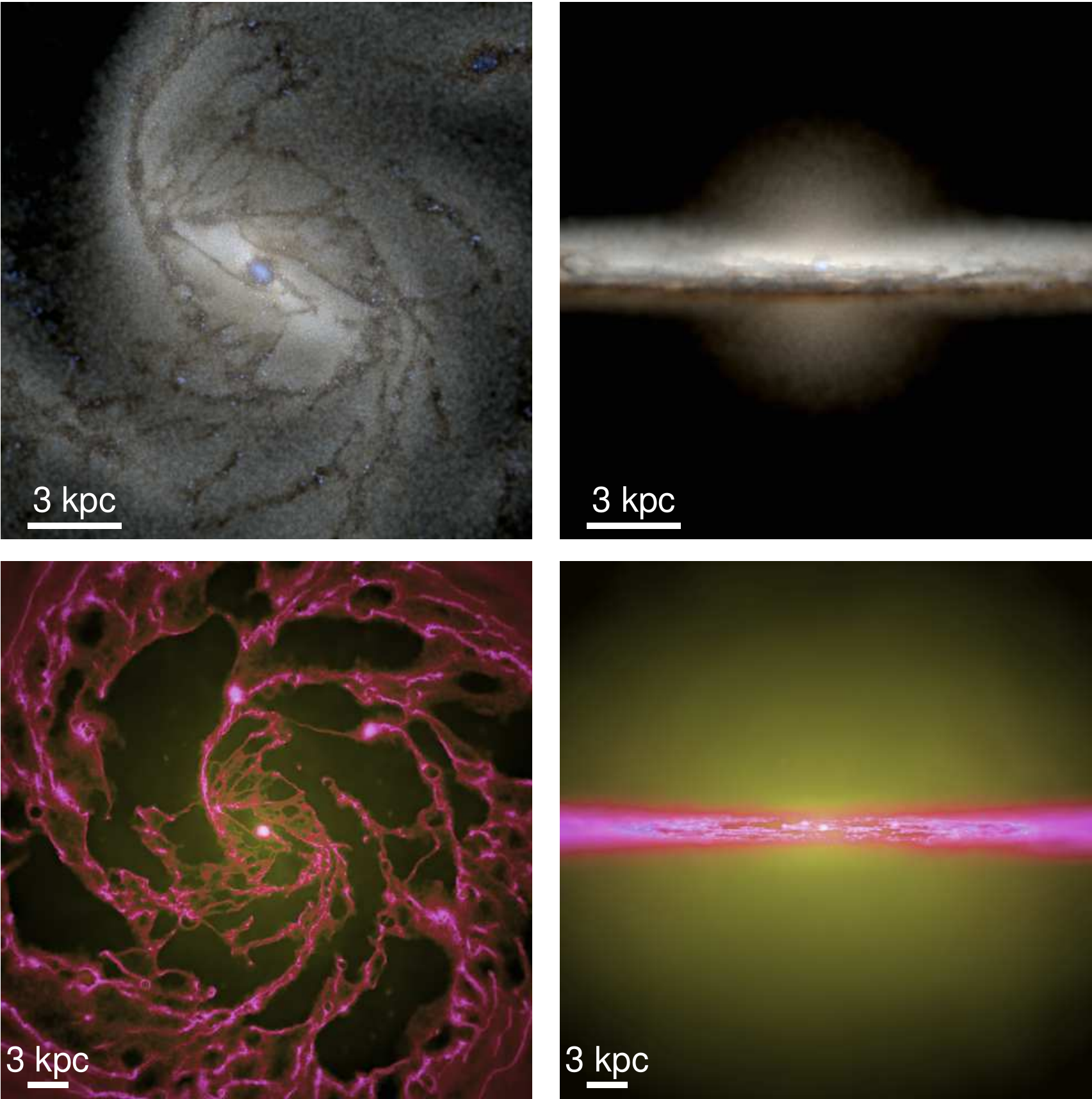}
    \caption{Stellar and gas images for our standard Milky Way model after it develops a strong bar (compare with Fig.~\ref{fig:morph.mw.spiral}, which is our MW model for a somewhat higher dark matter halo mass, which does not develop a bar).  The morphology is that of a typical 
    barred spiral (compare e.g.\ NGC1097). Star cluster formation is 
    enhanced along the bar, with a small nuclear concentration of star formation 
    driven by inflows along the bar. 
    \label{fig:morph.mw.bar}}
\end{figure}

\begin{figure}
    \centering
    \plotone{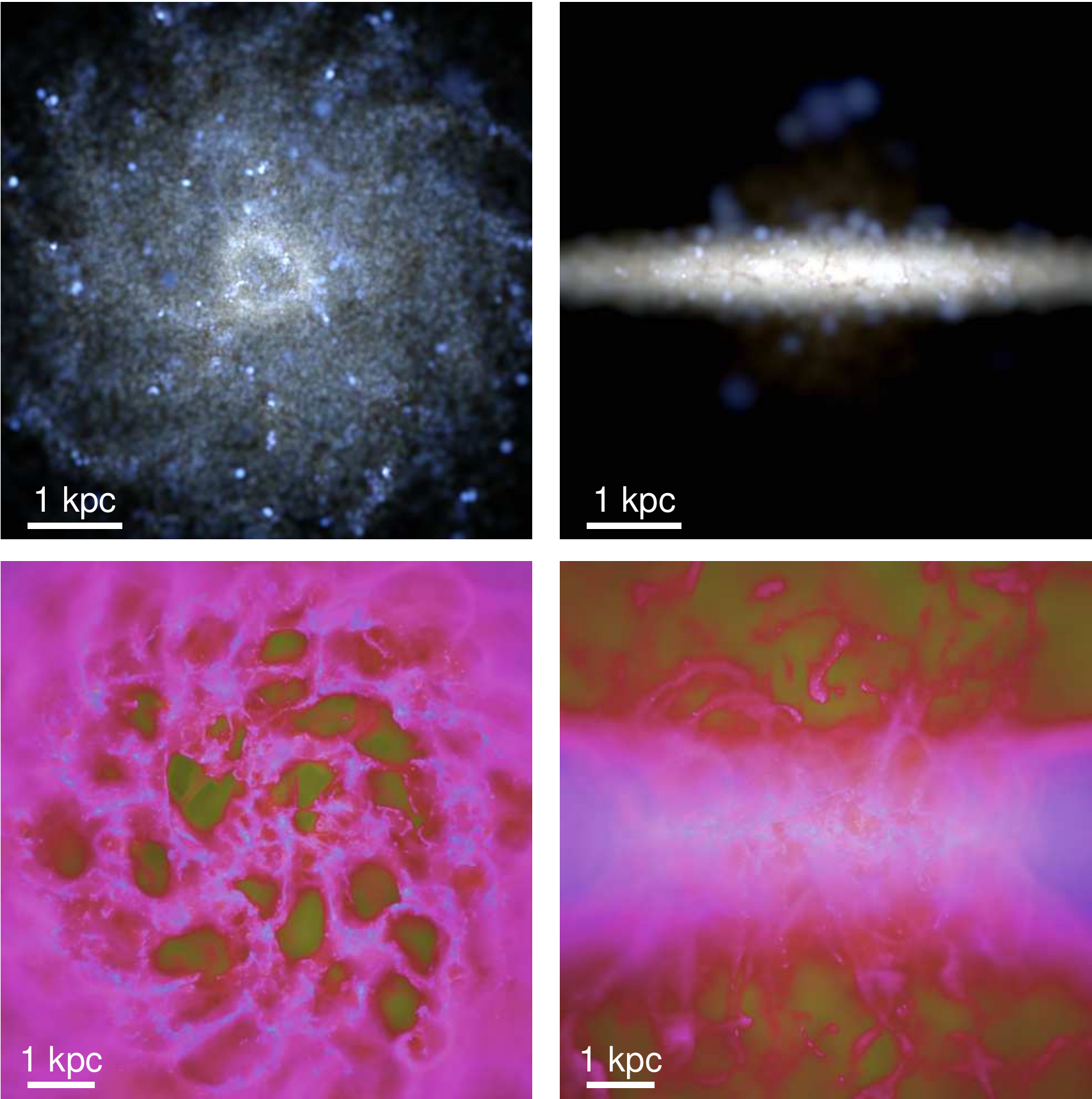}
    \caption{Stellar and gas images as in Figure~\ref{fig:morph.hiz}, but for the SMC model, an 
    isolated SMC-mass dwarf galaxy.    The morphology is here that of an irregular galaxy, with a puffy/thick disk of  mostly ionized gas, irregularly distributed star formation, 
    and large SNe bubbles driven by overlapping SNe explosions. 
    Despite a very low absolute star formation rate, a wind is plainly evident 
    with a mix of hot gas and entrained cold/warm material, with various filaments/loops/arcs 
    driven by the hot gas bubbles as well as directly venting hot gas. 
    \label{fig:morph.smc}}
\end{figure}

%

\begin{footnotesize}
\ctable[
  caption={{\normalsize Feedback Physics for Simulations Used in This Paper}\label{tbl:sims}},center,star
  ]{lccccccc}{
\tnote[ ]{Summary of the physics included or excluded in the simulations shown throughout this paper. 
The implementation of each set of physics is described in \S~\ref{sec:sims}.  
A \tableyes\ means the given ``module'' was included, \tableno\ means it was excluded.
Mod$=$``Modified'' runs are described in the text. 
The Enforced SED models modify the calculation of the long-range radiation pressure 
and are discussed in the Appendix. 
\paperone\ simulations use the SF model of Molecular Model A.
}
}{
\hline\hline
\multicolumn{1}{l}{Simulation Name} &
\multicolumn{7}{c}{Physics Included} \\
\hline
\multicolumn{1}{c}{\scriptsize{\ }} & 
\multicolumn{1}{c}{\scriptsize{Star Formation}} & 
\multicolumn{1}{c}{\scriptsize{Gas}} & 
\multicolumn{1}{c}{\scriptsize{Mass Recycling}} & 
\multicolumn{1}{c}{\scriptsize{Radiation Pressure}} & 
\multicolumn{1}{c}{\scriptsize{SNe (Types I \&\ II)}} & 
\multicolumn{1}{c}{\scriptsize{Stellar Winds}} & 
\multicolumn{1}{c}{\scriptsize{HII Photo-}}\\
\multicolumn{1}{c}{\scriptsize{\ }} & 
\multicolumn{1}{c}{\scriptsize{\&\ Molecular Gas}} & 
\multicolumn{1}{c}{\scriptsize{Cooling}} & 
\multicolumn{1}{c}{\scriptsize{(SNe \&\ Winds)}} & 
\multicolumn{1}{c}{\scriptsize{(Local/Long-Range)}} & 
\multicolumn{1}{c}{\scriptsize{(Energy/Momentum)}} & 
\multicolumn{1}{c}{\scriptsize{(Energy/Momentum)}} & 
\multicolumn{1}{c}{\scriptsize{Ionization}} \\
\hline
{\bf Standard} & \tableyes & \tableyes & \tableyes & \tableyes/\tableyes & \tableyes/\tableyes & \tableyes/\tableyes & \tableyes \\ 
{\bf No Feedback} & \tableyes & \tableyes & \tableyes & \tableno/\tableno & \tableno/\tableno & \tableno/\tableno & \tableno \\ 
\hline
{\bf No Feedback or SF} & \tableno & \tableyes & \tableno & \tableno/\tableno & \tableno/\tableno & \tableno/\tableno & \tableno \\ 
{\bf No Heating}  & \tableyes & \tableyes & \tableyes & \tableyes/\tableyes & \tableno/\tableyes & \tableno/\tableyes & \tableno \\ 
\hline
{\bf No Rad.\ Mom.} & \tableyes & \tableyes & \tableyes & \tableno/\tableno & \tableyes/\tableyes & \tableyes/\tableyes & \tableyes \\ 
{\bf Rad.\ Mom.\ Only} & \tableyes & \tableyes & \tableyes & \tableyes/\tableyes & \tableno/\tableno & \tableno/\tableno & \tableno \\ 
\hline
{\bf No Stellar Wind} & \tableyes & \tableyes & \tableyes & \tableyes/\tableyes & \tableyes/\tableyes & \tableno/\tableno & \tableyes \\ 
{\bf No HII Heating} & \tableyes & \tableyes & \tableyes & \tableyes/\tableyes & \tableyes/\tableyes & \tableyes/\tableyes & \tableno \\ 
\hline
{\bf Molecular Model A} & Mod & \tableyes & \tableyes & \tableyes/\tableyes & \tableyes/\tableyes & \tableyes/\tableyes & \tableyes \\ 
{\bf Molecular Model B} & \tableyes & \tableyes & \tableyes & \tableyes/\tableyes & \tableyes/\tableyes & \tableyes/\tableyes & \tableyes \\ 
{\bf Molecular Model C} & Mod & Mod & \tableyes & \tableyes/\tableyes & \tableyes/\tableyes & \tableyes/\tableyes & \tableyes \\ 
\hline
{\bf Enforced SED (\S~A)} & \tableyes & \tableyes & \tableyes & \tableyes/Mod & \tableyes/\tableyes & \tableyes/\tableyes & \tableyes \\ 
\hline
{\bf \paperone\ Simulations} & Mod & \tableyes & \tableno & \tableyes/\tableno & \tableno/\tableno & \tableno/\tableno & \tableno \\ 
\hline\hline
}
\end{footnotesize}

\vspace{-0.2in} 
\section{The Simulations}
\label{sec:sims}

The initial conditions and numerical details of the 
simulations used here are described in detail in 
\paperone\ (see their Section~2 and Tables~1-3). 
However we briefly summarize some 
basic properties of the models here. 
The simulations were performed with the parallel Tree-SPH code {\small
  GADGET-3} \citep{springel:gadget}.  They include stars, dark matter, and gas, 
with cooling, shocks, star formation, and stellar feedback. 

Gas follows a standard atomic cooling curve but in addition can cool to 
$<100\,$K via fine-structure cooling. This allows it to collapse to very high 
densities, and star formation occurs in dense regions above a threshold $n>1000\,{\rm cm^{-3}}$, 
with a rate $\dot{\rho}_{\ast}=\epsilon\,\rho/t_{\rm ff}$ 
where $t_{\rm ff}$ is the free-fall time and $\epsilon=1.5\%$ is an efficiency 
taken from observations of star-forming regions with the same densities 
\citep[][and references therein]{krumholz:sf.eff.in.clouds}. 
In \paperone\ we show that the equilibrium structure and 
SFR are basically independent of the small-scale star formation law 
(robust to order-of-magnitude variations in the 
threshold and efficiency, and changes in the power-law index 
$\dot{\rho}\propto \rho^{1-2}$) 
because it is feedback-regulated (so clumps simply form stars until sufficient
feedback halts further collapse). 

The feedback models are implemented in four distinct initial disk models, designed to 
represent a range of characteristic galaxy types. Each initial disk has a 
bulge, stellar and gaseous disk, halo, and central BH (although to isolate the 
role of stellar feedback, models for BH growth and feedback are disabled); 
they are initialized in equilibrium so that in the absence of cooling, star formation, and 
feedback there are no significant transients. The gaseous disk is initially 
vertically pressure-supported, but this cools away in much less than a dynamical 
time and the emergent vertical structure depends on feedback (changing 
the initial support to turbulent dispersion does not change our results). Our 
``low'' resolution runs (used to evolve the simulations for 
several Gyr, to ensure steady-state behavior) use $\approx3\times10^{6}$ particles, with 
$\approx10^{6}$ particles in the disk, giving SPH smoothing 
lengths of $\sim10$\,pc in the central few kpc of a MW-like disk (scaling 
linearly with the disk size/mass scale). 
Our ``standard'' resolution cases use $\sim30$ times as many particles, 
and correspondingly reach up to $\sim1-5\,$pc smoothing lengths 
and particle masses of $500\,\msun$, and are run for a few orbital times each. A couple of ultra-high resolution 
runs used for convergence tests employ 
$\sim10^{9}$ particles, with sub-pc resolution on kpc scales. 

The disk models include: 

(1) SMC: an SMC-like dwarf, with baryonic mass $M_{\rm bar}=8.9\times10^{8}\,\msun$ 
and halo mass $M_{\rm halo}=2\times10^{10}\,\msun$ (concentration $c=15$), 
a \citet{hernquist:profile} profile bulge with mass $m_{b}=10^{7}\,\msun$, and exponential 
stellar ($m_{d}=1.3\times10^{8}\,\msun$) and gas disks ($m_{g}=7.5\times10^{8}\,\msun$) 
with scale-lengths $h_{d}=0.7$ and $h_{g}=2.1$\,kpc, respectively. 
Table 1 gives the maximum circular velocity. The initial scale-height is $z_{0}=140$\,pc and the stellar disk is initialized such that the 
Toomre $Q=1$ everywhere. 

(2) MW: a MW-like galaxy, with 
halo $(M_{\rm halo},\,c)=(1.6\times10^{12}\,\msun,\,12)$, 
and baryonic 
$(M_{\rm bar},\,m_{b},\,m_{d},\,m_{g})=(7.1,\,1.5,\,4.7,\,0.9)\times10^{10}\,\msun$ 
with scale-lengths 
$(h_{d},\,h_{g},\,z_{0})=(3.0,\,6.0,\,0.3)\,{\rm kpc}$. 

(3) Sbc: a LIRG-like galaxy (i.e.\ a more 
baryon-dominated gas-rich spiral characteristic 
of those observed at low redshifts), 
with 
halo $(M_{\rm halo},\,c)=(1.5\times10^{11}\,\msun,\,11)$, 
and baryonic 
$(M_{\rm bar},\,m_{b},\,m_{d},\,m_{g})=(10.5,\,1.0,\,4.0,\,5.5)\times10^{9}\,\msun$ 
with scale-lengths 
$(h_{d},\,h_{g},\,z_{0})=(1.3,\,2.6,\,0.13)\,{\rm kpc}$. 

(4) HiZ: a high-redshift massive starburst disk, chosen to match the 
properties of the observed non-merging but massively star-forming SMG 
population, 
with 
halo $(M_{\rm halo},\,c)=(1.4\times10^{12}\,\msun,\,3.5)$ and a 
virial radius appropriately rescaled for a halo at $z=2$ rather than $z=0$, 
and baryonic 
$(M_{\rm bar},\,m_{b},\,m_{d},\,m_{g})=(10.7,\,0.7,\,3,\,7)\times10^{10}\,\msun$ 
with scale-lengths 
$(h_{d},\,h_{g},\,z_{0})=(1.6,\,3.2,\,0.32)\,{\rm kpc}$. 

The salient galaxy properties are summarized in Table~\ref{tbl:sim.ics}, 
reproduced from \paperone. The feedback and ISM physics we include in different runs is 
summarized in Table~\ref{tbl:sims}.

\subsection{Stellar Feedback}
\label{sec:stellar.fb}

\subsubsection{Local Momentum Flux from 
Radiation Pressure, Supernovae, \&\ Stellar Winds}
\label{sec:stellar.fb:mom}

First, we consider the radiation pressure produced on a sub-GMC scale from 
young star clusters. This feedback mechanism is that presented and discussed 
extensively in \paperone, but we briefly summarize its properties here. 
At each timestep, gas particles identify the nearest density 
peak (via a friends-of-friends 
algorithm averaged over a scale of a couple smoothing lengths), representing 
the center of the nearest star-forming ``clump'' or GMC-analog. 
The gas then lies at some point ${\bf R}_{\rm cl}$ from the cloud center. 
The summed luminosity from the simulation star particles inside 
the radius $L(<{\bf R}_{\rm cl})$ is determined from each star's known age 
and metallicity according to the STARBURST99 model \citep{starburst99}.
Assuming this is equally distributed to all gas inside the cloud, 
the rate of momentum deposition is then just 
$\dot{P}_{\rm rad}\approx (1+\tau_{\rm IR})\,L_{\rm incident}/c$, 
where the term 
$1+\tau_{\rm IR}$ accounts for the fact that most of the initial optical/UV radiation is 
absorbed and re-radiated in the IR; $\tau_{\rm IR}=\Sigma_{\rm gas}\,\kappa_{\rm IR}$ 
is the optical depth in the IR, which allows for the fact that the momentum is boosted 
by multiple scatterings in optically thick regions. 
Here $\Sigma_{\rm gas}$ is calculated as the local optical depth 
towards the identified clump, with $\kappa_{\rm IR}\approx5\,(Z/Z_{\sun})\,{\rm g^{-1}\,cm^{2}}$ approximately 
constant over the relevant physical range of dust temperatures. 
The imparted acceleration is directed away from the clump, along ${\bf R}_{\rm cl}$. 
There is {\em no} ``free-streaming'' or ``turnoff'' of cooling or other physics for 
particles affected by the feedback. 
For a detailed discussion of the normalization, resolution effects, 
effects of photon ``leakage,'' 
and choice of whether the momentum is applied as a continuous 
acceleration or discrete ``kicks,'' 
we refer to \paperone, but note that essentially all of 
our conclusions are robust to these and other 
small variations in the model implementation. 

There is also direct momentum injection from supernovae, stellar 
winds, and protostellar jets. We include this in the same manner, but with the appropriate 
momentum injection rates 
using $\dot{P} = \dot{P}_{\rm rad} + \dot{P}_{\rm SNe} + \dot{P}_{\rm w}$, 
where $\dot{P}_{\rm SNe}(t,\,Z)$ and $\dot{P}_{\rm w}(t,\,Z)$ are tabulated as a 
function of time and metallicity for each star particle from STARBURST99. 
We emphasize that this is the direct momentum from ejecta only -- 
where SNe are important, they typically impart much greater momentum from 
pressure-driven expansion of hot regions, which we wish to model directly. 
The direct ejecta momentum of SNe and stellar winds are usually 
small relative to the radiation pressure (they have a magnitude 
$\sim L/c$, but do not receive the ``boost'' of $\sim \tau_{\rm IR}$, which is 
typically at least a few in GMCs). 
The contribution from protostellar jets can in principle also be added explicitly, but when 
integrated over the IMF, their contribution is $\sim10-20\%$ that from stellar winds alone; protostellar jets are probably important in regions of low-mass star formation, but they have little global 
effect relative to the contribution of massive stars.

\subsubsection{Supernova Shock-Heating}
\label{sec:stellar.fb:sne}

In addition to providing momentum in ejecta, the gas shocked by 
supernovae can be heated to high temperatures, generating bubbles and 
filaments of hot gas; the resulting over-pressurized regions can then 
hydrodynamically accelerate nearby gas. The model in 
\S~\ref{sec:stellar.fb:mom} already  accounts for the direct
{\em momentum} of SNe ejecta; but not necessarily the full post-shock 
heating/energy budget (or the momentum generated by 
the pressure-driven expansion of bubbles). We therefore add an explicit model for this 
thermal heating. 

We know the age, mass, and metallicity for each star particle in the simulation, so
for a given timestep $t\rightarrow t+\Delta t$, we can calculate the 
total mechanical energy produced by Type-II SNe explosions. We do so using
a pre-tabulated $E(t)$ \citep[which we compile from the STARBURST99 
package, for a \citealt{kroupa:imf} IMF;][]{starburst99}. 
This gives an expected input $\Delta E$. 
However, on very short time/mass scales SNe are discrete events with a 
roughly fixed individual energy, so we actually determine the expected number of SNe per star particle in a given timestep assuming a canonical $E_{0}=10^{51}\,{\rm erg}$ per SNe, 
and then randomly determine the number of SNe in the timestep drawing from a Poisson distribution with the calculated expectation value; this ensures the total SNe energy comes out correctly. In fact, the time steps in our simulations are typically short enough ($\sim100-1000$\,yr) to resolve individual SNe events. We also include SNe Type-I, for which we adopt an empirical fit to a simple ``prompt+delayed'' model fitted to the observed SNe rates in \citet{mannucci:2006.snIa.rates}:   the prompt component is a narrow Gaussian at a few $10^{7}$\,yr, and the delayed component has a constant rate.
However, 
since the systems we consider here are all star-forming, Type-1 SNe make little 
difference and the energy input is  dominated by the Type-II population.

We wish to explicitly resolve the energy-conserving (Sedov) 
phase of the resulting SNe expansion. We therefore directly 
deposit the SNe energy in the local 
SPH smoothing kernel (nearest $\sim32$ neighbors), weighted by the 
kernel -- essentially the smallest 
meaningful length in the simulation. In this case, there is no implicit or sub-grid assumption about 
even the earliest SNe expansion.\footnote{Some other simulations \citep{katz:treesph,stinson:2006.sne.fb.recipe} inject the energy instead into a larger 
``blastwave radius'' around star particles, 
with  
\begin{equation}
R_{E} = 55\,E_{51}^{0.32}\,n_{0}^{-0.16}\,\tilde{P}_{04}^{-0.20}\,{\rm pc}
\end{equation}
\citep{chevalier:1974.sne.breakout.conditions,mckee.ostriker:ism}, 
where $E_{51}=\Delta E/10^{51}\,$erg, $n_{0}$ is the density in cgs, 
and $\tilde{P}_{04}=10^{-4}\,P/[{\rm erg\,cm^{-3}}]\,k_{\rm B}^{-1}$, 
intended to reflect the radius of the energy-conserving (Sedov) phase of the blastwave 
(often much larger than our softening length). 
Because we wish to resolve this expansion as opposed to assuming it in 
sub-grid fashion, we do not use this approximation. However, we have 
re-run our simulations using this approximation instead of coupling the 
energy inside of $h_{\rm sml}$.
For some individual blastwaves it can make a significant difference 
to their expansion history, but on average the behavior 
is similar where the densities are low and bubbles can expand.
The final statistical properties of our simulations do not depend on the injection radius 
within a modest range, 
either with this formula or adopting fixed physical radii $\sim1-10\,$pc 
and/or fixed neighbor number $=16-320$.}

Unlike most models of supernova feedback, we do {\em not} 
artificially turn off gas cooling after injecting thermal energy into the ambient gas. 
Because other feedback processes (radiation pressure, photoionization heating, etc.) typically expel the dense gas  in star-forming regions back out into the more diffuse ISM prior to when SNe explode, the effects of SNe on the ISM are significantly larger than in models that only include thermal SNe feedback.

\subsubsection{Gas Return and Stellar Wind Shock-Heating}
\label{sec:stellar.fb:gasreturn}

Stellar evolution returns gas to the ISM from SNe and stellar winds. 
This is implemented following the methodology of
\citet{torrey:2011.metallicity.evol.merger}, with a few minor modifications. 
Every star has a known age, mass, and metallicity; 
for a given timestep $\Delta t$  we  
determine the total mass-loss rate of the particle (treated as a single 
stellar population) $\dot{M}(t)$ 
(tabulated from STARBURST99 for both SNe and 
stellar wind mass losses assuming a Kroupa IMF), 
which gives an expected recycled mass $\Delta M = \dot{M}(t)\,\Delta t$. 
This mass is then distributed among local gas particles in the SPH smoothing 
kernel, weighted by the kernel. 
Gas recycled in SNe ejecta is returned discretely only in SNe (to the same 
gas heated by those events); gas recycled by stellar winds is returned 
continuously (in every time step).\footnote{Models in which the gas is 
stochastically returned by allowing an individual star particle to be 
turned back into a gas particle (``recycled'') with probability $p=\Delta M/m_{i}$
give nearly identical results, but are subject to greater shot noise, especially 
on sub-GMC scales.}
Integrated over 
a Hubble time, the total mass fraction returned according to the 
tabulated models is $\approx0.3$. 

The injected/recycled gas  has the same gross properties (momentum, 
location, etc) as the star from which it spawns. However, the initial thermal 
state must be specified. For the SNe ejecta the gas carries the ejecta energy 
discussed above.
We assume that the stellar wind gas shocks, so that it has an initial 
specific energy (and corresponding entropy/temperature) set by the wind properties as a function of time.
This is again tabulated from STARBURST99 as a function of age and 
metallicity. Integrated over their history, the winds have an energy-flux-weighted 
average temperature $T_{w}\approx 3\times10^{7}\,K$, reflecting the fact that 
most of the energy is in early ``fast'' winds with speeds 
$\gtrsim 500\,{\rm km\,s^{-1}}$; at late times, however, the energies and temperatures are much lower because AGB winds with 
velocities $\sim 10\,{\rm km\,s^{-1}}$ dominate.
As before, we do not modify the cooling or hydrodynamics of these gas particles. 

\subsubsection{Photoionization-Heating of HII Regions}
\label{sec:stellar.fb:HII}

On small scales around young stars, 
direct photoionization-heating by ionizing photons 
maintains gas in a warm $\sim10^{4}\,$K 
ionized state, which can in principle be an important source of pressure 
in low-density systems. 
A rigorous solution for the effects of 
this heating would involve on-the-fly radiation transport.   This is probably important for the 
study of the formation of the first stars and for detailed studies of GMC disruption in lower mass molecular clouds, but is more detailed than is possible given the goals of the feedback model we are developing.      Moreover, as we shall show, photoionization feedback is important, but not dominant, for  the more massive, evolved systems we model.   We approximate the key effects of photoionization on ambient gas with the following simple model. 

Each star particle produces ionizing photons at a rate 
$\dot{N}_{i}$, which we again tabulate as a function of the
age and mass of the star particle from the STARBURST99 models. 
We then allow these ionizing photons to ionize a region around the star as follows. 
First consider the gas particle nearest the star. If it is already ionized (either 
because it is at sufficiently high temperature $>10^{4}\,K$ or because it is already a 
part of another HII region), we move to the next nearest gas particle. If it is 
not ionized, then we calculate the rate of ionizing photons needed to maintain the 
gas as fully ionized, as $\Delta \dot{N} = N(H)_{j}\,\beta\,n_{e}$, where $N(H)_{j}=M_{j}/\mu\,m_{p}$ 
is the number of atoms in the particle, $\beta\approx3\times10^{-13}\,{\rm cm^{3}\,s^{-1}}$ 
is the recombination rate coefficient, and $n_{e}$ is the average election 
density of the particle when fully ionized. 
If $\dot{N}_{i}>\Delta \dot{N}_{j}$, 
the particle is tagged  
as being within an HII region, and the remaining photon production rate is 
tabulated as $\dot{N}_{i}\rightarrow \dot{N}_{i}-\Delta \dot{N}_{j}$. 
The process is then repeated for the next nearest particle. 
Once we reach a point where $\dot{N}_{i}>0$ and $\dot{N}_{i}<\Delta \dot{N}_{j}$, 
we determine whether or not to ionize the particle randomly, 
with probability $p=\dot{N}_{i}/\Delta \dot{N}_{j}$, and consume the remaining 
photon budget, ending the chain. 

This procedure  determines the Stromgren sphere around the particle, 
but allows for the resolved inhomogeneous density distribution and overlapping 
ionized regions. In low-density regions, the ionized material can extend to very large radii; because we also account for a uniform extragalactic ionizing background, we 
also truncate the ionizing region if we reach a radius at which the 
flux density from the ionizing source falls below the ionizing background 
(for background flux density $u_{\nu}$, this is the $r$ where $L_{\nu}/4\pi\,r^{2}\,c=u_{\nu}$). 
Gas particles flagged as being within an HII region are not allowed to cool below 
an effective temperature $t_{\rm HII}\approx 10^{4}\,$K in that timestep (i.e.\ until they 
are flagged as outside of such a region), and are immediately heated to 
$t_{\rm HII}$ if their temperature lies below this value. 

Although the HII regions estimated for individual star particles can be quite small 
(as the stellar masses can be as low as $100\,\msun$), this method allows for nearby star 
particles to generate overlapping HII regions, creating large HII bubbles of 
sizes $\sim0.1-1\,$kpc, similar to what is observed in many galaxies. 
Of course, if the densities around stars are much higher than what is 
resolved, this model could artificially over-ionize the surrounding region; 
however, given our ability to resolve dense clumps with 
$n>10^{6}\,{\rm cm^{-3}}$ (inside of which the Stromgren radii are negligible), we 
do not see much resolution dependence until we go to quite low resolutions 
(lower than our ``standard'' cases).   Moreover,  a non-trivial fraction of photons escape their birth environments, particularly late in the life of the most massive stars, as they disrupt their natal GMCs.

\subsubsection{Long-Range Radiation Acceleration}
\label{sec:stellar.fb:continuous.accel}

Stellar feedback on small scales can disrupt GMCs via the mechanisms summarized above. On larger scales, galactic-scale outflows can be continuously accelerated as well. 
There are two basic mechanisms that can do this. First, pressure 
forces from hot gas can drive outflows. This should be captured explicitly in our modeling, as part of the hydrodynamics, given the heating from photoionization, stellar winds, and supernovae.   In addition, 
gas can be accelerated outwards by the net radiation pressure produced by all of the stars in the galaxy.   (Cosmic rays can also contribute to such acceleration but are not modeled here.)

In principle, the physics of radiative acceleration by the cumulative stellar radiation field is very similar to that associated with the local radiation pressure force  discussed in \S~\ref{sec:stellar.fb:mom}; but in practice, the model in \S~\ref{sec:stellar.fb:mom} reflects the acceleration of gas particles by radiation pressure from only the nearest star cluster.   These are often dense regions in which even the infrared photons are trapped so that the {\em local} momentum deposition is $\sim \tau_{\rm IR}\,L/c$.   

Properly accounting for the full radiative acceleration by all of the stars is a formidable calculation, in particular because of the possibility of diffusion/scattering in the infrared.  Here we make the approximation that the dominant acceleration on relatively large scales (ie., outside of the dense environments of star clusters and their natal GMCs) is due to the optical and UV photons, whose flux we can estimate using simple attenuation along the line of sight between the gas and star particles.  Altogether, the total momentum imparted via this absorption/scattering will often be 
less than the momentum imparted locally from the trapped IR photons in the 
star-forming cloud, but the escaping UV/optical photons have the ability to 
accelerate gas at arbitrarily large distances from their origin.     Moreover, gas above the midplane of the stellar disk "sees" the entire surface flux from the disk.    Thus the light from many star clusters acts together, and can produce a large vertical acceleration, in principle launching the gas parcel out of the galaxy entirely.

For a single stellar source the flux incident on a given SPH particle is  \begin{equation}
{\bf F} = \frac{L}{4\,\pi\,r^{3}}\,{\bf r}
\end{equation}
where ${\bf r} = {\bf r}_{\rm gas}-{\bf r}_{\rm source}$ and we have temporarily ignored absorption and re-radiation.   This radiation flux corresponds to a momentum flux per unit area of $d{\bf P}/dA\,dt = {\bf F}/c$. 
This is incident on a mass $m_{i}$, 
area $\pi\,h_{i}^{2}$ and surface density 
$\Sigma_{i}=m_{i}/[\pi\,h_{i}^{2}]$ (here $h_{i}=1.07\,N_{\rm sml}^{-2/3}\,h_{\rm sml}$ 
is related to the SPH smoothing 
kernel, assuming all $N_{\rm sml}$ particles in the kernel are distributed uniformly over the 
volume and averaging over the kernel mass profile). 
The  momentum flux imparted to the gas
is just the area times the fraction of the flux absorbed or 
scattered per unit area, i.e.\ ${\bf \dot{P}}_{i}=(d{\bf P}/dA\,dt)\,\pi\,h_{i}^{2}\,(1-\exp{(-\kappa\,\Sigma_{i})})$. 
The acceleration is then $m_{i}\,{\bf a}={\bf \dot{P}}_{i}$, or 
\begin{equation}
\label{eqn:rad.accel}
{\bf a} =  \frac{L}{4\,\pi\,r^{3}\,c}\,{\bf r}\, \frac{\pi\,h_{i}^{2}}{m_{i}}\,(1-\exp{(-\kappa\,\Sigma_{i})})
\end{equation}
This generalizes to a range of wavelengths using:
\begin{equation}
L\,(1-\exp{(-\kappa\,\Sigma_{i})}) \rightarrow \int{L_{\nu}\,(1-\exp{(-\kappa_{\nu}\,\Sigma_{i})})\,{d\nu}}
\end{equation}
If the column through any individual particle is 
optically thin (typical at high resolution), this simply reduces to 
the standard optically thin force term $d{\bf P}/dt = \kappa\,F/c$, and 
if the column is optically thick, it reduces to the assumption that all radiation incident 
on the surface $\delta A = \pi\,h_{i}^{2}$ is absorbed.

In general, $L_{\nu}$ seen by a given gas particle is not simply the un-absorbed spectrum from each star. It is, however, prohibitive to solve the full radiative transfer equations on the fly.   The key approximation we make is to  assume that most of the absorption and 
re-emission happens in the vicinity of each source. The optical depth towards a star scales as $\tau \propto \int \rho\,d{\ell} \sim \Sigma$; for realistic density 
profiles the optical depth is often dominated by the contribution from  small radii near the star, justifying our local approximation.

In this limit, we can replace $L_{\nu}$ in equation~\ref{eqn:rad.accel} 
with an effective $\tilde{L}_{\nu}$, which represents the emission of a given source after 
 local absorption and reprocessing has been accounted for. The revised $\tilde{L}_{\nu}$ is modified by
 local radiative transfer effects on the SED and by the fact that reprocessing alters the effective size and geometry of the emitting region. To represent this geometric effect, we spread the emission out over 
a region of length scale $\epsilon_{L}$ with the softening kernel 
(we choose $\epsilon_{L}$ to be the maximum of the local gas smoothing 
or gravitational softening, so it represents both the local gas distribution and 
implicit stellar mass distribution). We have checked that this approximation, in terms of the {\em locations} of the flux origin, gives good agreement with 
the results of a full radiative transfer calculation including 
scattering and re-emission performed with SUNRISE \citep{jonsson:sunrise.release}. 

The reprocessing and attenuation also effects the emergent SED of a given source $j$ via:
\be
\label{eqn:attenuation}
\tilde{L}_{\nu}^{j} = L_{\nu}^{j}\,\exp{(-\kappa_{\nu}\,\Sigma_{\rm column}^{j})}
\ee
where $L_{\nu}^{j}$ is the intrinsic (un-obscured) spectrum from the 
stellar population (tabulated as a function of age 
from the STARBURST99 models), 
and $\Sigma_{\rm column}$ is the total column density (integrated to the absorbing particle) 
from the source. We show in \S~\ref{sec:appendix:num.longrange.calibration} that we can 
approximate the total column reasonably well by 
\be
\label{eqn:columnproxy}
\langle \Sigma_{\rm column}^{j} \rangle
\equiv \rho_{j} \,h_{\rm eff}
\approx \rho_{j}\,{\Bigl[}\tilde{\alpha}\,h_{j} + \frac{\rho_{j}}{|\nabla{\rho}_{j}|}{\Bigr]}
\ee
where $\rho_{j}$ is the local gas density evaluated at the stellar source, $h_{j}$ is the local smoothing 
length, and 
the characteristic scale-height of the density is given by ${\rho_{j}}/{|\nabla{\rho}_{j}|}$. 
Note that this second term makes the result for $\Sigma_{\rm column}$ independent of resolution.   If
the gas density around each source fell off with a smooth profile 
equation \ref{eqn:columnproxy} would be exact, with $\tilde{\alpha}\sim1$ a geometric factor that 
depends on the density profile shape (we adopt $\tilde{\alpha}=0.95$ which 
corresponds to a linear density profile smoothed over the SPH smoothing kernel).   

Because we are only interested in the integrated absorbed momentum,  it is not necessary to treat $L_{\nu}$ as a finely-resolved spectrum. 
Instead, we integrate the SED over three broad frequency intervals 
in the UV ($\lambda<3600\,$\AA), 
optical/near-IR ($3600\,$\AA$<\lambda<3\,\mu$), 
and mid/far-IR ($\lambda>3\,\mu$). 
The ``initial'' luminosity of a star particle $L_\nu$ is 
given by the corresponding $L_{\rm UV}$, $L_{\rm Opt}$, $L_{\rm IR}$ 
tabulated from the STARBURST99 models as a function of 
age and metallicity. 
The opacities $\kappa_{\nu}$ are then 
replaced by the corresponding flux-mean opacities 
$\kappa_{\rm UV}$, $\kappa_{\rm Opt}$, $\kappa_{\rm IR}$, 
which we take to be constant 
$=(1800,\,180,\,5)\,Z/Z_{\sun}$, respectively.\footnote{In detail, the 
effective $\kappa$  for each band will depend on the exact SED 
shape, but we choose these frequency intervals in part because 
the resulting $\kappa$ are insensitive (changing 
by $<10\%$) to the spectral shape 
for a wide range of stellar population ages $1-1000\,$Myr and 
degrees of dust reddening. 
The specific values used here are the average flux mean opacity of a solar-composition 
dust distribution in full radiative transfer calculations of the 
galaxy types modeled here (see Appendix~\ref{sec:appendix:num.longrange}).}  In general, the opacity will also be a function of the thermodynamic state of the gas, e.g., because dust can be destroyed by sputtering in hot gas.  These effects are not taken into account in our model.

Given the above model, it is relatively straightforward to implement the resulting radiation force on all of the gas in the simulation. Because 
the acceleration follows an inverse-square law with 
sources of luminosity $L$ instead of mass $M$, 
the computation of the acceleration is implemented with 
exactly the same tree structure as the existing gravity tree. 
Of course the force acts only on gas particles, with 
the final normalization coefficient including the terms dependent on 
particle properties (eq.~\ref{eqn:rad.accel}). 

In Appendix~\ref{sec:appendix:num.longrange} we present several tests 
and calibration studies of this model.  For example, we show that
the approximate treatment just described recovers the results of full (post-processing) radiative transfer 
calculations of the SED reasonably well. We also show that if we were to simply model the radiation field as that empirically observed in typical galaxies of 
each type we consider, we obtain very similar results for the momentum absorbed.

We stress that our model of the acceleration by the diffuse radiation field does 
not ``double count'' when combined with the 
radiation pressure included locally in \S~\ref{sec:stellar.fb:mom}; 
the radiation pressure term there was $(1+\tau_{\rm IR})\,L/c$. 
The factor ``1'' comes from the initial extinction applied here 
in equations \ref{eqn:attenuation} and  \ref{eqn:rad.accel}, 
$\exp{(-\kappa_{\nu}\,\Sigma_{\rm column}^{j})}$, i.e.,\ the 
local absorption of the initial UV/optical photons. In equation (\ref{eqn:rad.accel}), this factor does not add momentum to the gas,
rather it attenuates the ``initial'' spectrum---effectively reducing the momentum carried by the UV photons emitted by young stars. It is true 
that we slightly over-count when we approximate the absorbed fraction of un-reprocessed radiation by unity in \S~\ref{sec:stellar.fb:mom}, particularly if the actual optical depth 
$\kappa_{\nu}\,\Sigma_{\rm column}^{j} < 1$.   However, since the dust opacity to the initial radiation (which is mostly in the optical/UV) is large, this is rarely the case for the models we consider.  In fact, we have implemented a more accurate model that allows for UV optically thin sight lines in the `local' (to star clusters) radiation force calculation,
and find that it produces no measurable differences. 

\vspace{-0.2cm}
\subsection{Molecular Chemistry}
\label{sec:chemistry}

Another potentially important element in modeling star formation is the determination 
of where gas can or cannot become molecular. Stars form only from 
molecular gas -- albeit 
not all molecular gas, only that at the very highest densities.
We have therefore also implemented several models 
which allow for explicit estimates of the molecular gas fraction $f_{H_{2}}$ at each point 
in the simulation. 

In \S~\ref{sec:appendix.chemistry} we compare three different models 
for the molecular fraction $f_{H_{2}}$ and molecular chemistry: 

{Molecular Model A}: 
assume $f_{H_{2}}=1$ always in regions above the SF density threshold.

{Molecular Model B}: 
the model in \citet{krumholz:2011.molecular.prescription}, which gives a simple fitting function for $f_{H_{2}}$ as a function of the local column density ($\approx\rho\,h_{\rm sml}$) and metallicity.

{Molecular Model C}: 
the model in \citet{robertson:2008.molecular.sflaw}, where both the molecular fraction $f_{H_{2}}$ and the cooling function $\Lambda(T)$ at temperatures $\lesssim10^{4}\,$K are varied as a function of the local density, metallicity, and radiation field to match the results of CLOUDY calculations. 

In each case we allow only molecular gas to form stars by multiplying the calculated SFR in each particle by $f_{H_{2}}$. 

We show in \S~\ref{sec:appendix.chemistry}  that the choice of models (A)-(C) 
has no effect on our results. 
This is because we only allow star formation in very dense regions 
(above a threshold $n>100-1000\,{\rm cm^{-3}}$), and the 
galaxies we model here are already substantially metal-enriched 
(even the SMC model has $Z\sim0.1\,Z_{\sun}$). 
Thus the more complex models ultimately return the result that the 
gas in potential star-forming regions
is overwhelmingly molecular -- typical optical depths 
to photo-dissociating radiation are $\sim10^{4}$, so they are 
strongly self-shielding. 

At sufficiently low metallicities, however, $Z \lesssim 10^{-3}-10^{-2}\,Z_{\sun}$, molecule formation 
will be inefficient even at high densities and cooling times will become 
comparable to dynamical times. The formation of the first stars, and 
the lowest-mass dwarf galaxies (galaxy stellar masses $<10^{6}\,\msun$),
are in this regime, where the explicit treatment of the molecular fraction above may be important 
 \citep{kuhlen:2011.mol.reg.h2.dwarfs}.

\vspace{-0.4cm}
\section{Galaxy Morphologies}
\label{sec:morph}

\subsection{With Feedback} 
\label{sec:morph:all}

Figures~\ref{fig:morph.hiz}-\ref{fig:morph.smc} show 
the morphologies of each galaxy model, 
in our standard case where all feedback mechanisms are present. 

We show both the gas and stars, face-on and edge-on, at times 
after the systems have reached a quasi-steady state equilibrium. 
The gas maps show the projected gas density (intensity) 
and temperature (color, with blue representing cold molecular gas at 
$T\lesssim 1000\,$K, pink representing the warm ionized gas at $\sim10^{4}-10^{5}$\,K, 
and yellow representing the hot, X-ray emitting gas at $\gtrsim10^{6}\,$K. 
The stellar maps show a mock three-color observed
image, specifically a $u/g/r$ composite. 
The stellar luminosity in each band is calculated from each star particle 
according to the STARBURST99 model given its age, mass, and metallicity 
(and smoothed over the appropriate kernel). We then attenuate the stars 
following the method of \citet{hopkins:lifetimes.letter}: we calculate the total dust 
column (from the simulated gas) along the line-of-sight to each star particle 
for the chosen viewing angle (assuming a constant 
dust-to-metals ratio, i.e.\, dust-to-gas equal to the MW value 
times $Z/Z_{\sun}$), and apply a MW-like extinction and 
reddening curve \citep[as tabulated in][]{pei92:reddening.curves}. 

The dramatic differences between the different galaxy models are obvious.
The HiZ models result in disks with massive ($>10^{8}\,\msun$) 
$\sim$kpc-scale clumps, which form massive star cluster complexes. 
The gas densities are sufficiently high that extinction can lead to a clumpy optical 
morphology as well (by entirely extincting out regions in the galaxy). 
The clumpiness is more prominent in the gas, but remains visible in the optical 
images. Seen edge-on, the gas can resemble a ``clump-chain'' morphology. 
Violent outflows arise from throughout the disk, driven by a massive starburst 
with SFR $>100\,\msun\,{\rm yr^{-1}}$. 

The Sbc models also produce clumpy disk structure, but the individual 
clumps and star forming regions are relatively much smaller, and the 
global structure more clearly traces spiral arms and global gravitational 
instabilities. As a low-mass starburst (the galaxy has a SFR of $\sim1-5\,\msun\,{\rm yr^{-1}}$, 
despite being an order of magnitude less massive than the MW), the irregular gas morphology and thick disk 
more closely resemble dwarf starbursts in the local Universe such as 
NGC 1569 or 1313. A clumpy, multi-phase super-wind is clearly evident, arising from the 
high specific star formation rate. 

The MW models have a familiar spiral and barred spiral morphologies. But 
with our high resolution and realistic modeling of the ISM, individual star clusters 
and structure within the spiral arms -- dust lanes, feathering, and 
SNe-powered bubbles are evident. Edge-on, the dust lanes and moderate/weak 
outflows of hot gas are also evident, but without the violence of the HiZ case. 

The SMC models resemble typical irregular galaxies. 
The ISM is extremely patchy, with distributed molecular clouds and large, overlapping 
SNe bubbles. Edge-on, shells of cold material accelerated by winds out of the disk 
appear as blue ``loops'' at large scale heights, while direct venting of the hot gas is 
similarly apparent. The disk is ``puffy'' and thick, and the stars are distributed in 
the characteristic irregular fashion that defines the morphological class. 
The outer disk is clearly not star-forming; being low-density, this is maintained as an extended ionized 
disk by HII heating (from both the inner stellar disk and the UV background).

\subsection{Effects of Each Feedback Mechanism in Turn} 
\label{sec:morph:fbmodel}

\begin{figure*}
    \centering
    \plotside{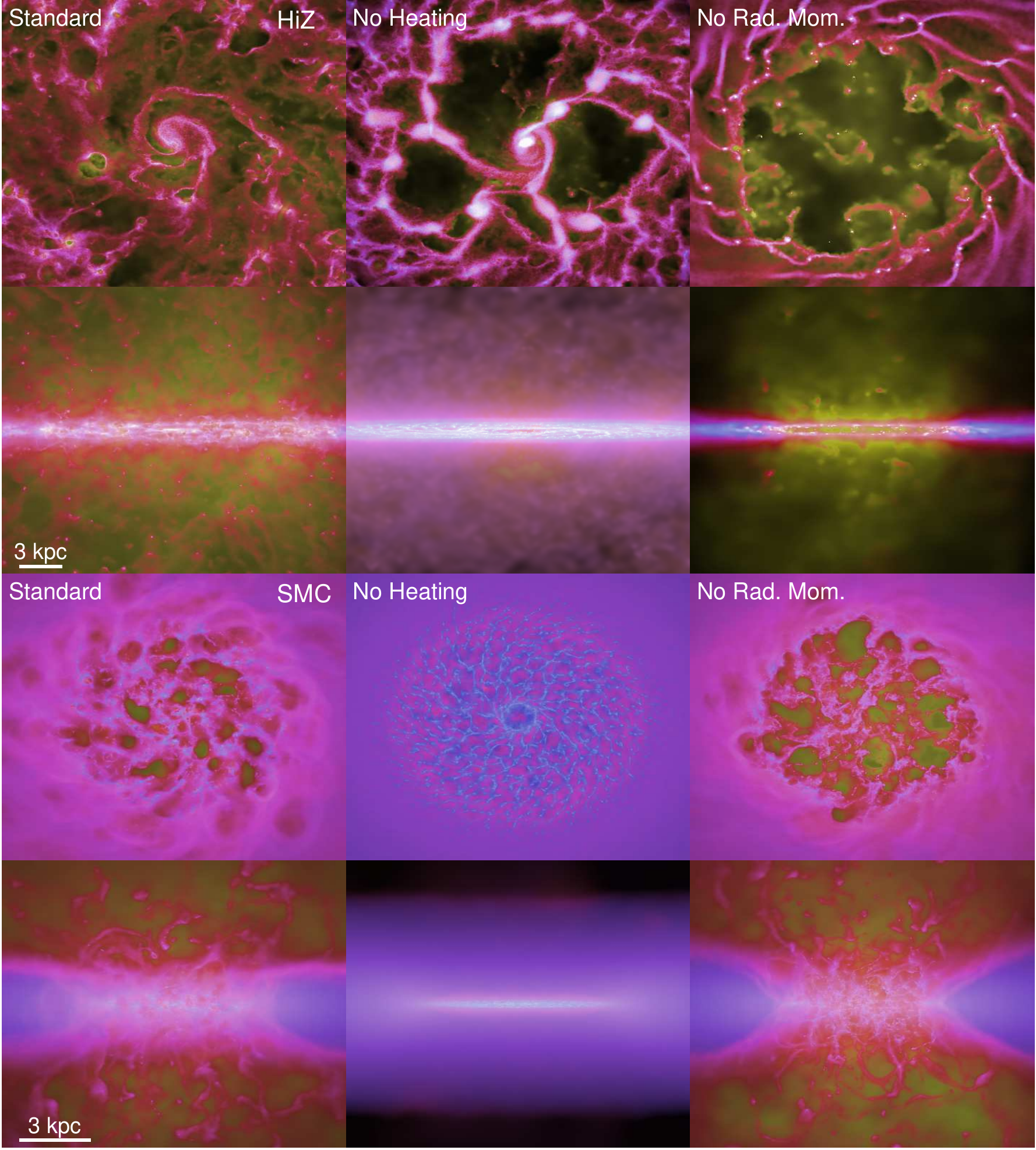}
    \caption{Face-on and edge-on gas morphologies of 
    the HiZ ({\em top}) \&\ SMC ({\em bottom}) galaxy models with different feedback mechanisms enabled and disabled;  the stretch and color scale are the same as in Figure~\ref{fig:morph.hiz}.
    {\em Left:} Standard model (all feedback on). 
    {\em Middle:} No heating: energy injection from SNe, shocked stellar 
    winds, and HII photoionization-heating is disabled (gas recycling from these 
    mechanisms remains, but is ``cold''). In the HiZ case, the ``hot'' volume-filling ISM is greatly diminished 
    but the global morphology remains similar.
    In the SMC case, the morphology is dramatically different with the heating mechanisms turned off -- there is no wind and the 
    morphology lacks the characteristic irregular/patchy structure.
    {\em Right:} No radiation pressure (local or long-range). In the HiZ case, 
    the hot medium is present, but individual GMC complexes ($\sim0.1-1$\,kpc-scale 
    white blobs at {\em left}) collapse without limit (to single pixels here) and 
    much of the cold wind mass is absent. In the SMC case, 
    GMCs and dense ionized regions also collapse further (leading to the 
    apparently larger filling factor of hot gas), but the wind and disk 
    thickness are not strongly affected. 
    \label{fig:sfhwind.morph.vs.fb.hiz}}
\end{figure*}



Having seen the typical galaxy morphologies in the case where feedback {\em is} 
present, we now consider the galaxy morphologies (and star formation 
histories, discussed later) with different feedback mechanisms de-activated in turn. 

Figure~\ref{fig:sfhwind.morph.vs.fb.hiz} shows this for the HiZ and SMC models. 
We show the morphologies (at fixed time and spatial scale) and star formation 
histories of a few models which are otherwise identical but include 
or exclude different feedback mechanisms. 
Qualitatively, in the HiZ model, the clumpy morphology is present in 
all cases -- since the galaxy is violently unstable, it is primarily gravity (not 
feedback) that dictates the gas morphology. But there are major 
differences between different feedback models. Most strikingly, 
without local momentum deposition from radiation pressure, 
the characteristic sizes of the individual clump cores are much smaller 
(this is difficult to see by eye -- the central regions of each clump that appear near-white in the other images 
are single white pixels in this case!)
This occurs because the long-range photon momentum, 
supernovae momentum, and stellar wind momentum are not sufficiently strong to 
resist the collapse of clumps beyond a critical threshold. 

The effectiveness of the radiative feedback in the HiZ model results from the 
fact that, as shown in \paperone, the IR optical depths reach $\tau_{\rm IR}\sim50$.
The photon momentum $L/c$ (neglecting the boost of $\tau_{\rm IR}$), or the 
direct SNe momentum (also $\sim L/c$) are lower than that produced by the IR 
radiation by an order of magnitude. 
Hot gas (from the SNe, stellar winds, and HII regions) is clearly insufficient to prevent collapse -- not surprising since the average density in the clumps is $>100\,{\rm cm^{-3}}$, 
so the cooling time is extremely short ($\sim100\,{\rm yr}$). 

Turning off sources of hot gas, the morphologies in the HiZ galaxy is
similar to the standard, all feedback case -- the differences being mostly in the diffuse inter-clump gas 
and diffuse (volume-filling) phase of the wind. 
This gas is much cooler (note the absence of yellow gas), and without 
SNe and stellar winds the clumps and spiral arms in the gas appear much 
smoother (the irregular edge morphology and occasional bubbles are largely due to 
SNe). But the absolute wind mass is fairly similar -- most of it is in the warm-phase gas clumps. 
Heating from HII regions is negligible -- unsurprising since $\sim10\,{\rm km\,s^{-1}}$ 
is significantly less than the turbulent velocities generated by the other sources 
of feedback. 

We do not show the Sbc and MW cases explicitly (their salient properties are shown below), 
but they are qualitatively similar to the HiZ result. In the MW-analogue without radiation pressure, 
GMCs and star-forming ``knots'' again collapse to significantly smaller sizes, 
although the runaway is not quite as severe (early ``fast'' stellar winds and HII heating are 
able to somewhat suppress runaway collapse). Removing the sources of heating 
does not much change the global gas morphology, but 
eliminates the volume-filling ``hot'' gas and leaves the ISM 
as a whole much more cold, with a larger fraction of material in dense clouds. 
The Sbc case is lower-mass than the MW but more strongly self-gravitating and 
gas rich, so lies as expected somewhere between the MW and HiZ cases in these properties.

However, when we repeat the experiment  for the SMC-like model (in the bottom half of Figure~\ref{fig:sfhwind.morph.vs.fb.hiz}), 
the results are strikingly different. The key difference from the other three types of galaxy is that the average gas density is much lower, 
$n\lesssim 0.1\,{\rm cm^{-3}}$,  approaching the regime where 
the cooling time is non-negligible even near young stars. 
Moreover, with a lower column density and metallicity, the IR optical 
depths can be much lower so photon momentum is boosted little beyond 
$\sim L/c$. Turning off the sources of photon momentum (both local and long-range) 
has a relatively weak effect on the morphology (some GMCs do collapse further, 
but many are disrupted by SNe and stellar winds). 

Turning off all of the ``heating terms'' from SNe, stellar winds, and HII photoionization 
however, has a large effect -- the volume filling 
hot gas ``bubbles'' disappear and the gas clearly collapses along more 
global gravitational structures into a filamentary morphology reminiscent of  
fragmenting very large-$m$ spiral arms, quite different from 
what is observed in real dwarf galaxies. 
Considering the three heating terms in turn, any one of them 
is able to prevent this complete over-cooling, leading to a morphology 
more reminiscent of the ``standard'' case. Quantitatively, however, SNe appear to be the most important 
heating mechanism on large scales: 
without SNe, there are still some hot regions created by stellar winds, but the integrated 
energy input in the winds is a factor $\sim8$ lower. 
Turning off stellar wind heating  has similar but less noticeable effects. 
HII regions and the warm gas pressure they provide are most important inside of 
individual GMCs; they can produce some suppression of structure but no ``hot'' gas, 
and when SNe and stellar winds are present, the addition of photoionization-heating has
 little effect on the global gas morphology in the star-forming disk.

\section{Star Formation Histories}
\label{sec:sfh}

\begin{figure*}
    \centering
    \plotside{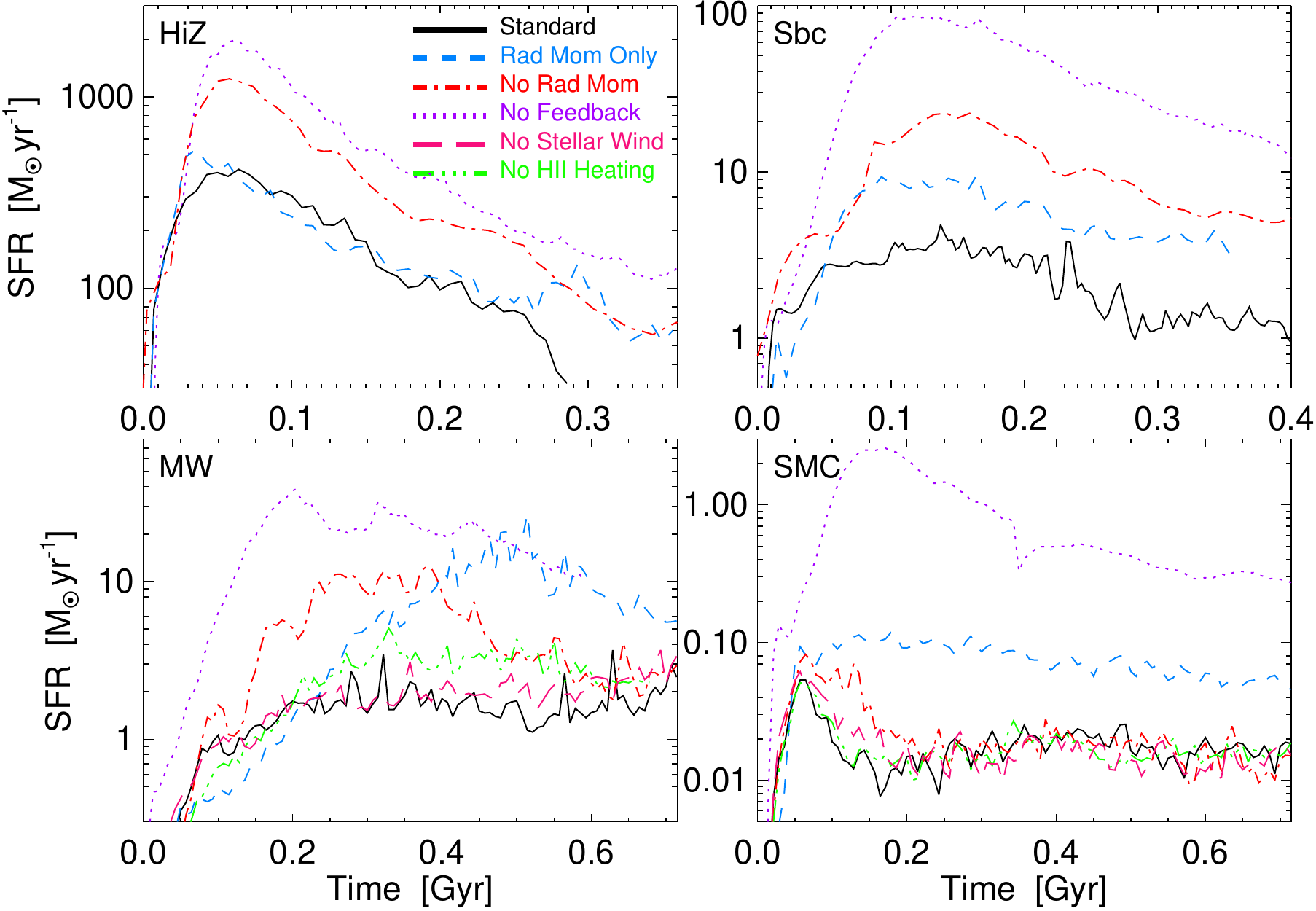}
    \caption{Star formation histories (total SFR versus time) for each of the galaxy 
    models discussed in \S~\ref{sec:sims}, with no feedback (dotted line), 
    all feedback mechanisms enabled (the ``standard'' model, solid line), and 
    various mechanisms turned on and off in turn. 
    In every galaxy type, with no feedback the disk collapses into GMCs that turn into 
    stars in runaway fashion, leading to enormous SFRs  exceeding the 
    Kennicutt-Schmidt relation by $\sim10-100$. These enormous SFRs
     decline over time, but only because most of the gas is consumed into stars.
    By contrast, with all feedback enabled, the disks reach an equilibrium SFR  and maintain a quasi steady, feedback-regulated equilibrium (the star formation still declines with time in the HiZ and Sbc models because of gas exhaustion).   
    {\em Top Left:} HiZ disk. Radiation pressure dominates the 
    regulation of the SFR; turning off all heating mechanisms (dashed line)
    makes a small $\sim20\%$ impact on the SFR. 
    {\em Top Right:} Sbc starburst disk. Radiation pressure is still very important, 
    but hot gas pressure makes a significant contribution to self-regulation. 
    Removing all the explicit heating mechanisms (dashed line) leads to factor $\sim3$ higher equilibrium SFR; removing  radiation pressure (dot-dashed line) leads to factor $\sim10$ higher SFR.
    {\em Bottom Left:} MW-like disk.  The interplay between the different feedback mechanisms is more subtle in this case -- see the main text (\S \ref{sec:sfh}) for a detailed discussion.   {\em Bottom Right:} SMC-like isolated dwarf. Here, SNe heating is the most 
    important mechanism regulating the galaxy-scale star formation.  \label{fig:sfh.vs.fb}}
\end{figure*}

Figure~\ref{fig:sfh.vs.fb}  
shows the corresponding star formation history for each model 
discussed in \S~\ref{sec:morph:fbmodel} 
with different feedback effects included or excluded. 
For reference, we show both no-feedback models (in which case nothing 
stops catastrophic  collapse) and the standard all-feedback models for each galaxy type.

In each galaxy type, with no feedback, the disk experiences a runaway collapse
leading to the 
consumption of a large fraction of the gas in a single dynamical time 
($\dot{M}_{\ast}\sim M_{\rm gas}/t_{\rm dyn}$). This gives much higher 
peak SFRs ($\approx2500,\,70,\,40,\,3\,\msun\,{\rm yr^{-1}}$ for the 
HiZ, Sbc, MW, and SMC models, respectively) 
than observed in the real systems the simulations are designed to 
model, and violates the observed Kennicutt relation by factors of $\sim10-100$ 
(see \paperone). The SFR declines only owing to gas exhaustion.

With the standard feedback model enabled, each galaxy type reaches
a much lower equilibrium SFR where feedback balances gravity. These
models remain in 
quasi steady-state for many dynamical times. In \paperone\ we 
show these SFRs agree well with the observed Kennicutt relation, 
as is clear from the fact that they are in close agreement with the star formation rates observed in galaxies of 
each type, i.e.,  $\dot M_*\sim100-300,\,1-5,\,1-2,\,0.01-0.03\,\msun\,{\rm yr^{-1}}$, respectively. 

How does this self-regulation depend on the inclusion or 
exclusion of different feedback mechanisms? In most cases the 
different SFHs parallel the differences in morphology 
discussed in \S~\ref{sec:morph:fbmodel}. 
For the HiZ model, including or excluding heating from SNe, stellar 
winds, and HII regions has a negligible effect on the SFH 
(removing all three together is similarly negligible). 
Excluding either of the local momentum or 
long-range radiation pressure forces increases the 
peak SFR by a factor of $\sim2-3$. Removing both leads to a 
runaway burst similar to the no-feedback case, 
with a peak SFR $\sim1000\,\msun\,{\rm yr^{-1}}$ and 
then a falloff from gas exhaustion. The system is then an order-of-magnitude 
above the SFR predicted by the observed Kennicutt-Schmidt relation.
Clearly the critical physics are the cold radiation sources of momentum.

In the MW-like case, 
the SFHs are again most strongly affected by the local momentum 
deposition and long-range radiation pressure, but the effects of SNe 
heating are non-negligible. 
Removing both local momentum deposition and long-range 
radiation pressure forces leads to a near-runaway consumption 
of gas with a peak SFR $>10\,\msun\,{\rm yr^{-1}}$ in the earlier stages 
of the simulation evolution. This is again in excess of the observed Kennicutt-Schmidt relation.

Removing {\em all} nuclear powered sources of heating (SNe, shocked stellar winds, 
and HII photoionization) is somewhat more complex: the SFR for the first 
few dynamical times remains low, but then rapidly rises above the 
``standard'' case by a factor $\sim3-10$. 
The effect is quite non-linear, and in particular non-local.  It arises from the fact that stellar mass loss is present in the model, 
but is now ``cold'' (the returned mass is injected at $\lesssim10^{4}$K and cools rapidly). 
Because the MW is quite gas-poor, this cold recycled gas can easily double the 
available (cold) gas mass supply 
at small radii (the central couple kpc) over the first $\sim0.5$\,Gyr; what we see is actually 
that the gas lost by stars in the bulge (where the MW is observed and by 
construction initialized to have very little gas) builds up until a $\sim$kpc-scale 
gas bar appears and drives a nuclear 
starburst at about $\sim0.5\,$Gyr. Unlike the ``no-feedback'' case, the 
radiation pressure terms are sufficient to ensure that the model remains on the 
observed Kennicutt-Schmidt relation 
at all times: the SFR is high because that excess gas becomes compact and 
builds up a large mass, so the system moves along, rather than off, the relation. 
This explains why, in \paperone, with no recycling of stellar gas, we saw that the ``radiation pressure alone'' case was sufficient to maintain the MW at a steady-state SFR $\sim1\,\msun\,{\rm yr^{-1}}$ 
for several Gyr.

Among the sources of gas heating, SNe are the most important. If we 
just turn off shock-heating from stellar winds (``No Stellar Wind''; recall this retains stellar 
mass loss but it is cold), the equilibrium SFR is higher but only by a factor $\sim1.2-2$. 
This is not surprising, since the integrated SNe energy deposition is $\sim8$ times 
larger than that in stellar winds. 
Turning off HII heating also has only a small impact 
on the equilibrium SFR (when other mechanisms are still present). The HII heating has 
some effect on the evolution of individual star-forming regions (aiding in their dissociation 
at early times, if they are relatively low-density), but does not strongly affect the global 
balance of feedback versus collapse. 

As in \S~\ref{sec:morph:fbmodel}, the Sbc-like case 
generally falls between the MW and HiZ cases. Heating from SNe feedback has a 
significant effect, but weaker than radiation pressure by a factor of $\sim3$ or so. 

In the SMC-like case, we see the importance of hot gas 
reflected in the SFHs. Removing HII heating has a negligible effect 
on the SFH, and removing hot stellar winds increases the peak 
SFR by a small factor. Removing both the local 
momentum and long-range radiation pressure forces only increases 
the median SFR by a significant but not dramatic factor $\sim2$ 
at early times. 
In contrast, removing SNe increases the typical SFR by a 
factor of $\sim5-8$. It is worth noting that even after this increase, the 
SFR is still much less than it would be with no feedback 
($\sim0.1\,\msun\,{\rm yr^{-1}}$ versus $\sim3\,\msun\,{\rm yr^{-1}}$), 
so there are important contributions to self-regulation from {\em all} 
sources of feedback, but SNe are the most important among them.

\section{The (Negligible) Role of Molecular Chemistry}
\label{sec:appendix.chemistry}

\begin{figure}
    \centering
    \plotone{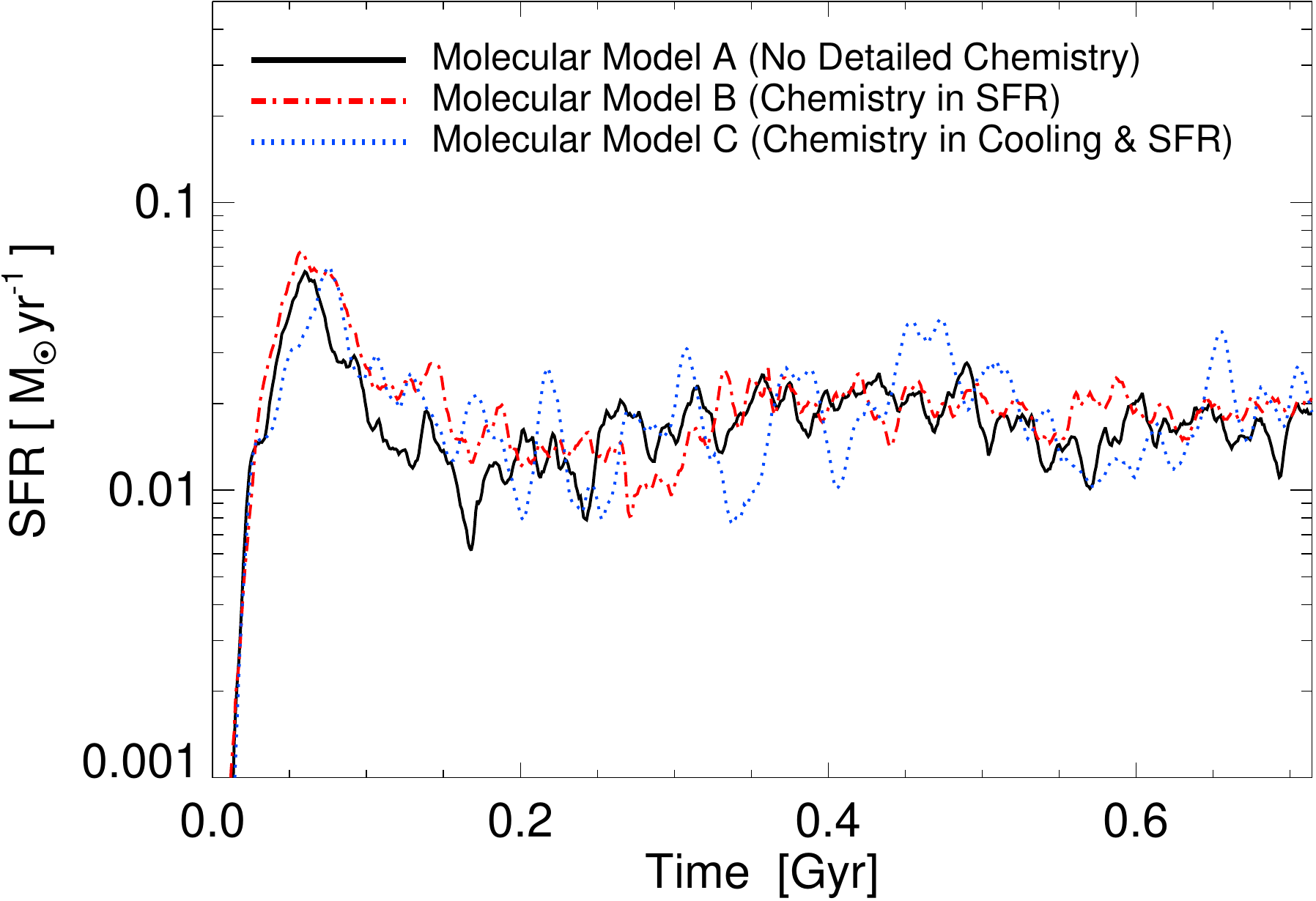}
    \caption{Star formation history for SMC models with different treatments of the molecular chemistry of the gas; all of the simulations have all of our  feedback mechanisms enabled.         The SMC model is the most sensitive of our galaxy models to the molecular chemistry because of its lower surface density and metallicity.
    {\em (A):} We use our standard cooling curve (see \S \ref{sec:sims}) and assume that all gas above the star-forming threshold ($n\gg100\,{\rm cm^{-3}}$) 
    is molecular (the $H_2$ molecular fraction $f_{H_{2}}=1$) and so can 
    turn into stars at an assumed efficiency per dynamical time of $1.5\%$.
    {\em (B):} We use the fitting functions in \citet{krumholz:2011.molecular.prescription} 
    for $f_{H_{2}}$ as a function of local column density and metallicity 
    for each gas particle and only allow star formation in molecular gas. 
    {\em (C):} We use the model in \citet{robertson:2008.molecular.sflaw}, 
    in which the molecular fraction $f_{H_{2}}$ and the cooling function $\Lambda(T)$ 
    at low temperatures $\lesssim10^{4}\,$K are a function 
    of the local gas density, metallicity, and radiation field (based on CLOUDY calculations); 
   star formation again only occurs in molecular gas.    Overall, the treatment of the molecular chemistry has a negligible effect on the star formation history.   This is because star
    formation only occurs in high density gas, $n\gtrsim100-1000\,{\rm cm^{-3}}$, and at the SMC's metallicity of $Z\sim0.1\,Z_{\sun}$, the molecular fractions in the 
    dense, star-forming gas are always near unity (average $f_{H_{2}}>0.998$ for models 
    {\em (B)} and {\em (C)}).   In test calculations, only at extremely low metallicity, $Z\ll 10^{-2}\,Z_{\sun}$, does the molecular     chemistry begin to have a measurable effect on the star formation history.   
    \label{fig:molecules}}
\end{figure}

It is well-established that stars form only from molecular gas -- albeit 
not {\em all} molecular gas, but only a small fraction at the highest densities. 
Hence a number of groups have attempted to model galactic star formation using 
various schemes to estimate the molecular 
fraction in a large volume of gas in a simulation \citep[see][]{robertson:2008.molecular.sflaw,
ostriker:2010.molecular.reg.sf,kuhlen:2011.mol.reg.h2.dwarfs}. 
Given our inclusion of a wide range of feedback physics, does chemistry matter?

We test this by re-running our standard models 
with all feedback present, but with different explicit models 
for the molecular fraction in gas. 
First (Molecular Model A), we make the crude assumption that the molecular fraction is always unity in 
all regions above our density threshold for star formation.

Second (Molecular Model B), we adopt the prescription in \citet[][eqn.~1-4 therein]{krumholz:2011.molecular.prescription}, which 
approximates the results of full radiative transfer calculations with a simple 
analytic fitting function that gives $f_{H_{2}}$ (the fraction of 
gas mass in a given particle which is molecular) 
as a function of the local column density and metallicity. 
We approximate the local column as the integral across the SPH 
smoothing kernel ($\propto \rho\,h$, analogous to what is done with the 
AMR experiments therein).\footnote{This is a conservative 
assumption because, if the integral column over large spatial scales is 
much larger, then it will further self-shield the gas, which 
only strengthens our conclusions below.} 
We then allow only the molecular gas to form stars -- 
multiplying the SFR our standard code would estimate by $f_{H_{2}}$. 

Third (Molecular Model C), we adopt a more detailed treatment motivated by that in 
\citet{robertson:2008.molecular.sflaw}. 
We allow the cooling function $\Lambda$ 
to be a function not just of temperature but also the local gas density, 
metallicity, and radiation field, where the latter is approximated 
by the sum of the cosmic background ($z=0$ value) 
and the local SFR (taking the instantaneous SFR in the local gas and assuming the 
continuous SF limit) with the SEDs tabulated in STARBURST99. 
Specifically the cooling functions are tabulated from 
Figure~1 of \citet{robertson:2008.molecular.sflaw} (on a simple 
grid in density, metallicity, and radiation field) and we interpolate logarithmically. 
We then calculate the local molecular fraction in each particle 
as a function of column density, metallicity, and also ionizing field strength 
from the fitting function in \citet{krumholz:2011.molecular.prescription}. As before, we then only
allow SF from the ``molecular'' gas. 

Figure~\ref{fig:molecules} shows that modeling the molecular gas fractions 
has no effect on our results. If we plot the morphology of 
each system (or any other measurable quantity presented in this paper), 
we find the same.

Why do we find that detailed chemistry is essentially irrelevant, while other 
authors have found it is critical? 
The key is the scales that we resolve and on which we allow star formation 
to proceed. In previous models where the molecular fraction is estimated and 
makes a large difference, it is usually the molecular fraction in a large volume of the 
ISM that is being estimated -- radii at least $>100\,$pc and as large as several kpc, 
with bulk volume densities of  $\sim0.1-10\,{\rm cm^{-3}}$. 
Essentially, then, the ``molecular'' model is really an implicit model for the 
fraction of dense gas within the patch. 
In contrast, our simulations explicitly resolve dense regions, and only allow star formation in them; our typical threshold for star formation is a local density of $n>100-1000\,{\rm cm^{-3}}$. 
At these densities, the gas is overwhelmingly molecular, so 
explicitly accounting for a molecular fraction and allowing only that gas to 
form stars is essentially identical to our existing prescription. 
In fact, we know the typical optical depths through the star-forming 
patches of gas (from our local momentum-driving model) and in the 
optical and UV wavelengths, they are typically $\sim10^{4}$, so there is 
no question that the gas can self-shield. 
In fact, when we implement the model of \citet{krumholz:2011.molecular.prescription}, 
we find the average estimated molecular fraction in the 
gas we would allow to form stars is $>99.8\%$. 

Furthermore, in \paperone, we show explicitly that the {\em global} 
star formation history is independent of the small-scale star formation law, 
because the star formation rate is set by the requirement that feedback 
from young stars maintain marginal stability. 
This requirement of vertical pressure support implies that allowing only molecular gas to form stars on a constant timescale within these high density regions will give the same star formation rate.

We also have tested, in Appendix~A of \paperone,
that changing the cooling curve shape dramatically -- for example, 
changing its normalization by a factor of several, or 
changing its shape to a simple constant from $10-10^{4}\,$K --  
has no impact on our results, because in each case the cooling time 
in the gravitationally collapsing dense regions is still shorter than the 
dynamical time by a factor $\approx 10^{4}-10^{5}$ at 
densities near the threshold for star formation (at least down to $\approx100\,$K, at which point the 
thermal pressures are totally negligible compared to gravity, 
and so subsequent cooling has only weak effects). 

\citet{glover:2011.molecules.not.needed.for.sf} study star formation 
on small scales with a variety of different chemical 
and cooling models, and reach the same conclusion we do here.
They show that the molecular fraction under otherwise identical 
densities and metallicities is causally irrelevant to the collapse of dense gas, since 
cooling via C$^{+}$ fine-structure and dust emission is nearly as efficient as CO -- the presence of CO is very tightly correlated with, but does not {\em drive} the collapse of gas to high densities.

Thus it is only if cooling and the formation of metals, dust, and 
molecules are dramatically 
suppressed that the explicit chemistry will become important 
for any of the quantities we model. 
This occurs at metallicities below $Z \sim 10^{-3}-10^{-2}\,Z_{\sun}$. 
Specifically, this is the metallicity at which the 
molecular fraction begins to fall below 
unity in gas near our threshold density; this is also the metallicity at which the cooling time can become comparable to, or longer than, the dynamical time \citep[see e.g.][]{robertson:2008.molecular.sflaw}. The formation of the first stars, and 
the lowest-mass dwarf galaxies 
(galaxy stellar masses $<10^{6}\,\msun$) 
are within this regime and will be strongly influenced by the ability of the 
gas to form molecules even at high densities. 
But the more massive systems we model here 
(even the SMC) are safely in the regime in which 
the dense gas is almost all molecular and thus so long as star formation is restricted to the dense gas, an explicit treatment of the molecular chemistry is not important.

\section{Global Properties: Disk Velocity Dispersion and Support}
\label{sec:dispersion}

We now examine the structural properties of our model disks, as a function of 
time and radius, with different feedback mechanisms enabled and disabled. 

\begin{figure*}
    \centering 
    \scaleup
    \plotside{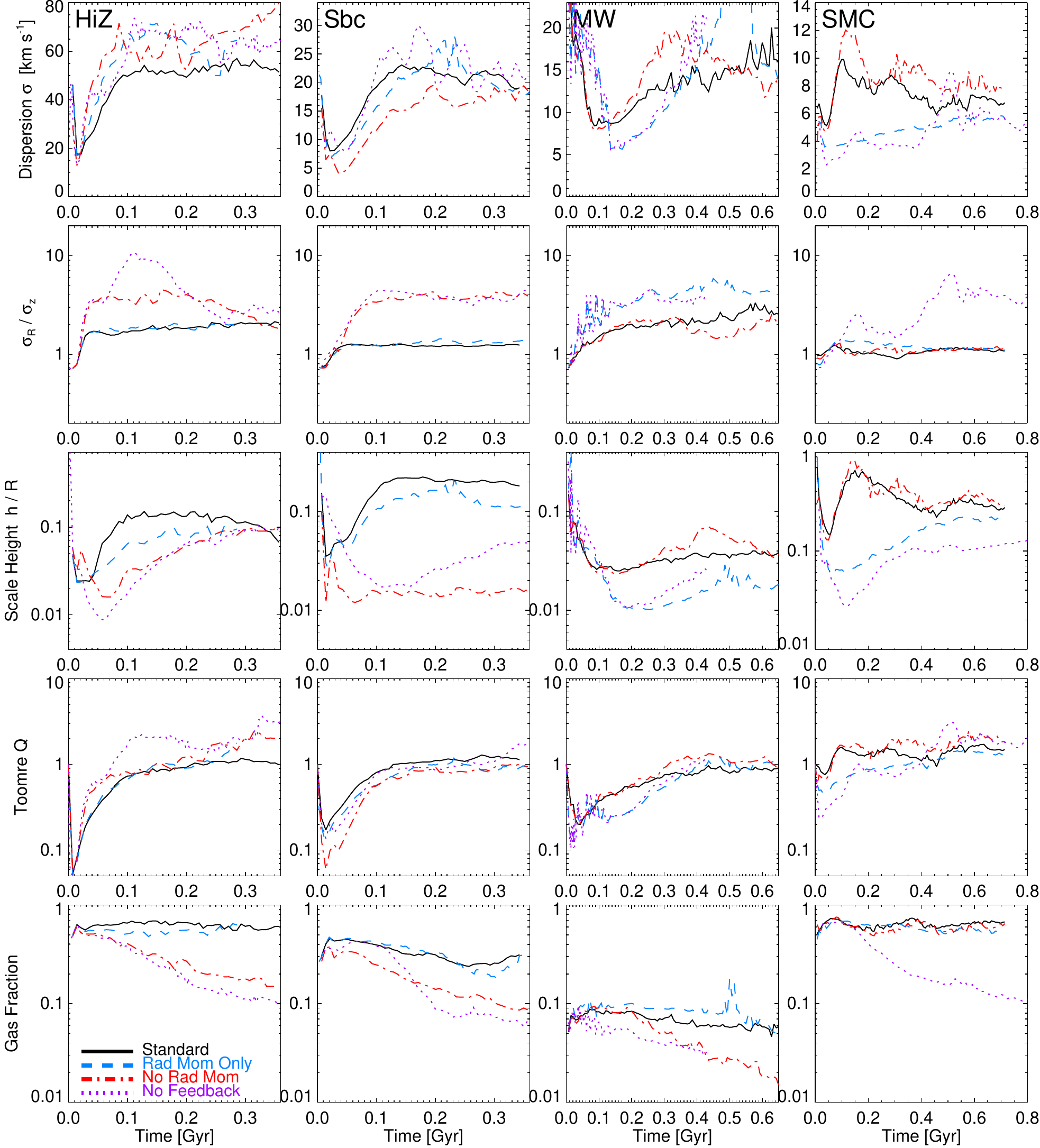}
    \caption{Global galaxy structural properties in {\em gas} (not stars) as a function of time, 
    in otherwise identical simulations of each galaxy model with various feedback effects enabled  (different lines -- see Table 2). 
    Each quantity is the weighted average over the star-forming disk.
    {\em Top:} Sightline-averaged one-dimensional gas velocity dispersion $\sigma$. 
    {\em Second from Top:} Velocity (an)isotropy:  ratio of in-plane radial dispersion $\sigma_{R}$ 
    to vertical dispersion $\sigma_{z}$.
    {\em Middle:} Vertical scale height-to-radius ratio $h/R$. 
    {\em Second from Bottom:} Toomre Q in the gas ($\sigma_{R}\,\kappa/\pi\,G\,\Sigma$).
    {\em Bottom:} Gas fraction $M_{\rm gas}/(M_{\rm gas} + M_{\ast})$. 
    With all feedback enabled, values agree well with those observed: $f_{\rm gas}$ 
    evolves slowly, feedback preserves stability ($Q\approx1$, with $\sigma$ driven to the 
    appropriate observed value for each galaxy) and isotropizes the dispersions 
    ($\sigma_{R}\approx\sigma_{z}$, giving $h/R\approx\sigma_{z}/V_{c}$). 
    With no feedback, runaway star formation consumes the gas. Dispersion and $Q$ 
    still increase, but only because all high-density (rapidly cooling, $Q<1$) gas is 
    totally exhausted.  In the HiZ and Sbc cases, removing radiation pressure yields results similar 
    to the ``no feedback'' case (SNe, stellar winds, and photoionization-heating cannot prevent 
    runaway dissipation in dense gas). In the SMC case, gas heating by SNe is 
    more important than radiation pressure for the disk structure. 
    In the MW case, all mechanisms contribute comparably.     \label{fig:hr.vs.t}}
\end{figure*}

\begin{figure}
    \centering 
    \scaleup
    \plotonesmall{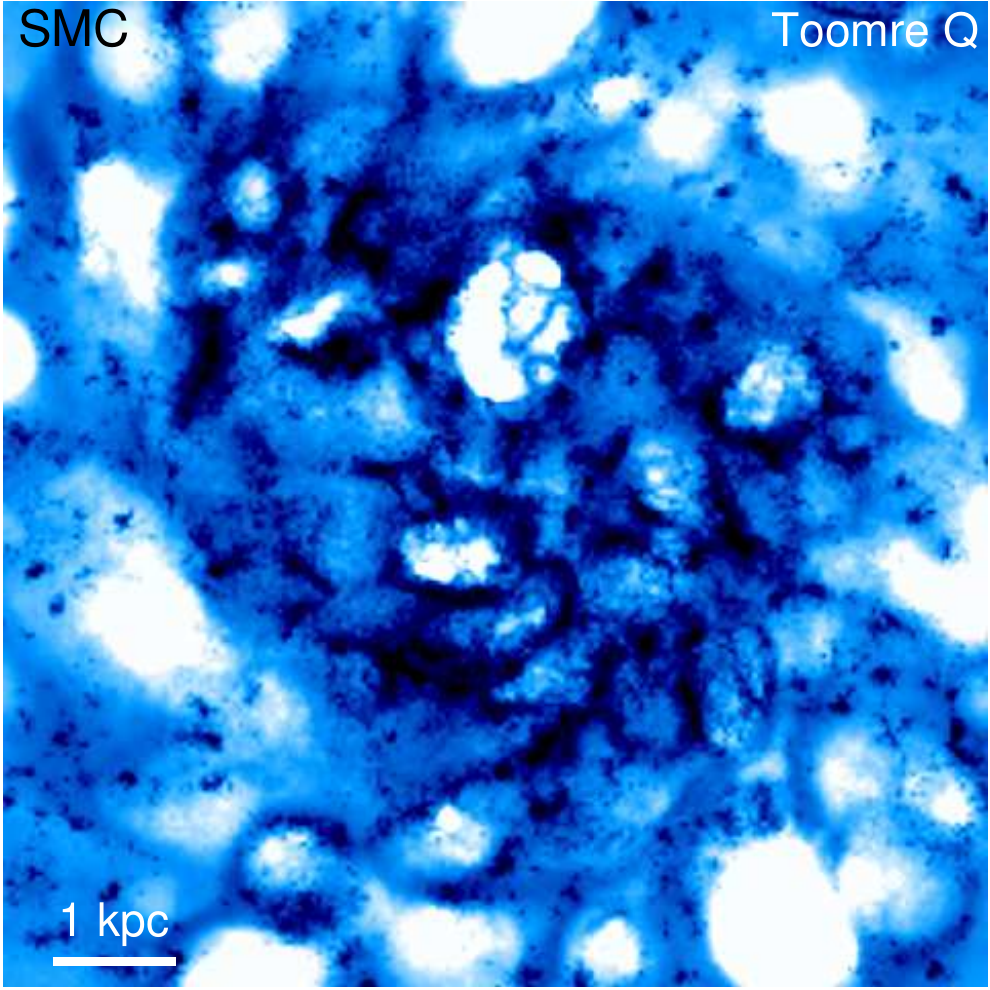}
    \caption{Face-on map of the Toomre $Q$ parameter (eqn.~\ref{eqn:toomreq}) for the gas within the disk averaged over $\sim10\,$pc annuli. Intensity is scaled logarithmically with $Q$ from $Q\le0.1$ (black) to $Q\ge10$ (white). We show the SMC model as Fig.~\ref{fig:morph.smc}. Despite the robust global convergence to $Q\sim1$ in Fig.~\ref{fig:hr.vs.t}, there are large local variations in $Q$, as some regions have dissipated their support and are collapsing (on their way to forming stars) while others have just experienced strong local bursts of stellar feedback and are rapidly expanded and/or heated. 
    \label{fig:qmap}}
\end{figure}

Figure~\ref{fig:hr.vs.t} plots a number of disk structural properties as a function of 
time. 
We are particularly interested in the properties of the star-forming disk, 
as opposed to e.g.\ the extended ionized HI disk (which is obviously not 
strongly affected by feedback), so we 
calculate all values as gas-mass weighted averages in  
each radial annulus at each time, then plot the SFR-weighted 
average over radial annuli (the results reflect the value near 
$\sim R_{e}$, but do not vary rapidly in radius). 
First, the resolved vertical velocity dispersion $\sigma_{z}$ 
(specifically, half the $15-86\%$ interval in $v_{z}$, to minimize the bias 
from large ``tails'' of outflowing material at high-$v_{z}$). 
Second, the ratio $\sigma_{R}/\sigma_{z}$, between the in-disk plane radial velocity 
dispersion and vertical dispersion (the azimuthal dispersion 
$\sigma_{\phi}\approx\sigma_{R}$ in all cases). 
Third, the (fractional) disk scale height $h/R$, 
where the scale-length $h$ 
of the disk fit to a vertical ${\rm sech}^{2}$ profile. 
Fourth, the Toomre $Q$ parameter, 
\be
Q\equiv \frac{\delta v_{\rm R}\,\kappa}{\pi\,G\,\Sigma}
\label{eqn:toomreq}
\ee
where $\delta v_{\rm R}$ is the effective radial dispersion of the gas, 
$\equiv \sqrt{\sigma_{R}^{2}+c_{s}^{2}}\approx \sigma_{R}$ 
(because the thermal sound speed 
$c_{s}\ll \sigma_{R}$ in the dense gas, in all cases), 
$\kappa$ is the epicyclic frequency ($\approx \sqrt{2}\,\Omega$ for 
nearly constant-$V_{c}$ disks here), and $\Sigma$ is the 
gas+stellar disk surface density. 
Finally, the disk gas fraction, defined locally as 
$f_{\rm gas}\equiv \Sigma_{\rm gas}/(\Sigma_{\rm gas}+\Sigma_{\ast})$. 

The very early time behavior is a result of the initial conditions: these 
include large vertical support for the disks, which 
quickly vanishes as the disks cool, then dense GMCs form and begin to 
produce stars. After about an orbital time, the systems reach a quasi-equilibrium state, 
which can clearly be seen (at least in cases with feedback) in the flat or slow 
subsequent time dependence. Once in this regime, details of the initial conditions 
(e.g.\, whether the initial disks are supported by turbulent or thermal velocities) 
have no significant effects.

\subsection{Dispersions and Gas Fractions}
\label{sec:dispersion:vsigma}

\subsubsection{Comparison to Observations}
\label{sec:dispersion:vsigma:obs}

For the cases with all feedback mechanisms 
enabled (our ``standard'' model), the values generally agree with observed 
disks at similar mass and redshift to the analogs in each model. 
The HiZ disks have SFRs of $\sim70-300\,\msun\,{\rm yr^{-1}}$, 
a morphology of massive clumps, 
purely vertical velocity dispersions of $\sim 30-40\,{\rm km\,s^{-1}}$ 
and sightline-averaged dispersions of $\sim 50-100\,{\rm km\,s^{-1}}$, 
giving typical $V/\sigma\sim5$ or so given their maximum 
circular velocity $V_{\rm max}\approx 230\,{\rm km\,s^{-1}}$, 
and high gas fractions $\sim30-60\%$ for most of their lifetimes,\footnote{Recall 
this is essentially $\fgas$ at the effective radius of star formation, which 
evolves weakly in time as 
gas can be depleted at small radii (giving the slow decline in SFR
seen in Figure~\ref{fig:sfh.vs.fb}).} all in 
good agreement with observed properties of 
the most massive star-forming disks \citep{genzel:highz.rapid.secular,
forsterschreiber:z2.disk.turbulence,erb:lbg.gasmasses,
tacconi:high.molecular.gf.highz}. 

The Sbc disks also maintain high gas fractions, low $V/\sigma\sim5-8$, 
with large dispersions and puffy scale heights. 
In both of these cases, the specific star formation rate is high, and 
the disks are clumpy and self-gravitating; feedback and gravity therefore 
maintain large dispersions in the gas. 

The MW-like runs maintain the canonical observed $\sim8-15\,{\rm km\,s^{-1}}$ 
dispersion, which with $V_{\rm max}\approx220\,{\rm km\,s^{-1}}$ gives 
$V/\sigma\sim20-30$, and gas fractions $\approx10\%$ -- because they 
are much more stable (with large dark matter halos), and have low gas fractions (and low 
specific star formation rate), these disks are thinner, relative to the 
starburst systems. 

The SMC-like runs show a similar moderate dispersion $\sim6-8\,{\rm km\,s^{-1}}$, 
giving $V/\sigma\sim5-15$, and maintain very high gas fractions $\sim70-80\%$; 
these properties all agree very well with typical properties for 
$\sim10^{9}\,\msun$ dwarf galaxies \citep[see e.g.][]{belldejong:tf,mcgaugh:tf}. 
Here, the self-gravity, star formation rate, and corresponding 
dispersion are low, but the potential is very shallow, so even small 
dispersions maintain relatively ``puffy'' gas disks, as observed. 
Gas fractions are generally larger at large radii as the disks have some bulge and a 
more extended gas versus stellar disk; there are some small flares at large 
radii but these are weak. There is a rise in $V/\sigma$ at large radii 
as the gas at these radii is very low-density and has 
a low specific star formation rate, hence low feedback support and low $\sigma$. 
In these outer regions the disk is supported in part supported by thermal pressure, maintained by external radiation, either from the interior of the disk or from the cosmic background ultraviolet radiation.


We have also measured the full turbulent velocity power spectrum in these simulations, 
and find its shape -- essentially consistent with equal power on all scales (with a break 
below the disk scale-height where the system transitions from being effectively two-dimensional 
to three-dimensional) -- 
varies little across simulations (with normalization just set by 
$\sigma$ and the scale-lengths here). This is expected: since the bulk gas motions are super-sonic 
and gravity is the dominant force, the turbulence should be largely scale-free.
This is true even if motions are ``pumped'' by stellar feedback, because the fraction of the time that 
{\em direct} feedback forces are dominant over gravity and bulk hydrodynamic 
forces as the force on a given gas parcel is small. For this reason, we do not see strong imprints of the input feedback 
injection or velocity scale -- quite unlike a case where gas is ``held up'' in some hydrostatic sense by feedback forces.

\subsubsection{Failings of Feedback-Free Models}
\label{sec:dispersion:vsigma:nofb}

There are important differences if we compare runs with feedback to the no-feedback cases. 
Most obviously, without feedback the gas fractions plummet extremely quickly, as essentially the 
entire galaxy gas supply collapses into dense regions and is turned into stars in 
a single global dynamical time. 

The {\em vertical} dispersions (and as a consequence, scale heights) 
are also significantly lower (at least for most of each run) in the cases without feedback.
This is despite the fact that the rapid gas depletion should make it much easier for 
non-feedback (and non-nuclear powered) mechanisms to maintain large dispersions in the remaining gas.
In the HiZ and Sbc cases, the systems are strongly gravitationally unstable, 
with large clumps scattering off each other and massive inflows along bars 
and spiral arms. However, almost all of this gravitational potential energy goes into maintaining the in-plane dispersions.
The in-plane dispersions can be maintained at large values by global gravitational 
instabilities (sinking clumps, bars, and spiral arms) that avoid rapid shocks and 
dissipation precisely because they are global structures. In contrast, the 
vertical motions drive shock heating on the local dynamical time, so $\sigma_z$ rapidly 
drops to a low value.

At late times in both the HiZ and Sbc cases $\sigma_{z}$ begins to rise, but the 
models have a have a gas fraction of just a few percent at the time, and that gas is mostly 
being ``dragged'' by a few super-massive stellar clumps, and undergoing 
mergers at the galaxy center.

The SMC case is physically quite different from the HiZ and Sbc models: the system requires only
small dispersions, but it is so gravitationally {\em stable} (owing to the large dark matter 
to disk ratio) that it does not develop large turbulent motions in the absence of feedback. 
Again, though, it shows a large $\sigma_{R}/\sigma_{z}$ without feedback, 
from the moderate spiral structure in place. The MW model, interestingly, is in some ways 
a ``worst-case'' system for discriminating the effects of 
feedback (the differences are minimized) -- this is because it has a low gas fraction, 
requires a low gas dispersion for vertical pressure equilibrium, and is marginally unstable 
to e.g.\ spiral arm formation.

\subsubsection{Effects of Each Feedback Mechanism In Turn}
\label{sec:dispersion:vsigma:vsfb}

Now consider cases with different feedback mechanisms included or 
excluded. For the HiZ and Sbc-like cases, removing the sources of 
``hot'' gas (e.g.\ SNe feedback, stellar wind shock-heating, and HII heating) 
has relatively little effect on any of the quantities plotted in 
Figure~\ref{fig:hr.vs.t}. Removing the radiation 
pressure (momentum feedback acting on cold, dense gas) however, 
leads to much more rapid star formation and hence a runaway decrease 
in $f_{\rm gas}$. The full dispersion, $\sigma_{R}^{2}+\sigma_{\phi}^{2}+\sigma_{z}^{2}$, 
is essentially the same for any combination, or lack, of feedback. As we argued in 
the previous paragraph, global gravitational instabilities (powered by accretion 
through the disk) can maintain $\sigma_\phi$ and $\sigma_{R}$. 
But clearly the no-feedback and radiative momentum models are ``failed'' in the sense that 
they did not prevent runaway star formation  and, at a fixed gas fraction $f_g$, 
the vertical velocity dispersion $\sigma_z$ is too small.

For the SMC case, ``hot'' feedback is much more important, since the 
average densities are much lower and opacities and specific star formation 
rates (important for radiation pressure) much lower as well. 
Removing all the hot gas sources, but keeping radiation pressure in place,  
moderate but slightly lower vertical dispersions $\sim5-6\,{\rm km\,s^{-1}}$ can 
be maintained, and star formation efficiencies are still much lower 
than the no-feedback case (by a factor $\sim10$, see 
\paperone\ and \S \ref{sec:sfh}) but they are higher than the case 
with hot gas sources by a factor of several. Supernovae are most important source
of hot gas in the SMC model (as we saw for the SFR above 
and winds in \citealt{hopkins:stellar.fb.winds}). 
Removing stellar winds or HII heating has very little effect on the dispersions or 
other properties if SNe feedback is still included; however, if SNe are absent, 
they are able to maintain comparable dispersions.

Finally the properties of the MW-like model lie in between the HiZ and Sbc 
models on the one hand, and the SMC model on the other,
as expected. As noted above, in the MW model all the feedback mechanisms seem to 
contribute comparably, making it difficult to distinguish the effects of 
different feedback mechanisms.

\subsection{Toomre Q and Disk Stability}
\label{sec:dispersion:stability}

All of the systems modeled here equilibrate at a Toomre $Q\sim1$, 
and maintain this over the disk surface. 
In the feedback-regulated cases, this must happen: $Q<1$ regions will 
collapse and begin forming stars, which leads to a super-linear 
feedback (in that the further collapse proceeds, the more rapid the 
SF and hence momentum/energy injection becomes), until it arrests further 
collapse (which by definition is when it re-equilibrates to $Q\sim1$). 
If $Q>1$, then there is no collapse, hence no dense regions form and no 
star formation takes place; without feedback to hold up the disk, the 
cooling time is much less than the dynamical time, and so it radiates away 
its support in at most a crossing time (the dissipation time for turbulent support). 

Fig.~\ref{fig:qmap} illustrates how this global equilibrium is maintained, by showing a map of the local values of $Q$ (calculated within $10\,$pc annuli) for the SMC model (the other models are qualitatively similar). Although the disk maintains an average $Q\sim1$ there are large local variations from $Q\sim0.1-10$, corresponding at low-$Q$ to the regions which have begun to collapse, and at high-$Q$ to regions which have recently formed massive stars and are being disrupted and heated by feedback. The characteristic spatial and time scales, therefore, for these fluctuations, are closely associated with the Jeans scale and associated crossing times.

However, even the feedback-free systems also equilibrate at $Q\sim1$.
The reason is simple: nothing can prevent runaway collapse and star formation in 
$Q<1$ regions, so all the gas that can reach $Q<1$ does so, and then turns into 
stars, leaving only $Q\gtrsim1$ gas remaining.
Thus $Q\sim1$ can be maintained, but at the expense of runaway gas consumption.
Quantitatively, we make a crude analytic estimate of the dispersions
that can be maintained by inflow and instabilities. The vertical velocity 
dispersion dissipates on a vertical crossing time, $H/\sigma_z=r/V_c$, where 
the second expression follows from the assumption that the turbulence provides 
the vertical support against gravity required to maintain the vertical scale height $H$. 
The luminosity dissipated by turbulence is then $L_{turb}=\eta \sigma_z^2 \Omega M_g$, 
where $M_g$ is the gas mass of the disk and $\eta$ is a constant of order unity. 
In the absence of feedback, this luminosity must be supplied by global gravitational 
instabilities (sinking clumps, bars, 
spiral arms), ultimately powered by accretion of gas through the disk. (We assume that the
accretion energy is converted into isotropic turbulence). The accretion luminosity
is $L_{acc}= 2\dot M_{acc}V_c^2$.
\footnote{The factor of four ($2=4\times 1/2$) comes from the energy  transported by the torques causing the accretion; this factor is larger (by 4/3) in a constant-$V_c$ disk than that in a Keplerian disk.} 
We can find a lower limit to $\dot M_{acc}$ 
assuming the disk is in a quasi-steady state, so that the mass accretion rate must exceed the star formation rate, given by the Kennicutt-Schmidt relation, $\dot \Sigma_*=\epsilon \Sigma_{\rm gas}\Omega$, with $\epsilon=0.017$. Averaging over the disk, $\dot M_* = \epsilon \nu^{-1} M_g \Omega$ where $\Omega$ is evaluated at the disk effective radius, and $\nu\approx 2$ depends on the surface density profile. Combining these relations, we find
\be 
\frac{\sigma_{z}}{V_{c}} = \left(\frac{3\epsilon}{2\eta\nu}\right)^{1/2}\ .
\ee 
Using this estimate for $\sigma_z$ in the expression for the Toomre parameter,
\be
Q_{\rm grav} 
\approx {\Bigl(} \frac{\sigma_{R}}{V_{c}}{\Bigr)}\,\frac{2\,\tilde{\kappa}}{\nu\,f_{\rm gas}}
\approx f_{\rm gas}^{-1}\,{\Bigl (}\frac{2\,{\kappa}}{\nu\,\Omega} {\Bigr)}
\,{\Bigl(}\frac{3\epsilon}{2\,\nu\,\eta}{\Bigr )}^{1/2} 
\approx \frac{0.1}{f_{\rm gas}}
\ee
so we expect $Q\sim1$ as $f_{\rm gas}$ depletes below $\lesssim0.1$, 
which is roughly what the simulations show.

Even in simulations with no star formation, and hence no gas consumption
(``No Feedback or SF'' in Table~\ref{tbl:sims}), the systems 
quickly reach $Q\sim1$. In these models, the gas collapses and cools until 
it is essentially all in extremely dense clumps, $n>10^{6}\,{\rm cm^{-3}}$ in 
Fig.~3 of \paperone\ -- corresponding to $\sim$kpc-scale patches collapsing to $\sim$pc. 
This makes the gas effectively collisionless! The gas ``nuggets'' (which internally continue 
to dissipate, but are long-lived and rarely collide) scatter and develop $Q\sim1$ just as 
the stars do.

\begin{figure*}
    \centering
    \plotside{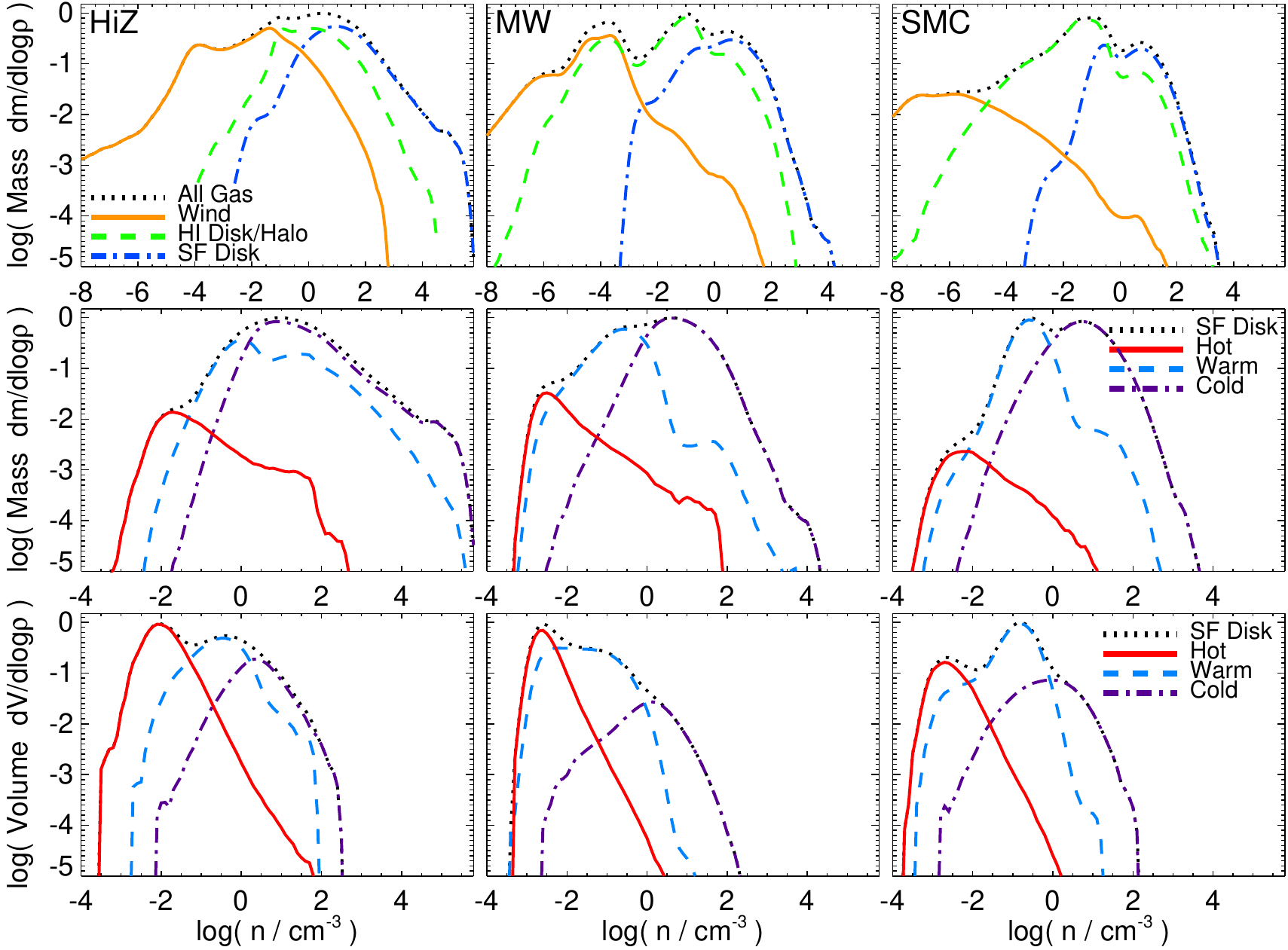}
    \caption{Density distribution of different ISM phases, 
    for different galaxy disk models (HiZ, MW, and SMC), at 
    times after they reach feedback-regulated steady-state. Variation 
    in time after the first orbital time is modest.     {\em Top:} 
    Mass-weighted density PDF (${\rm d}\,m_{\rm gas}/{\rm d}\log{n}$), 
    i.e.\ the mass fraction per logarithmic interval in density $n$. 
    We show the distribution for all gas in the simulation 
    ({\em black}), the gas within the (multi-phase) star-forming disk ({\em blue}), 
    the gas in the extended, ionized non-star forming HI disk and halo ({\em green}), 
    and wind/outflow material ({\em orange}). 
    These trace the material at low, intermediate, and high densities, 
    respectively (as expected). Each density PDF  has a very broad density distribution. 
    {\em Middle:} 
    Mass-weighted density PDF within the star-forming disk -- i.e.\ the traditional 
    multi-phase ISM. We show the density PDF for all of the gas in the disk 
    ({\em black}), the ``cold/molecular'' phase ({\em purple}; $T<2000$\,K), 
    the ``warm ionized'' phase ({\em cyan}; $2000<T<4\times10^{5}\,K$), 
    and the ``hot diffuse'' phase ({\em red}; $T>4\times10^{5}\,K$). 
    Most of the mass is in the cold phase (which 
    dominates at high densities), but with a non-negligible ($\sim20-30\%$) contribution 
    from the warm medium (the hot phase contributing $\sim5-10\%$). 
    There is also a weak trend towards the cold phase being more important in more massive, gas-rich disks (HiZ versus SMC).  Note that the multi-modal nature of the total density PDF is a consequence of the strong phase separation. 
    {\em Bottom:} Volume-weighted 
    density PDF (${\rm d}\,V_{\rm gas}/{\rm d}\log{n}$) for  the star forming disk. 
    The hot diffuse phase dominates, with a moderate 
    ($\sim20-40\%$) volume filling fraction for the warm phase, 
    and a small ($\sim1-5\%$) volume filling fraction of cold clouds. 
    \label{fig:rho.dist}}
\end{figure*}

\section{ISM Phase Structure}
\label{sec:ism.structure}

\subsection{Mass and Volume of Phases with Feedback}
\label{sec:ism.structure:global}

Figure~\ref{fig:rho.dist} shows the density distribution 
in three of our ``standard'' simulations with all feedback implemented normally.
For each, we present the distribution in three ways. 
First, we consider the total mass-weighted density distribution 
($dm_{\rm gas}/d\log{n}$), for {\em all} gas in the simulation. 
This covers a very wide dynamic range in both density and spatial scale, 
and is not directly comparable to the typical ``ISM density distribution'' 
from galaxy studies. To see this, we divide the gas into three 
categories: the star-forming disk (defined as gas within the 
projected radius that 
encloses $90\%$ of the total star formation and within one vertical scale height 
of the disk inside that radius), the 
extended disk+halo gas (gas outside the disk, but not in the wind defined below), 
and the wind/outflow material (gas above a scale height, with 
large radial outflow velocity $v_{r}/|{\bf v}|\ge 0.75$ and $v_{r}>100\,{\rm km\,s^{-1}}$).
The star-forming disk (radii 
$\sim8,\,4,\,10,\,3\,$kpc in the HiZ, Sbc, MW, and SMC 
case) extends out to a couple optical scale-lengths, and (unsurprisingly) includes 
almost all of the dense gas ($n\gtrsim1\,{\rm cm^{-3}}$). 
The extended disk and halo cover intermediate densities: this includes the initial extended 
gas disks, which extend to very low densities and tend to be ionized hydrogen disks with 
$T_{\rm eff}\sim 10^{5}\,K$, and material which is pushed out of the disk by feedback 
but does not contain enough energy/momentum to drive an outflow (the ``halo'' -- recall there is no 
initial extended gaseous halo in these simulations). 
This contains a large fraction of the gas mass in all cases, most of which is out at several 
optical disk scale lengths. 
The wind material extends to the lowest densities -- since there is no IGM, in principle escaped 
material can reach arbitrarily low densities -- and although it contains some cold, 
dense clumps (visible in the density distribution or in Figures~\ref{fig:morph.hiz}-\ref{fig:morph.smc}), 
it contains most of the hot gas with $T_{\rm eff}\gtrsim10^{6}\,K$. Much of this material 
is out at very large radii of order the virial radius. 

We next consider the phase structure of the star-forming disk in detail, 
since this is the traditional region in which the ISM phases are defined 
and where stellar feedback has the greatest impact. 
Specifically, we show the mass-weighted density distribution 
($dm/d\log{n}$) and volume-weighted density distribution 
($dV/d\log{n}$; where the volume of each particle is proportional 
to its smoothing length cubed), within the star-forming disk identified 
above (since there is no outer boundary to the simulation, the ``total'' volume-weighted 
density distribution diverges at low densities with the material at the furthest edges of the simulation). 
We divide the gas into three traditional phases with a simple 
temperature cut: ``cold atomic+molecular'' gas ($T<2000\,K$), ``warm ionized'' gas
($2000<T<4\times10^{5}\,K$) and ``hot'' gas ($T>4\times10^{5}\,K$). 
The exact temperature cuts are arbitrary but the qualitative comparison does not 
change if we shift them by moderate amounts. 
Unsurprisingly, the high-density gas ($>1\,{\rm cm^{-3}}$) 
is predominantly cold, the intermediate-density gas ($0.01-1\,{\rm cm^{-3}}$) 
primarily warm, and the low-density gas ($<0.01\,{\rm cm^{-3}}$) hot. 
The temperature structure also explains the features seen in the 
total density distribution, arising from the separate components -- 
the total distribution is crudely log-normal, but each component individually 
is more clearly so. 

By mass, the cold/molecular phase dominates, but with a comparable contribution 
from the warm/ionized phase, and the ratio of the warm-to-cold gas increases 
as we move to lower-mass, low-average density systems 
($M_{\rm warm}/M_{\rm cold}\sim0.3,\,0.5,\,0.7$ in the HiZ, MW, and SMC-like 
cases, respectively). 
The ``hot'' phase constitutes at most a few percent of the total mass budget 
within the star-forming disk ($\approx1-3\%$ in each case). 
By volume, however, the results are reversed. 
The ``hot'' phase has a near-unity volume filling factor in each case. 
The ``warm'' phase has a non-trivial filling factor 
($\sim0.3,\,0.5,\,0.7$ in the HiZ, MW, and SMC-like cases), but 
it can be significantly less than unity. 
Recall from Figures~\ref{fig:morph.hiz}-\ref{fig:morph.smc}  
that the ``warm'' phase is still often concentrated in 
dense structures such as the spiral arms (or in the HiZ case, in giant cloud complexes) 
as well as loops and shells blown by hot bubbles (although again, most of the warm 
gas is in the extended ionized disk). 
The cold phase has a very small filling factor 
of a few percent ($\sim3-10\%$). 

The warm and hot phases of the ISM have similar thermal pressure 
$P\approx 5\times10^{-13}\,{\rm erg\,cm^{-3}}$, which for the 
warm material reflects gas with $n\sim0.1-0.4$ and $T\sim1-5\times10^{4}$ 
and for the hot gas $n\sim0.003$ with $T\sim10^{6}\,$K.
Both include a tail of material with pressures up to $\sim100$ times larger, 
much of which is recently-heated material that is venting out of the disk, 
but some of which is recently photo-ionized dense gas that is ``breaking out'' of 
GMCs. As expected, the thermal pressure of the cold gas is lower than the ambient medium 
by a factor of $\sim10-30$. However, the turbulent pressure of the clouds is 
much larger, $P\sim 10^{-10}-10^{-9}\,{\rm erg\,cm^{-3}}$ -- this is consistent with their 
being marginally bound (discussed below), for which $P\sim G\,\Sigma_{\rm cloud}^{2}
\sim 10^{-9}\,{\rm erg\,cm^{-3}}\,
(\Sigma_{\rm cloud}/100\,\msun\,{\rm pc^{-2}})^{2}$. 
The warm/hot pressure increases as well if we include the turbulent ram pressure, 
but by order-unity factors; they satisfy $Q\sim1$ hence  
$P\sim G\,\langle \Sigma \rangle^{2}$, so the pressure ratio is approximately 
the squared ratio of GMC to disk-average surface densities. 
The cold clouds are therefore clearly not typically pressure-confined, but 
held together by self-gravity.

\begin{figure*}
    \centering
    \plotside{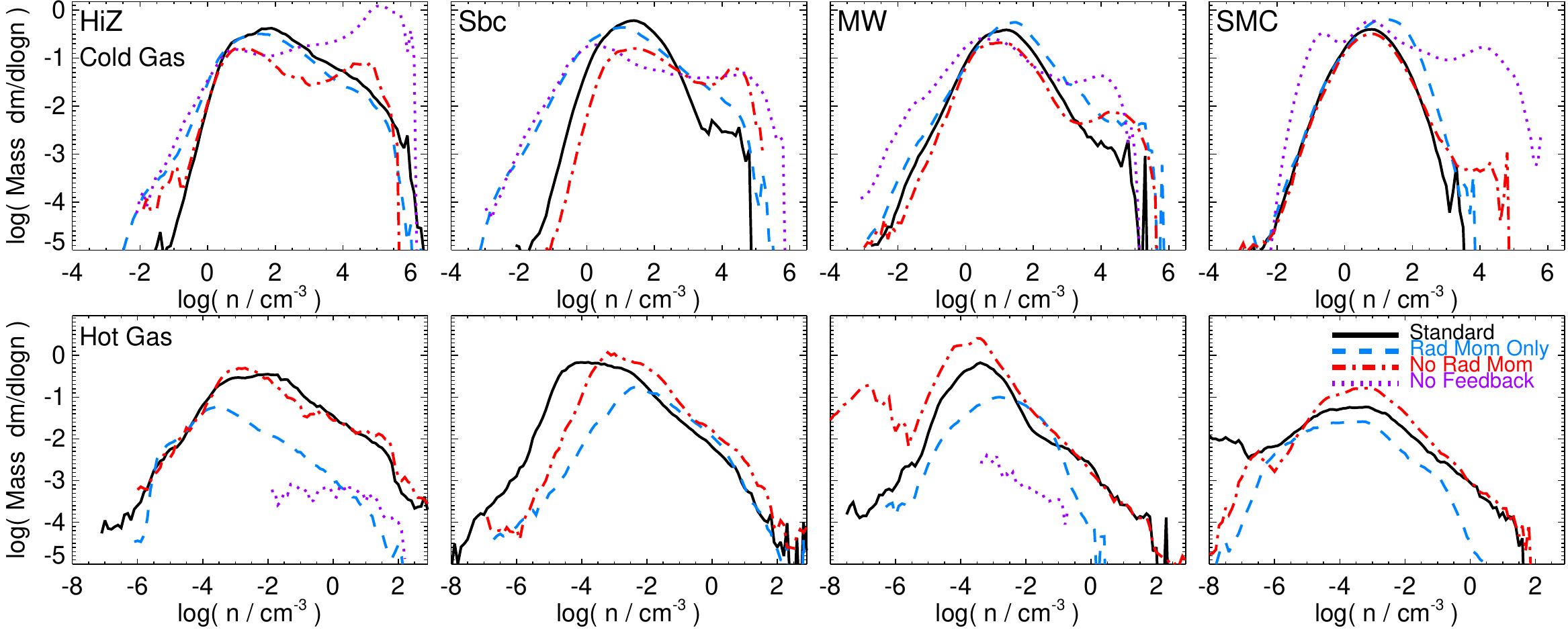}
    \caption{{\em Top:} Mass-weighted 
    density distribution of ``cold'' gas ($T\lesssim1000\,$K)for models with different feedback enabled (as in Fig.~\ref{fig:sfh.vs.fb} and Table 2). 
    {\em Bottom:} Mass-weighted 
    density distribution for the ``hot'' gas ($T\gtrsim10^{6}\,$K).
    Without feedback, gas piles up at $\gtrsim10^{5}\,{\rm cm^{-3}}$, 
    and there is essentially no hot gas.
    Removing radiation pressure leads to a 
    similar pile-up at very high densities, even in the MW \&\ SMC cases 
    (in which SNe heating alone can regulate {\em global} properties). 
    Removing gas heating (by SNe and stellar winds) leads to more low-density 
    cold gas and dramatically suppresses the hot gas mass.
    \label{fig:rho.dist.all}}
\end{figure*}

\subsection{Dependence on Feedback Mechanisms}
\label{sec:ism.structure:vsfb}

It is useful to identify observational tests which 
discriminate between different feedback mechanisms, 
and in this section we explore some candidates. 
In Figure~\ref{fig:rho.dist.all}, we consider how the 
distribution of cold and hot gas depends on the various feedback mechanisms 
implemented. 
With no feedback, there is runaway collapse in the cold gas with a 
large secondary peak at $n\rightarrow10^{6}\,{\rm cm^{-3}}$ 
(the highest densities we can resolve).
Turning off {\em all} heating mechanisms (SNe, stellar winds, and HII photoionization-heating) 
together leads to a modest increase in the amount of 
high-density $n\gtrsim10^{4}\,{\rm cm^{-3}}$ 
material in the MW and SMC models (these mechanisms are also responsible for 
heating the small tail of ``cold'' material at $n\sim0.01-1\,{\rm cm^{-3}}$
which appears when feedback is turned off). 
But in all models, turning off radiation pressure yields a much more dramatic 
increase in the amount of very dense material. 
In other words, even where global 
self-regulation can be set by SNe heating, the dense material 
at $n\gtrsim 10^{4}\,{\rm cm^{-3}}$ is regulated by radiation 
pressure. This should not be surprising -- even in a low-density 
galaxy like the SMC, the optical depths at these densities 
are large in the infrared; and the cooling times for a SNe remnant in 
such high-density material are $\sim10^{4}$ times shorter than 
the dynamical time, so pure gas heating is ineffective.

With no feedback, there is essentially no hot gas, in any model -- 
pure gravitational turbulence and shocks alone can maintain 
some gas at $\sim10^{4}\,$K or so, but not $>10^{6}\,$K. 
Without gas ``heating'' from SNe and shocked stellar winds 
the mass of hot gas drops by a 
large factor of $\sim10-100$, even in the HiZ and Sbc 
cases. The turbulent motions in the disk driven by 
radiation pressure have velocities comparable to the 
escape velocity from massive star clusters -- 
tens of ${\rm km\,s^{-1}}$, so shocks between them produce 
little ``hot'' material. HII photoionization-heating obviously 
does not produce temperatures in this range.
As we have shown above, most of the hot gas comes from 
SNe (the integrated hot gas energy in stellar winds being 
smaller by a factor $\sim8$, even though their instantaneous 
luminosity in the OB phase can be higher). 

The large variation in the amount and average density of the dense 
gas in Figure~\ref{fig:rho.dist.all}, and the deviations from 
log-normality, are striking, given that 
we showed in Figure~\ref{fig:hr.vs.t} that the different feedback models all 
produce nearly identical velocity dispersions and self-regulation at $Q\approx1$. 
They {\em all} maintain large-scale turbulence, in other words, even though 
some experience runaway collapse (which is then reflected in the 
much larger SFRs for these models). But it is commonly assumed in 
various star-formation models that 
maintaining large-scale turbulence alone 
is sufficient to prevent runaway collapse \citep[e.g.][]{ballesteros:gmc.core.lifetimes,
tasker:2009.gmc.form.evol.gravalone,krumholz.schmidt}. 
Moreover many high-resolution studies of idealized, turbulent boxes have 
found that ``pumping'' the global modes in the box with momentum 
equivalent to $Q\sim1$ establishes a lognormal density PDF 
\citep[with a dispersion that scales very weakly with mach number 
$\propto \sqrt{\ln{(1+3\,\mathcal{M}^{2}/4)}}$; e.g.][]{vazquez-semadeni:1994.turb.density.pdf,
padoan:1997.density.pdf,scalo:1998.turb.density.pdf,
ostriker:1999.density.pdf,federrath:2008.density.pdf.vs.forcingtype,price:2011.density.mach.vs.forcing}
and prevents runaway collapse/star formation 
\citep{vazquez-semadeni:2003.turb.reg.sfr,li:2004.turb.reg.sfr,li:2006.turb.reg.sfr}.

However, there are two major differences between our simulations 
and many idealized turbulence models. 
First, the typical idealized box simulation realizes only a sub-region of 
one cloud: with a mass of a few tens typically measured in 
units of the thermal Jeans mass ($\sim c_{s}^{3}\,G^{-3/2}\,\rho^{-1/2}$); 
as such the statistics only extend to the $\sim10\%$ level (i.e.\ to 
$\sim1\,\sigma$ in the distribution), and ``lognormality'' 
is seen only on a linear scale. Indeed, even in these simulations, hints 
have often appeared of non-lognormal ``tails'' in the distribution 
\citep[see][]{ballesteros-paredes:2011.dens.pdf.vs.selfgrav}.
But with such sampling, it is not possible for a large sub-region (itself containing 
many Jeans masses) to detach from the global flow, dissipate, and collapse.
In contrast our simulations here 
sample $>10^{7}$ thermal Jeans masses, and so can easily probe 
such fluctuations. 

Second, we include cooling and self-gravity, 
which are not always followed self-consistently in 
idealized simulations. This is critical for ``runaway.'' 
Turbulence does tend to drive the 
distribution to a lognormal. But the tails converge very slowly (with the Poisson error) -- according to the 
central limit theorem, convergence to lognormality in a given portion of the 
distribution requires a number of ``events'' such that $N^{-1/2}\ll1$, i.e.\ for each Jeans mass region 
(here more properly defined in terms of the turbulent Jeans mass), 
$N_{\rm crossing\ times}^{-1/2}\ll 1$. But if cooling is fast relative to the dynamical time, 
a self-gravitating region collapses on a single crossing time. Outside of $\sim1-2$ standard deviations from the 
core of the lognormal, then, a region cannot be ``held up'' or dissociated by cascades of turbulence from large scales 
``mixing'' it back to low densities before it collapses.
Regions will instead be continuously ``pumped'' to 
randomly cross this threshold, at which point they detach and collapse, building up the excess at 
large densities. It is therefore critical to invoke a feedback mechanism that can act {\em directly} 
on the gas at very high densities to dissociate these collapsing regions before they completely 
run away, and re-mix them into the larger-scale medium. 

And this is exactly what we see: the fact that all the simulations maintain similar velocity 
dispersions is the reason that the median/peak density (around $\sim10-100\,{\rm cm^{-3}}$) 
in the dense gas is similar in all cases, and within 
$\sim1\,\sigma$ of this peak, they almost all appear reasonably well-behaved lognormal distributions.
But the radiation pressure and other feedback 
acting in the dense gas is critical to suppress the runaway ``tails'' seen in the no-feedback runs,  and in the ``No Radiative Momentum'' runs in the three more massive galaxy models. 
It is these tails which in turn contribute the 
gas that is actually forming stars (densities $>10^{4}\,{\rm cm^{-3}}$), and 
hence lead to runaway star formation.

\begin{figure}
    \centering
    \plotonesmall{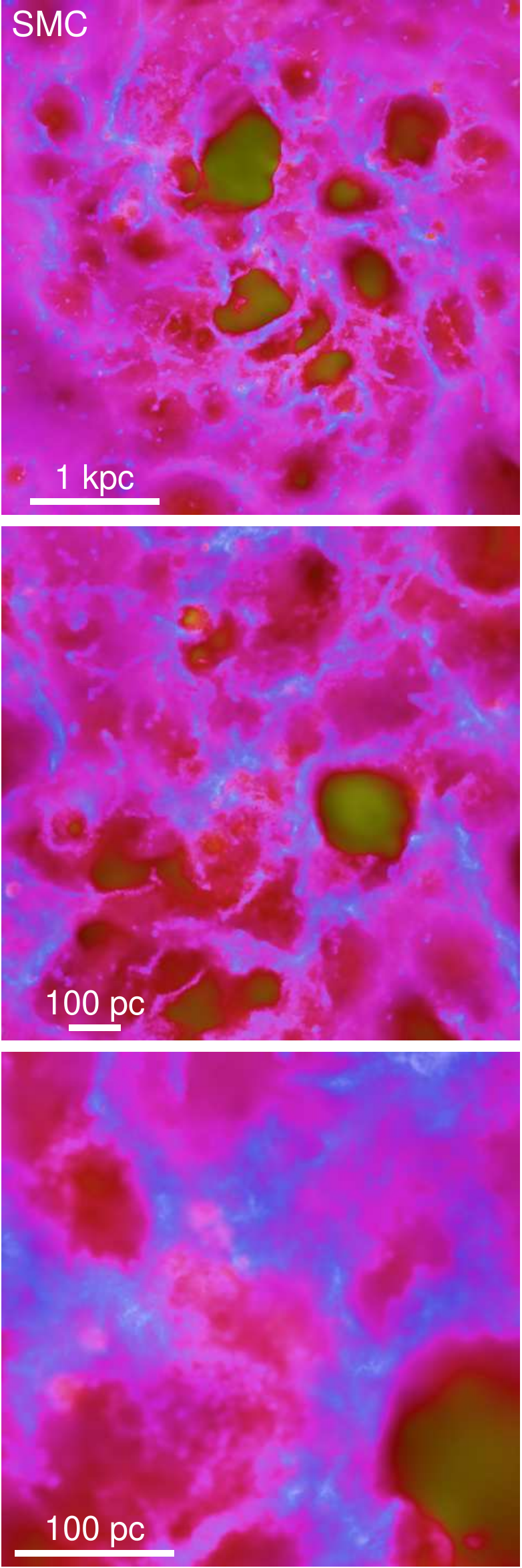}
    \caption{Zoom-in of regions with a large 
    concentration of GMCs in the SMC simulation.   As in Figure~\ref{fig:morph.hiz}, brightness encodes projected gas density (logarithmically scaled with a $\approx4\,$dex stretch) and color encodes gas temperature, with blue being $T\lesssim1000\,$K molecular gas, pink $\sim10^{4}-10^{5}$\,K 
    warm ionized gas, and yellow $\gtrsim10^{6}\,$K hot gas.  The filamentary nature of clouds, with dense cores (bright white-blue regions) and diffuse 
    molecular material (diffuse blue regions), 
    and $<10\,$pc-scale structure are evident.
    \label{fig:morph.demo.d}}
\end{figure}

\section{The Properties of GMCs}
\label{sec:clumps}

Having considered the global properties of the ISM above, 
we now turn to the sub-structure predicted: specifically, the properties of giant 
molecular clouds. Figure~\ref{fig:morph.demo.d} shows an 
illustrative example of a region near the center of the SMC simulation, 
with GMC structure resolved down to $\sim$pc scales. It is clear that 
the GMCs formed exhibit considerable structure, with 
filamentary large-scale molecular gas and dense knots and clumps 
(the star-forming regions) embedded throughout. The irregular, clumpy, 
and triaxial morphology of these 
clouds is very similar to typical observed GMCs. They are {\em not} 
``small disks,'' which is the typical case when ``GMCs'' are formed in 
simulations with weak feedback (a point we will discuss in more detail below). 

To identify individual clouds (especially given their 
complex morphology), we follow a procedure motivated by observational 
studies, but taking advantage of the full simulation data.
Specifically, we consider each simulation output snapshot and 
apply the sub-halo finder SUBFIND, which employs a friends-of-friends 
linking algorithm with an iterative unbinding procedure to robustly identify 
overdensities \citep[for details and tests, see][]{springel:cluster.subhalos}. 
The routine is modified to run on the disk gas (instead of dark matter) 
and with softenings and linking lengths appropriately rescaled for the mean 
gas densities within the disk effective radius. 
Visual inspection confirms that this correctly identifies the 
obvious clumps, and binding criteria (discussed below) also support the identifications. 
Changing the linking lengths and other numerical quantities has a weak effect 
on most properties, but we explicitly note where the GMC properties are
sensitive to the algorithm. This tends to occur in the most massive 
clumps in each simulation, as these are often located at global 
critical points (e.g.\ along spiral arms or bars) and have their own sub-structure, 
so deciding whether or not to break them into sub-units is sometimes 
ambiguous. This same uncertainty, however, applies to observational 
studies as well -- we discuss its possible effects where they may be 
important below, but these few cases do not affect our qualitative 
conclusions.

\subsection{GMC Mass Functions}
\label{sec:clumps:mass:gmcs}

\begin{figure*}
    \centering
    \scaleup
    \plotside{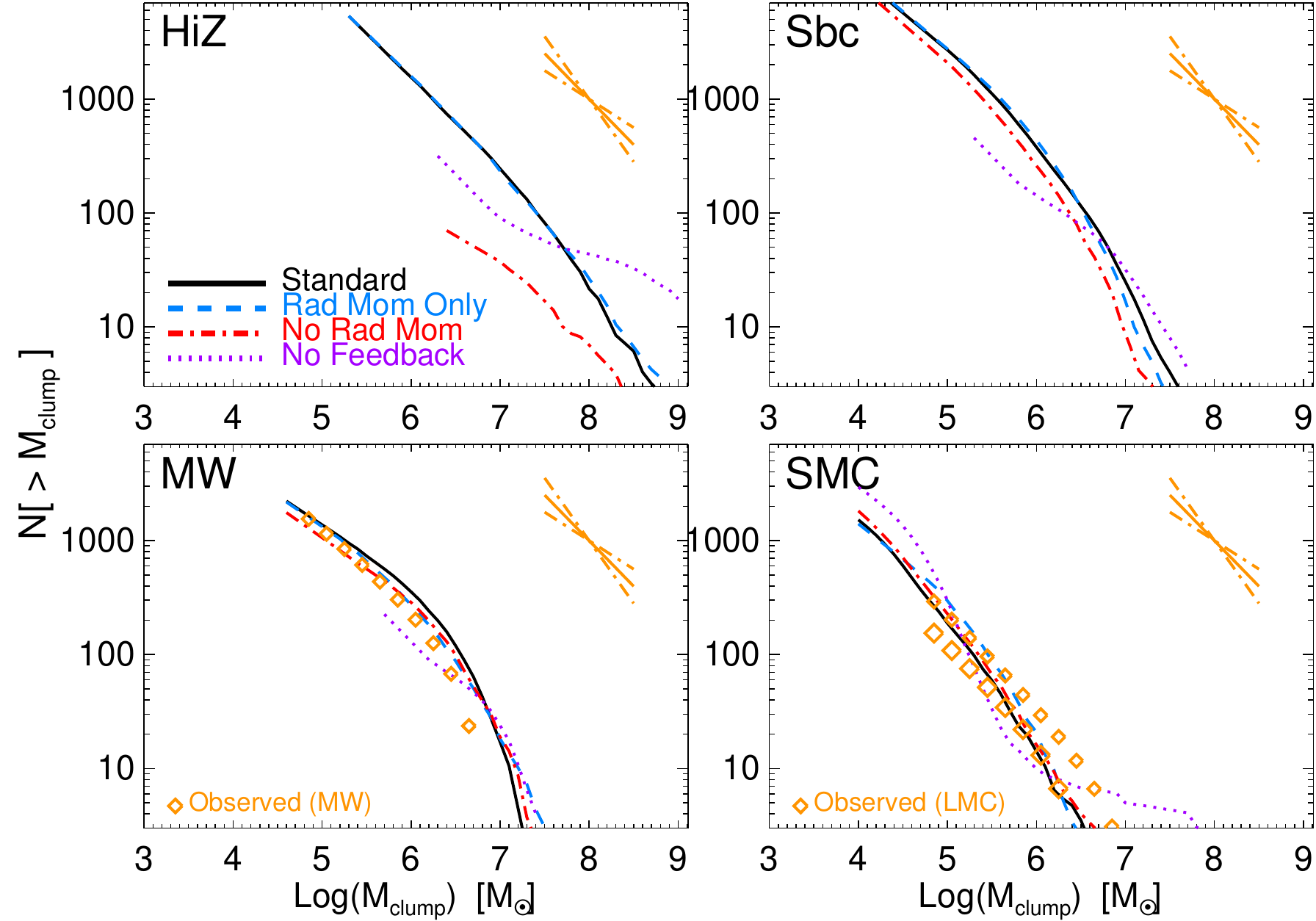}
    \caption{Predicted mass function of GMCs in the simulations (see \S \ref{sec:clumps} for a discussion of how we identify GMCs).
    We show the different galaxy models, each with runs that include 
    different feedback mechanisms (see Table 2).   The distributions can all be fit by a power-law 
    with a cut-off at high mass (eqn.~\ref{eqn:nclumps}).      For the MW model and dwarf/SMC model, we compare with the observed MW and LMC mass functions, respectively 
    \citep[][orange diamonds]{williams:1997.gmc.prop,
    rosolowsky:gmc.mass.spectrum}. Characteristic slopes in {\em observed} systems 
    are also shown for comparison (orange lines for 
    $\alpha=-1.5,\, -1.8,\,-2.1$, where $N(>M) \propto M^{-\alpha + 1}$). 
    In all models, the predicted slope is similar to that observed: it follows generically 
    from gravitational collapse (with super-sonic flows).
    The cutoff mass traces the Jeans mass in each model 
    (eqn.~\ref{eqn:mjeans}); since all reach $Q\sim1$, the high mass cutoff is weakly sensitive to feedback, primarily through more/less efficient gas exhaustion ($f_{\rm gas}(R)$). 
    With weak/absent feedback, the very high-mass tail is enhanced even though typical clouds are less massive; this is because massive clouds along global ISM structures (e.g., spiral waves)
    do not disrupt but accrete continuously.
    \label{fig:clump.mf}}
\end{figure*}

Figure~\ref{fig:clump.mf} shows the resulting GMC mass function, 
for our different galaxy models with feedback enabled.
The results converge to a quasi-steady state, despite the fact that (we will show) individual 
clumps are short-lived, so the mass function reflects continuous dissociation and re-formation. 
The shape in all cases is an approximate power law with a steep (exponential or log-linear) cutoff 
at some high mass, so one could fit the usual truncated power-law or Schechter functions: 
\be
\label{eqn:nclumps}
N(>M) = N_{0}\,{\Bigl (}\frac{M}{M_{0}} {\Bigr)}^{-\alpha+1}\,\exp{(-M/M_{0})}
\ee
we choose this form so that $\alpha$ corresponds to the 
logarithmic slope of $dN/dM$, standard in observational studies.
Observed values tend to lie within the range $-2.1\lesssim \alpha \lesssim -1.5$ 
shown in the Figure~\ref{fig:clump.mf}, 
with the ``canonical'' value of $-1.8$ very close to the typical 
slope we find (fitting the function above to all simulation timesteps, we find 
an average $\alpha\approx-1.75\pm0.2$). 

We show this explicitly by comparing the observed GMC mass functions 
for the MW \citep[from][corrected to represent the total MW population, 
since we consider all GMCs in the simulation]{williams:1997.gmc.prop}, 
appropriate to compare to our MW-like simulations, 
and for the LMC \cite[from][]{rosolowsky:gmc.mass.spectrum}, 
appropriate to compare to the SMC-like simulation. 
Of course the normalization $N_{0}$ of the observed and simulated populations 
is arbitrary and depends on the total gas mass, but for these comparisons the total 
gas mass is similar. We expect the Sbc case 
to be qualitatively similar to these but with different normalization and zero point.
For the high-redshift case, the total gas mass and most 
massive clump mass are much larger than local systems, so there 
is no direct comparison available, but observations have indicated that 
typical systems similar to our HiZ model have approximately $\sim5-10$ clumps 
in the $\gg 10^{8}\,\msun$ range, similar to the number predicted 
\citep{forster-schreiber:2011.ifu.clump.optical.obs}. 

The mass function slope appears fairly generic. To see this, we 
compare models with different feedback mechanisms enabled 
and disabled, as in \S~\ref{sec:dispersion}. 
The characteristic or maximum masses may shift by small amounts without certain 
feedback included, but the slope $\alpha$ is similar in 
all cases. We have also checked this within each 
feedback model, for example, arbitrarily multiplying or dividing the 
strength of the momentum coupling from radiation pressure by a factor of $5$. 
A mass function slope close to $\alpha\sim-2$ tends to 
emerge generically in any system dominated by gravitational collapse 
\citep[e.g.][]{pressschechter,bond:1991.eps}; 
the location of the high-mass cutoff and exact deviation from 
$-2$ depend on details of e.g.\ the power spectrum, non-gravitational 
terms, and Jeans conditions (if the medium is gaseous), 
but the low-mass slope must be close to $-2$ because gravity is 
scale-free.\footnote{Implicit here is the assumption that motions 
are super-sonic, so the thermal properties do not set a preferred length 
scale, but this is easily satisfied in the simulations and observations, 
down to scales well below what we resolve ($\lesssim 0.03\,$pc).}
This explains why simulations with 
very different physics included recover similar behavior 
\citep[compare our runs and e.g.][]{audit:2010.gmc.massfunctions,
tasker:2011.photoion.heating.gmc.evol}.

The cutoff mass $M_{0}$ appears to be simply 
related to the turbulent Jeans mass in each disk:
\begin{align}
\label{eqn:mjeans}
\nonumber M_{J} &= \frac{\sigma^{4}}{\pi\,G^{2}\,\Sigma} = 4\,\pi\,\Sigma\,h^{2} \\ 
&\approx {\Bigl(} \frac{\nu\,Q}{2} {\Bigr)}^{2}\,f_{\rm gas}^{2}\,M_{\rm gas}(<R)
\end{align}
where the latter expression includes the Toomre $Q\sim1$ 
and the parameter $\nu=1-2$ which depends on the mass profile, 
and $f_{\rm gas}$ refers to the fraction of gas mass relative to the 
{\em total} enclosed mass.
Evaluated at $R_{e}$ for the typical conditions of each galaxy model 
seen in Figure~\ref{fig:hr.vs.t}, this gives 
$M_{J}\sim (1-5\times10^{6},\,0.5-2\times10^{7},\,1-4\times10^{7},\,0.3-1\times10^{9})\,\msun$ 
for the SMC, MW, Sbc, and HiZ cases. Of course, 
the exact values will change with time and the precise location in the disk, 
and mergers and agglomerations 
can grow clumps beyond this limit. 
But it appears to be a quite good approximation to the ``cutoff'' in the 
integrated mass functions.

When we compare models with different feedback mechanisms enabled 
or disabled, the slope is similar, but there are some small shifts 
in the ``cutoff'' mass. At low/intermediate masses, models with less 
efficient feedback are shifted to slightly lower GMC masses. 
This is most noticeable when we remove 
radiation pressure in the HiZ or Sbc cases, remove hot gas
in the SMC case, or remove all feedback in any case -- just 
what we expect, since we already saw these have the largest 
effects in Figure~\ref{fig:hr.vs.t}. 
Recall, we showed above that all the models tend to $Q\sim1$, 
so this primarily reflects somewhat more efficient gas exhaustion 
lowering $M_{\rm gas}$ and so the Jeans mass. 

However at the very highest masses the behavior is perhaps unexpected; 
despite these shifts the extreme high-mass ``tail'' of the mass function 
extends to higher maximum masses. 
Two effects drive this. First, 
the most massive clumps tend to be massive clump-complexes along 
e.g.\ spiral arms, which are linked by intermediate-density material. 
Without being able to fully dissociate this material, 
the friends-of-friends GMC finder tends to link these into single 
``super-clumps.'' This alone is ambiguous -- it is well known in observations 
that at high masses the assignment of ``clump mass'' is quite sensitive 
to the exact clump identification and linking criterion 
\citep[see][]{pineda:2009.clumpfind.issues}. We err on the side of over-linking rather than 
missing real clumps, but if this alone drove the effect seen we would 
not assign it much meaning. However there is also a robust physical effect: 
clumps survive longer with weak feedback. The same most massive 
clumps, located along global structures, can then accrete large quantities of 
gas, growing beyond the Jeans mass. This leads to the odd shapes 
of the high-mass mass function in these cases, where the high-mass 
end does not rapidly fall off, but ``turns up'' in a manner not observed to date.

\subsection{Density Distributions}
\label{sec:clumps:mass:density}


\begin{figure}
    \centering
    \scaleup
    \plotone{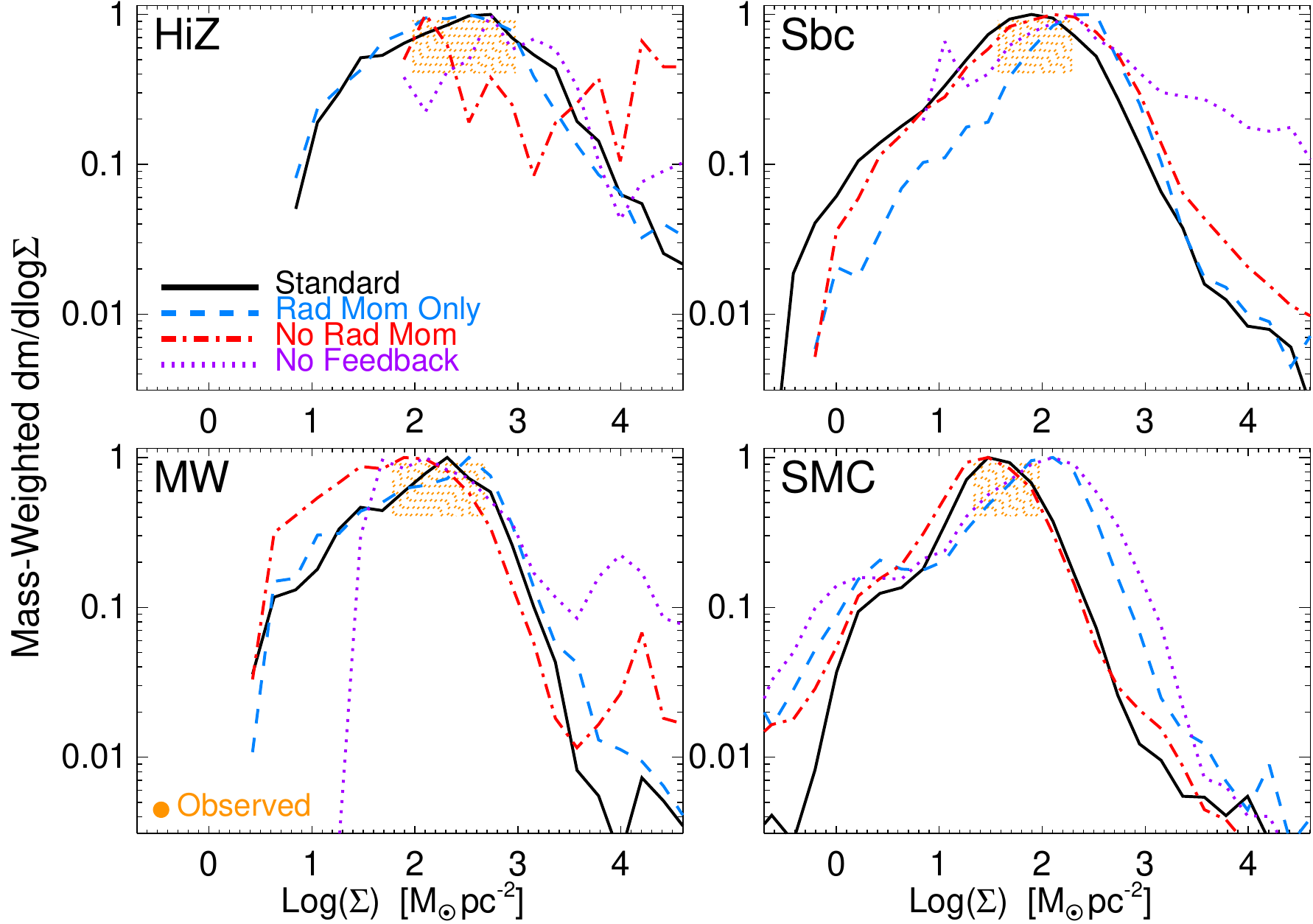}
    \caption{Distribution of average GMC surface densities $\langle \Sigma \rangle$, 
    calculated as the mass-weighted $\langle \Sigma \rangle$ averaged over viewing angles
    for each cloud. We show each 
    disk model with different feedback mechanisms enabled (see Table 2). 
    The distributions are close to log-normal in the core, with the median surface density
    $\sim50-300\,\msun\,{\rm pc^{-2}}$
     increasingly modestly as the 
    average gas surface density of the star-forming disk increases. The scatter in the surface density of GMCs is small ($\approx0.2$\,dex).
    For comparison, we show the observed median $\langle \Sigma \rangle$ and 
    $\pm1\,\sigma$ range (shaded orange) inferred 
    for SMC and MW clouds and the range for all dwarf galaxies observed for the Sbc-like run 
    \citep{bolatto:2008.gmc.properties}; for the HiZ run we 
    take the estimate for $z\sim2$ galaxies in \citet{forsterschreiber:z2.sf.gal.spectroscopy}.
    With no feedback, {\em local} collapse proceeds without limit (to high $n$), 
    which gives rise to the high-$\Sigma$ tail in the GMC surface densities seen here, but 
    does not necessarily change the median $\langle \Sigma \rangle$.
    With feedback, clouds are more triaxial and short-lived and so do not globally contract by a large amount.
    \label{fig:clump.sigma}}
\end{figure}

\subsubsection{Three-Dimensional Densities}
\label{sec:clumps:mass:density:3d}

The distribution of three-dimensional densities $n$ in clouds is 
more or less the distribution of ``cold gas'' shown in Figure~\ref{fig:rho.dist.all}. 
As discussed above, the distributions in our standard models are crudely log-normal 
with median $\sim10-100\,{\rm cm^{-3}}$ and scatter $0.25-0.30$\,dex,
within which most of the GMC mass resides. 
This value increases systematically with the average density of the galaxy; 
fitting a mode to each lognormal gives $\approx 25\,{\rm cm^{-3}}$ in the 
SMC model and $100\,{\rm cm^{-3}}$ in the HiZ model. This must happen: the HiZ 
model has an initial smooth disk-averaged $n\sim40\,{\rm cm^{-3}}$ inside $R_{e}$, 
so any clumps must have larger $n$. 
There are tails at the $<10\%$ level in the distributions which include the high-$n$ cores. 
This reproduces a key feature of observed GMCs: although 
most of the mass within e.g.\ the star-forming galactic disk is in ``dense'' 
gas, in the form of GMCs, most of the gas {\em within} those GMCs is 
at relatively low densities, {\em not} within star-forming cores 
that have densities $\gtrsim 10^{4}\,{\rm cm^{-3}}$ 
\citep[e.g.][and references therein]{williams:1997.gmc.prop,evans:1999.sf.gmc.review}.

As noted above, these distributions remain steady-state throughout the simulations. After the initial couple dynamical times, the median densities vary (randomly) within factors $\lesssim1.5$ and dispersions vary within $0.1$\,dex. This is non-trivial -- as time passes, there is not a one-way collapse to high densities but self-regulation at a constant density distribution. 

However, models without feedback do not have a stable density distribution, and ``pile up'' at the highest resolvable densities $n\gg10^{4}\,{\rm cm^{-3}}$ as seen in Figure~\ref{fig:rho.dist.all}. We saw there that this is primarily a function of the radiation pressure support. In the MW and SMC models, gas heating by stellar winds and HII photo-ionization is able to stave off complete runaway cloud collapse but without radiation pressure the number of ultra-dense cores is enhanced. We emphasize though, that 
the effect of feedback is primarily on the high-density tails in Figure~\ref{fig:rho.dist.all}; since most of the gas is at low (non star-forming densities), its distribution is similar even in the presence of runaway local collapse and star formation.

\subsubsection{GMC Surface Densities \&\ Size-Mass Relation}
\label{sec:clumps:mass:density:2d}

A closely related metric is the average cloud surface density $\langle \Sigma \rangle$, 
shown in Figure~\ref{fig:clump.sigma}.
Here we use 
the mass-weighted average surface density over sightlines through the cloud $\langle \Sigma \rangle$; 
this is not exactly the same as the area-average $\sim M/\pi\,R^{2}$ but is well-defined even for 
arbitrary cloud geometry.
In each model, the $\langle \Sigma \rangle$ distribution follows a relatively narrow 
log-normal. The median GMC surface density increases 
from $\approx 50\,\msun\,{\rm pc^{-2}}$ for the SMC, to $\approx500\,\msun\,{\rm pc^{-2}}$ 
in the HiZ model, as they must given the increasing background density.
The scatter is again roughly constant, only $\approx0.2$\,dex, and 
the distributions in this projection are closer to exact log-normal than $n$. 
This narrow distribution in $\Sigma$ means the clouds follow a size-mass relation 
approximately $R\propto M^{0.5}$, with a normalization consistent with the 
observed range (weakly increasing with galaxy surface density) and small scatter.

These surface densities (and corresponding size-mass relation) and their scatter 
are quite similar to those observed in MW-like and SMC-like local galaxies 
\citep{bolatto:2008.gmc.properties,fukui:2008.lmc.gmc.catalogue,
wong:2008.gmc.column.dist,goodman:2009.gmc.column.dist}. 
Observations have even found evidence for a systematic shift in the normalization of the 
size-mass relation ($\langle \Sigma \rangle$) towards lower densities in SMC clouds and 
higher densities in dense regions of massive galaxies \citep{bolatto:2008.gmc.properties,
heyer:2009.gmc.trends.w.surface.density}.
In fact, the properties of the most massive clouds in our HiZ simulation 
agree well with those inferred for the most massive star-forming molecular ``clumps'' 
in $z\sim2-3$ galaxies observed with integral field data and 
in some cases individually lensed \citep{forsterschreiber:z2.sf.gal.spectroscopy,
swinbank:clumps}, which are offset towards higher $\langle\Sigma\rangle$ 
relative to the extrapolation of the local relation.

Our predictions 
are also, interestingly, similar to the predictions in higher-resolution simulations of individual 
GMCs with quite different physics included \citep[e.g.\ ignoring radiation pressure 
but including magnetic fields; see][]{ostriker:2001.gmc.column.dist,
tasker:2011.photoion.heating.gmc.evol}. 
This suggests that they are fairly generic.
No-feedback cases do feature a larger 
tail towards high-$\Sigma$, but the median values are actually quite 
similar to the cases with feedback. This fits with the behavior we 
saw in \S~\ref{sec:dispersion:vsigma}, where dissipation is efficient in the vertical 
direction of the disk without feedback, but there is still large radial 
and azimuthal motion. In detail, what we see with no feedback is often 
clouds collapsing, dissipating all of their energy in a crossing time, but then 
being arrested by residual angular momentum. They become small, 
rapidly rotating disks, with clumpy sub-regions collapsing further. 
This has been seen in other high-resolutions disk 
simulations without feedback, as well \citep{bournaud:disk.clumps.to.bulge,
dobbs:2008.gmc.collapse.bygrav.angmom,
tasker:2009.gmc.form.evol.gravalone}. 
In other words, without feedback, GMCs do not collapse three-dimensionally, 
and so they can locally collapse to arbitrarily high densities, 
turning all their mass into stars in a couple crossing times, 
even though their surface densities are not very different from 
cases with efficient feedback. This behavior, and strong rotational support for 
highly ``flattened'' molecular clouds are, however, not observed 
\citep{rosolowsky:2003.gmc.rotation,imara:2011.gmc.ang.mom}. 
Much closer to observed systems are the models 
with strong feedback, which have similar local $\sigma_{z}$ and $\sigma_{R}$, 
leading to molecular clouds which, while certainly non-spherical, 
are not highly flattened \citep[they tend towards irregular, triaxial and 
filamentary shapes, with some preference for elongation in the galactic plane 
as observed; see][]{koda:2006.mol.cloud.shapes.vs.gal.plane} 
and not rotationally supported (see Figure~\ref{fig:morph.demo.d}).

With feedback included, clouds are three-dimensional; 
feedback has greatly slowed collapse. Moreover, 
we will show that with feedback, clouds are dissociated in just a few 
dynamical times. As a result, the clouds can undergo little global contraction 
(even though over-dense regions within the clouds can collapse to 
very high densities and form stars). The surface densities $\langle \Sigma \rangle$ 
are therefore just a reflection of the average surface densities of the disk 
from which they form. Specifically, the peak in typical cloud surface density 
in Figure~\ref{fig:clump.sigma} in each model is just $\approx2-5$ times the 
average disk surface density within the star-forming disk.
In other words, 
the remarkable independence of the global cloud densities on the details 
of feedback, and apparent narrow range in densities, is just a consequence of 
the fact that they are dissociated after just a few dynamical times, before 
they can globally contract (and are slowed in that contraction even during that 
time). This is quite similar to what is observed: although the 
star-forming {\em cores} within GMCs reach large 
columns $N_{H}\gtrsim 10^{24}\,{\rm cm^{-2}}$ (optically thick in the IR), 
MW GMC complexes are observed to have a surface-averaged column 
density of just $10^{21}-10^{22}\,{\rm cm^{-2}}$, 
similar to the average through the disk.


\begin{figure*}
    \centering
    \scaleup
    \plotside{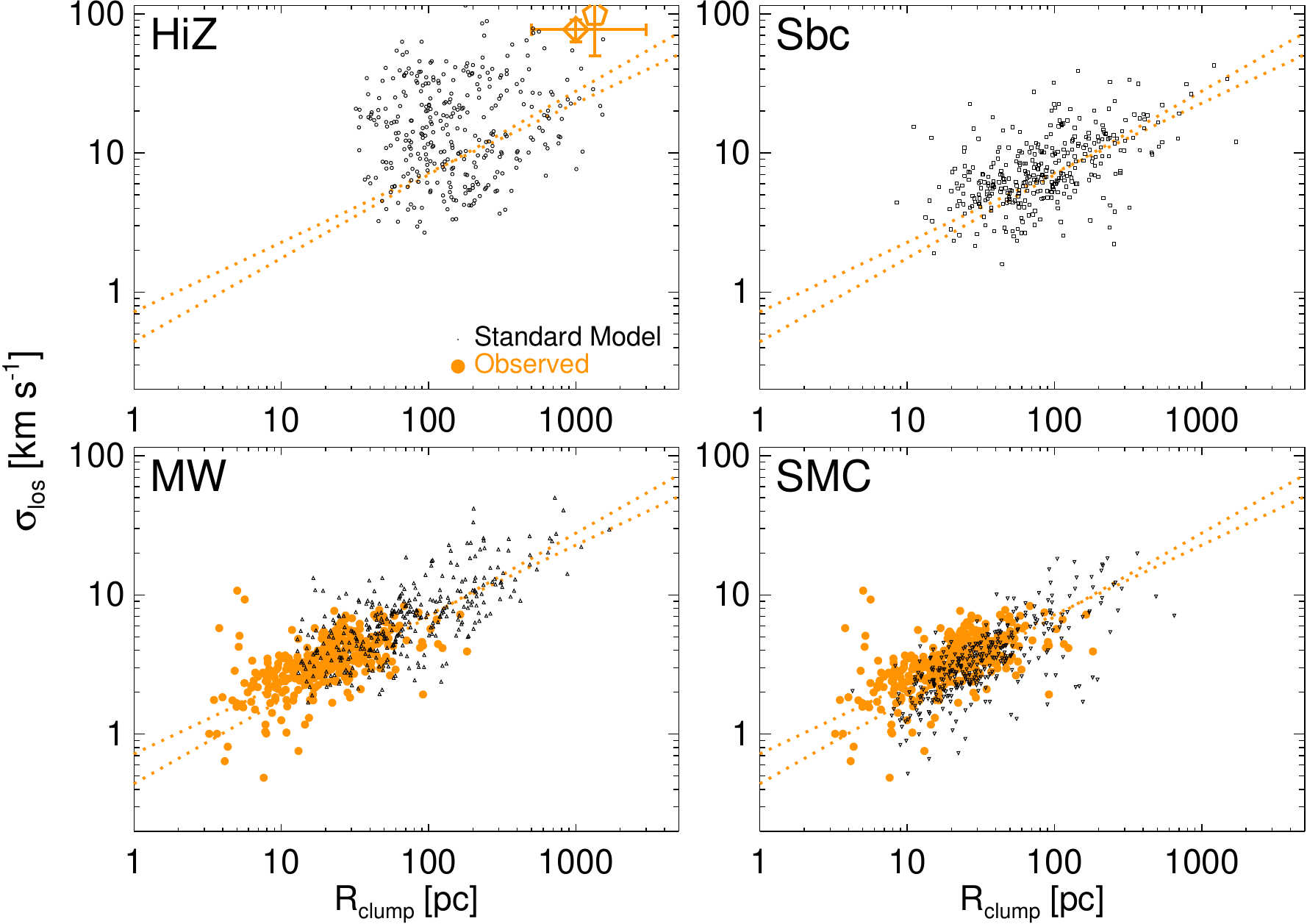}
    \caption{Linewidth-size relation of GMCs in our standard simulations of each galaxy model.
    We randomly sample clouds in time 
    for each disk model. We only show clouds with $>100$ member particles 
    and sizes $>5$ softening lengths (this determines the low-mass cutoff in the plots). 
    We compare observations from local-group galaxies
    \citet[][filled points]{bolatto:2008.gmc.properties}, 
    with typical best-fit power-laws $\sigma\propto R^{0.5-0.6}$ 
    (dotted lines). For the HiZ case, we compare typical properties of very 
    massive clumps observed in \citet{forsterschreiber:z2.sf.gal.spectroscopy,
    forster-schreiber:2011.ifu.clump.optical.obs} (open points). 
    Because GMCs are short-lived, these scalings 
    essentially reflect the initial collapse conditions, and so are relatively insensitive 
    to feedback.
    \label{fig:clump.linewidth}}
\end{figure*}

\subsection{Linewidth-Size Relation}
\label{sec:clumps:structure:linewidth}

Closely related to the typical densities and surface densities of clouds 
are the Larson's law-type scaling relations. We have already discussed 
the size-mass relation; in Figure~\ref{fig:clump.linewidth}, we plot 
the linewidth-size relation. For simplicity, we define the radii and velocity as the 
rms spherical radii and velocity dispersion; 
because observations generally de-project to 
three-dimensional quantities assuming an ellipsoid in 
projection, this is probably a reasonable approximation 
at the level we consider here.
We are not making a detailed mock observation of molecular gas in various states, 
so there will be some systematic uncertainties in the comparison, but the senses of the scalings 
are robust to various density limits and definitions. 
For clarity, we plot just a small randomly sampled subset of the 
clouds.
The predicted linewidth-size relation is broadly similar to that observed
with an approximate $\sigma\propto R^{0.5}$ scaling, in the range where 
there is overlap ($\sim5-300\,$pc), and continues the observed scalings beyond that to 
larger $\sim$kpc clump sizes in the massive systems \citep[we plot two characteristic points
from the observations of][for $z\sim2$ clumpy disks]{forsterschreiber:z2.sf.gal.spectroscopy,
forster-schreiber:2011.ifu.clump.optical.obs}. 
As expected from the behavior of $\langle \Sigma \rangle$, there is a 
normalization shift such that the HiZ systems 
obey a similar power-law but with higher $\sigma$ at fixed $R$ (reflecting their higher densities).
The predicted values, which exhibit this shift, appear consistent with the observed massive high-redshift systems.
For clarity, we have shown just our ``standard'' model, because the power-law scaling 
is similar for most of our models, with normalization offsets that follow simply from the 
differences in average density (above) and virial parameters (discussed below).

Again, models with a wide range of different physics included appear to capture 
the linewidth-size relations. As with the size-mass relation, the key point is that with 
feedback present, clouds are short-lived and collapse weakly. Moreover 
because the scaling above is based on second moments in velocity and size, 
it is especially sensitive to the less tightly-bound/collapsing material. 
As such, the observed scaling reflects the conditions of marginal 
self-gravity and Jeans collapse. The scale-length of Jeans collapse is just 
$\lambda\sim \delta v^{2}/\pi\,G\,\Sigma$. 
So if the systems have not been able to evolve very far from their initial collapse,
i.e.\ if they are short lived, then we simply 
expect $\delta v \propto \sqrt{\lambda}$ across regions with similar $\Sigma$ , 
with a residual normalization dependence 
\be
\frac{\delta v}{R^{0.5}} \propto \Sigma^{0.5}
\ee
similar to what is observed and simulated.

\begin{figure}
    \centering
    \scaleup
    \plotone{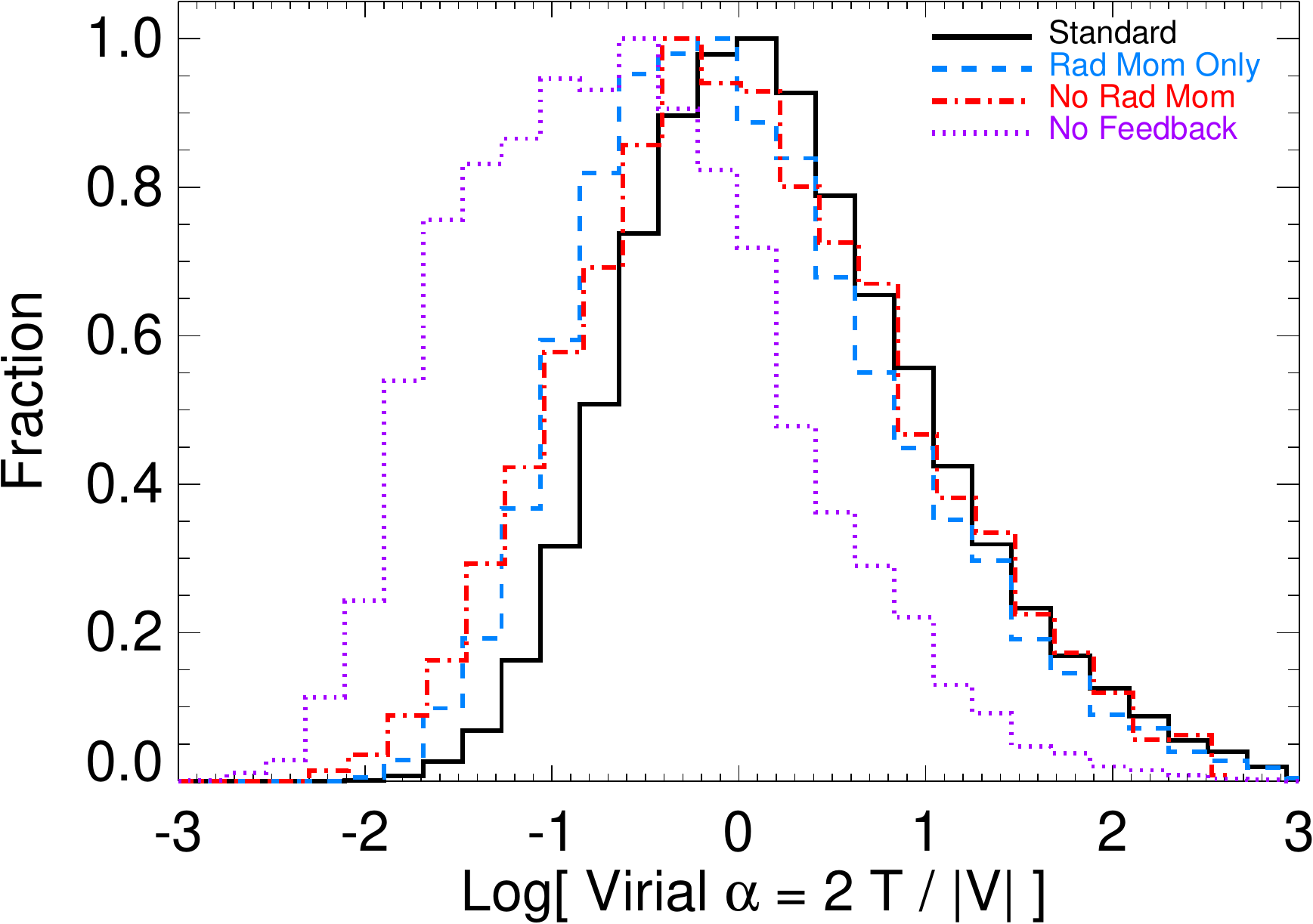}
    \caption{Predicted distribution of  GMC virial parameters 
    (the ratio of kinetic to potential energy; eqn.~\ref{eqn:alpha.vir}). 
    We plot the distribution for all disk models together, since the individual 
    results are very similar.   Once sufficient feedback exists to resist collapse and make clouds 
    relatively short-lived, all models equilibrate at marginal binding 
    ($\alpha \sim 1$ and is only weakly correlated with feedback strength); 
    With no feedback, 
    however, clouds are much more tightly bound, with global $\alpha < 0.1$.
      \label{fig:vir.param}}
\end{figure}

\begin{figure}
    \centering
    \scaleup
    \plotone{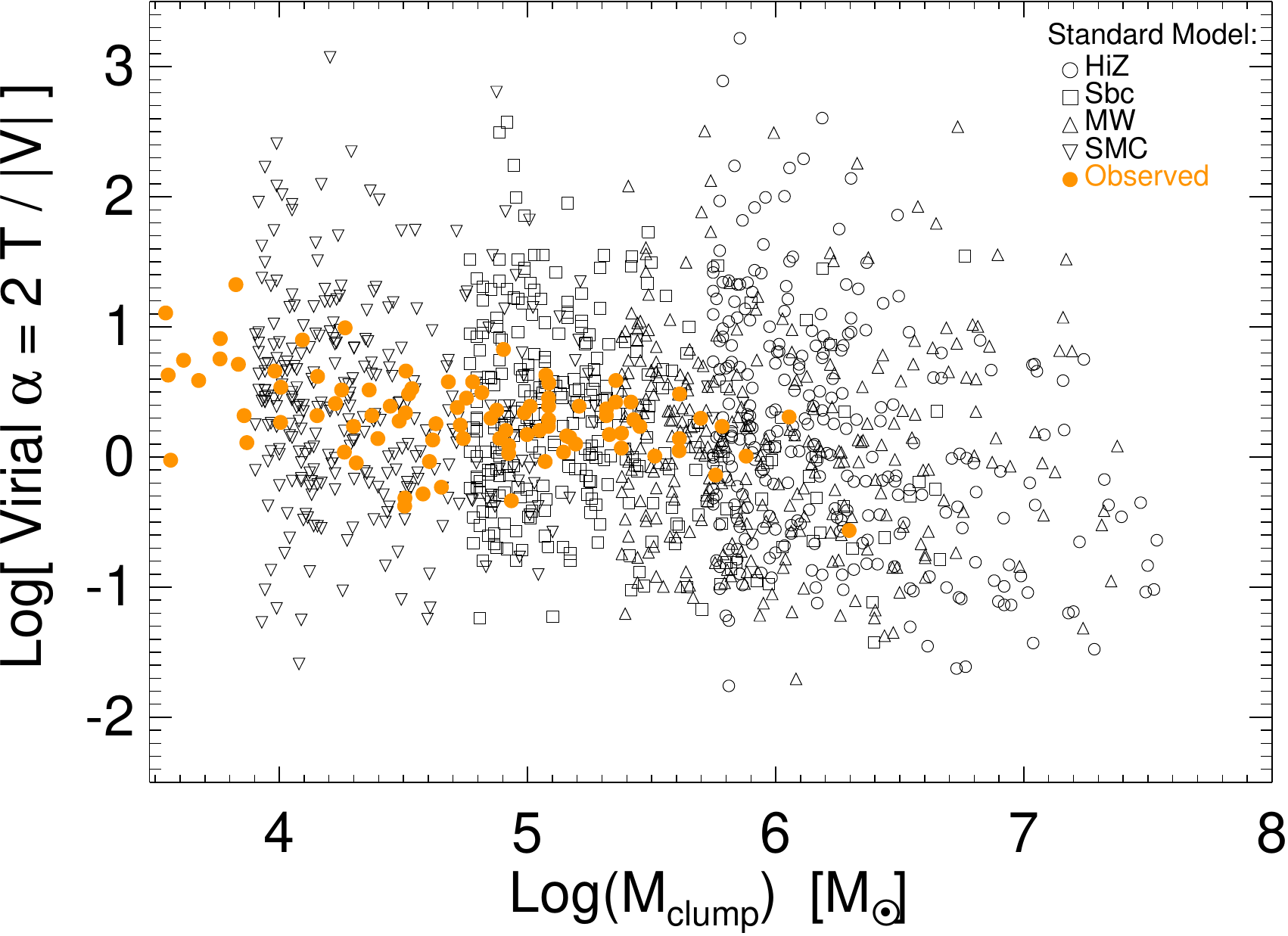}
    \caption{Correlation of the GMC virial parameter $\alpha$ (Figure~\ref{fig:vir.param}) 
    with  mass for our standard simulations with all feedback mechanisms enabled.   We compare the observed local group clouds from \citet{solomon:gmc.scalings}. 
    The model and observations both show a weak trend in which more massive GMCs tend to be 
    more tightly bound $\alpha\propto M_{\rm cl}^{-0.3}$.
    \label{fig:vir.param.vs.mass}}
\end{figure}

\subsection{GMC Virial Parameters}
\label{sec:clumps:structure:virial}

Figure~\ref{fig:vir.param} shows the 
distribution of virial parameters $\alpha$ of the simulated clouds, 
defined as the ratio of kinetic to potential energy. 
Observationally, this must be estimated from 
global cloud properties, so is commonly defined as
\be
\label{eqn:alpha.vir}
\alpha\equiv {\Bigl |}\frac{2\,T}{V}{\Bigr |} 
\approx 
\frac{a_{1}\,M_{\rm cl}\,\sigma^{2}/2}{a_{2}\,G\,M_{\rm cl}^{2}/R_{\rm cl}}
\approx \frac{5\,\sigma^{2}\,R_{\rm cl}}{G\,M_{\rm cl}} = 
\frac{M_{\rm vir}}{M_{\rm cl}}
\ee
where the factor of $5$ follows from common 
assumptions for the terms $a_{1}$ and $a_{2}$ 
(order-unity constants that depend on the true mass profile 
shape and projection), $\sigma$ and $R_{\rm cl}$ are the 
average dispersion and radius,\footnote{For consistency 
with the observations we compare to, we 
take $\sigma$ and $R_{\rm cl}$ as randomly-sampled 
line-of-sight and corresponding projected dispersion 
and half-mass-radii.} and 
$M_{\rm vir}\equiv5\,\sigma^{2}\,R/G$. 
With this definition (if the assumed coefficients 
are correct), $\alpha=1$ would be virial equilibrium.

The ``average'' clumps are consistent 
with being marginally gravitationally bound, $\alpha\sim1$. 
There is a broad dispersion of $\gtrsim0.5$\,dex about this median, 
reflecting both different states of various clouds, but also 
projection effects, since the clouds are highly non-circular. 
This is all expected, since we argued above that clumps dissociate 
before they can become strongly bound or globally contract much. 

Within each cloud, there is a broad dispersion as well of 
``local'' virial parameters. Much of the diffuse material 
at $\langle n \rangle$ is actually un-bound 
(at least $\sim$half the material is at $\alpha>1$, 
and can reach $\alpha\sim10-100$). The bulk velocity of this material, 
in the center of mass frame of the GMC, exceeds the cloud escape velocity. 
This material is associated with the clump only in a transient sense (e.g.\ from turbulent 
compression, shocks, outflows from dense regions, or convergent flows). 
But dense clumps with local $n>10^{4}\,{\rm cm^{-3}}$ (the regions 
that will actually form stars, but a small mass fraction) tend to be bound. 
This is similar to what has been inferred from observations of 
GMC complexes \citep[e.g.][]{heyer:2001.mol.cloud.vir.vs.mass}. 
Quantitatively, we can consider the ``true'' $\alpha_{\rm true}$ 
(calculated knowing the actual true kinetic and potential energy 
relative to the cloud center of mass for every gas particle) 
for two separate populations: the cloud ``cores'' 
(for convenience, gas inside the spherical half-mass 
radius with relative velocity below the median of the clump material), 
and the remaining ``non-core'' material; 
we find typical $\langle \alpha_{\rm true} \rangle \sim 0.05-0.2$ 
and $\langle \alpha_{\rm true} \rangle \sim 30-100$, respectively.\footnote{Note 
that because the ``diffuse'' material is a non-negligible fraction of the mass, 
calculating the ``true'' $\alpha$ for all cloud material often gives large values, 
dominated by the material which is transiently associated in shocks and 
has large kinetic energy.}

With no feedback, clouds are 
at least an order-of-magnitude more tightly bound, $\alpha\sim0.05-0.1$. This follows 
from their (largely vertical) collapse, giving smaller $R$ and $\sigma$ 
at the same mass (these quantities often collapse together in a way 
that keeps the clouds not far off the linewidth-size relation, but gives 
much smaller $\alpha$ at fixed mass). However, once sufficiently strong 
feedback is in place to slow/dissociate clumps, the properties 
are relatively insensitive to the feedback model. This is not surprising:
provided sufficient energy/momentum to slow Jeans collapse and/or 
dissociate a cloud, objects will either globally equilibrate at marginal 
binding, or spend most of their lifetime at the point where 
infall reverses to expansion, i.e., at the maximum value of $\alpha$. If the feedback
operates rapidly, this minimum $\alpha$ will be near the value where star formation
commences. This is why (as we saw with other cloud properties) models with very different 
assumptions and more simplified models for feedback have also 
seen similar characteristic $\alpha$ \citep{dobbs:2011.why.gmcs.unbound}.
For the HiZ case, we again find that radiation pressure momentum is most important 
(removing sources of hot gas has little effect). 
For the SMC-like case, in contrast, gas heating plays a large role in regulating 
the virial parameters. 

Figure~\ref{fig:vir.param.vs.mass}  plots the virial parameters 
of GMCs as a function of their mass. We show this only for our 
standard model, as the systematic offsets 
between models can easily be read off Figure~\ref{fig:vir.param}. 
For comparison, we plot the observed points from \citet{solomon:gmc.scalings}. 
The simulated and observed GMCs agree well.
Not only is the median $\alpha$ similar, but the simulations reproduce 
the observed weak scaling of $\alpha$ with $M_{\rm cl}$, 
approximately $\alpha\propto M_{\rm cl}^{-0.3}$, such that 
more massive clumps tend to be more bound 
\citep[see][]{heyer:2001.mol.cloud.vir.vs.mass}.
This is partly related to the fact we show below, that more massive 
clumps tend to have longer lifetimes. 
We note, however, that Figure~\ref{fig:vir.param.vs.mass} implies that 
large fractions of the population 
are really not self-gravitating -- it is better in this regime to consider the 
systems as simply ``molecular overdensities'' rather than ``clouds'' in the 
traditional sense.

\begin{figure*}
    \centering
    \scaleup
    \plotsidesmall{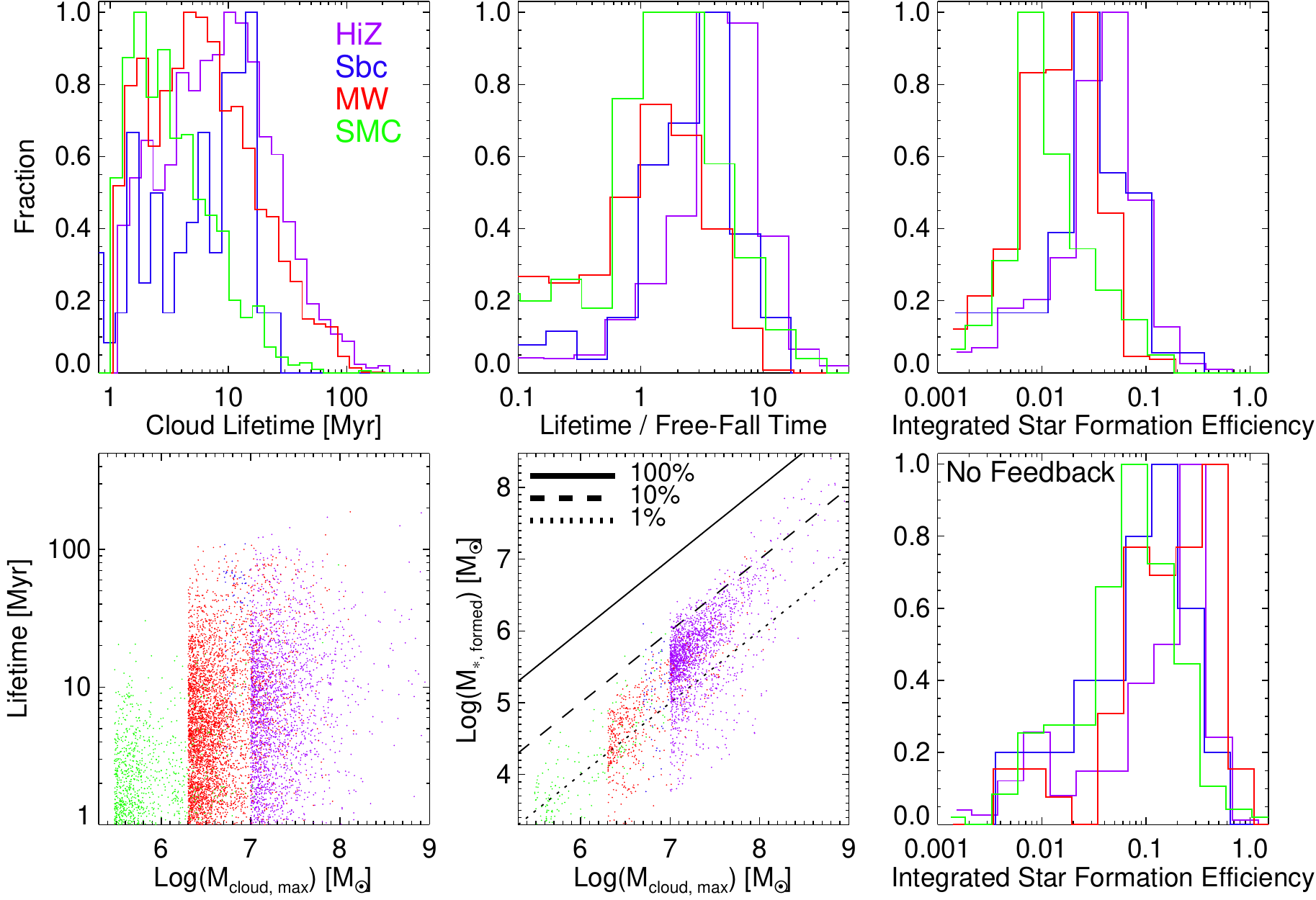}
    \caption{Statistics of cloud lifetimes and star formation efficiencies
    in our standard simulations with all feedback mechanisms enabled.    For each simulation, 
    we identify all clumps that can be well-resolved ($>100$ particles) down to 
    $10\%$ of the clump mass ($>1000$ particles at peak); lifetime is defined 
    as the time above $10\%$ the maximum cloud mass. 
    {\em Top Left:} Distribution of cloud lifetimes $t_{\rm cl}$ for each of our galaxy models.
    {\em Top Center:} Lifetimes relative to the free-fall time $t_{\rm cl}/\langle t_{\rm ff} \rangle$ 
    of each cloud.
    {\em Top Right:} Integrated star formation efficiency (stellar mass formed in clump 
    over maximum cloud mass). 
    {\em Bottom Left:} Lifetime versus maximum cloud mass.
    {\em Bottom Center:} Stellar mass formed versus maximum mass. 
    Lines correspond to the different efficiencies labeled. 
    Clouds live for a few dynamical times, and turn a few percent of their mass 
    into stars. There is also a weak trend for more massive systems to have higher 
    efficiencies and longer lifetimes. 
    {\em Bottom Right:} Integrated star formation efficiency (as {\em top right}), but 
    in models with no feedback. Without feedback, clouds persist until most of their dense gas 
    is turned into stars, giving star formation efficiencies of $\sim10-50\%$ on 
    timescales of $\sim10\,$Myr.
    \label{fig:clump.lifetimes}}
\end{figure*}


\subsection{Lifetimes and Star Formation Efficiencies}
\label{sec:clumps:evol}

We next consider typical cloud ``lifetimes'' and integrated 
star formation efficiencies. 
To define these, we need to link clouds in time between different snapshots. 
Given a specific clump $p$ in snapshot $i$, we identify the descendant $d$ of this clump 
(in snapshot $i+1$) as the clump which contains the most total mass in particles 
from the original clump $p$. If no clump in snapshot $i+1$ contains any particles 
from $p$, then the clump has no descendant (its mass becomes zero).
This defines a ``merger tree'' for clumps with time.\footnote{To 
properly sample this requires that the snapshot time spacing 
be less than typical clump lifetimes. We have experimented with spacings 
as short as $10^{5}\,$yr, and generally find converged results for the 
lifetime statistics presented here for spacings $\lesssim10^{6}\,$yr.} 
To consider a ``main branch'' of that tree (i.e.\ growth of the primary clump, which 
may grow by accretion/inflow, or by mergers of smaller clumps), we 
note that if a clump $d$ is identified with multiple progenitor clumps $p$, the 
``main progenitor'' is simply the most massive. 
The other progenitors are marked as having ``merged onto'' the main branch 
at this time. 

For each main branch, we then have a mass evolution versus time, 
$M_{\rm cl}(t)$. This generally rises exponentially as the clump 
starts forming, then (with feedback enabled) falls rapidly as feedback unbinds 
the gas. We can then define other important quantities: 
the maximum mass of the clump (just $M_{\rm max}={\rm MAX}(M_{\rm cl}[t])$), 
and the lifetime $t_{\rm cl}$. We take the latter to be the total time 
when the clump is above some cutoff threshold $\eta\approx 10\%$ of $M_{\rm max}$ -- 
the choice is arbitrary, but because the clumps tend to grow 
and dissociate quickly the lifetimes are not especially sensitive to the 
choice so long as $\eta\ll 1$. 

Figure~\ref{fig:clump.lifetimes} plots the distribution of the resulting clump lifetimes in 
our standard simulations. We plot it in absolute units and in units of the clump dynamical 
time, $t_{\rm dyn}=1/\sqrt{G\,\rho_{\rm cl}}$. 
We do not include clumps that are lost via merger onto more massive clumps, 
because we cannot know how long they would have survived, 
but their growth/decay curves tend to be no different from main-branch clumps 
of similar mass and spatial locations until their merger. 
And we find that the majority of clouds grow primarily 
by accreting ``smooth'' gas (i.e.\ gas not in another massive cloud), rather than via 
(at least major) hierarchical mergers, so this is not a large effect.

Our simulated GMCs are short-lived: with feedback present, most live $\sim10^{6}-10^{7}\,$yr, 
$\sim1-10$ free-fall times. Star formation within clouds should therefore 
be inefficient: we expect just a few percent of their mass will be 
turned into stars, but we can calculate this explicitly. 
We sum the SFR from all cloud particles at each 
time, integrated over the cloud lifetime to get the total stellar 
mass formed, and compare this to the maximum clump mass 
(defining the efficiency $M_{\ast,\,{\rm formed}} / M_{\rm cloud,\,max}$). 
The typical cloud converts $\sim1-10\%$ of its peak 
mass into stars, similar to what has been inferred from a range of 
observations \citep{zuckerman:1974.gmc.constraints,williams:1997.gmc.prop,evans:1999.sf.gmc.review,evans:2009.sf.efficiencies.lifetimes}. This does not necessarily mean, however, 
that this fraction of all mass that enters the cloud is converted into 
stars. Over the cloud lifetime, mass is continuously accreted and lost, 
so the total mass in gas that gets processed ``through'' the cloud 
can be even larger, implying an even lower net ``efficiency'' of converting 
GMC material to stars. 

What sets the efficiencies and lifetimes of GMCs?
As we discussed above, for the {\em dense} gas, the most important feedback 
mechanism is the local radiation pressure. 
Therefore consider just this, for simplicity, in a spherically symmetric cloud 
of mass $M$ and projected effective radius $R_{e}$.
The total gravitational force in a GMC (summing over mass) is 
\be
F_{\rm grav} = \frac{G\,M^{2}}{\beta_{1}\,R_{e}^{2}}
\ee
where $\beta_{1}\sim1$ depends on the mass profile 
The force from radiation pressure, integrated over the sphere, is 
\be
F_{\rm rad} = (1+\tau_{\rm IR})\,\frac{L}{c}
\ee
where the pre-factor assumes the cloud is optically thick in the 
UV/optical, but not necessarily in the IR. 
If the lifetimes are short, of order a few Myr, then 
$L\propto M_{\ast}$, because most SNe have not exploded; 
for a standard stellar population with the same IMF assumptions 
as our models, $L\approx 1137\,L_{\sun}\,(M_{\ast}/M_{\sun})$. 
Since the average surface density of clouds is within a narrow range, 
it is convenient to define $\langle \Sigma \rangle = \beta_{2}\,M/(\pi\,R_{e}^{2})$ 
($\langle \Sigma \rangle$ is the mass-weighted average $\Sigma$, $\beta_{2}$ 
again just depends on the mass profile). 
Setting $F_{\rm grav} = F_{\rm rad}$, 
we obtain 
\be
\frac{M_{\ast}}{M_{\rm cloud}} \approx
0.04\,\beta_{0.7}\,\frac{\langle\Sigma\rangle_{100}}{1+\tau_{\rm IR}}
\ee
where $\beta_{0.7}=(\beta_{1}\,\beta_{2})^{-1}/0.7$  
($=0.95,\,1.07,\,1.12$ for a constant-density, Plummer, or 
\citealt{hernquist:profile} profile) and 
$\langle\Sigma\rangle_{100}=\langle\Sigma\rangle/100\,\msun\,{\rm pc^{-2}}$.

This simple argument reproduces the efficiencies at low cloud 
masses, where these assumptions hold, and $\tau_{\rm IR}$ is small, 
for the typical observed average surface densities of such clouds 
($\langle\Sigma\rangle_{100}\sim1$). 
The predicted efficiency does increase 
at higher $\langle\Sigma\rangle$, which could in principle lead to a runaway collapse 
if there were no offsetting force. However, 
at sufficiently high $\Sigma\gtrsim1000\,\msun\,{\rm pc^{-3}}$, 
$\tau_{\rm IR}$ becomes larger than unity even for the smoothed average surface density. 
More accurately the typical $\tau_{\rm IR}\approx c^{2}\,\kappa_{\rm IR}\,\langle\Sigma\rangle$ where 
$c^{2}$ is a ``clumping fudge factor'' that accounts for the excess rise in density around each stellar cluster/core sub-region, a well as contributions from other feedback mechanisms (typical effective $c\sim2-5$).

Once $\tau_{\rm IR}\gg1$, the efficiency should asymptote to a maximum 
\be
\frac{M_{\ast}}{M_{\rm cloud}} \rightarrow 0.4\,c^{-2}\,\beta_{0.7}\,\kappa_{5}^{-1}\ \ \ \ \ 
(\tau_{\rm IR}\gg 1)
\ee
where $\kappa_{5}=\kappa_{\rm IR}/5\,{\rm cm^{2}\,g^{-1}}\sim1$. 
Given modest $c\sim$ a couple, and the fact that our 
simulations rarely reach such extreme densities, this matches reasonably well 
the ``upper limit'' to the star formation 
efficiencies we find; together, these values bracket the range in the simulations. 

Cloud lifetimes follow from these integrated efficiencies, as the time 
needed to form this number of stars, $\Delta t = M_{\ast} / \dot{M}_{\ast}({\rm cloud})$.
For an observed instantaneous SFR of a one or two percent per free-fall time, 
it takes a few free-fall times to form this mass and disrupt the cloud 
(in this limit, of course, we insert this efficiency in our SFR model, 
so the lifetime is not predicted in a fully a priori sense). However, in \paperone, we 
showed that the {\em global} star formation rate did not depend on the efficiency
employed in our model. Since it takes a free-fall time to assemble a GMC, and a free-fall time to disrupt it,  even if we assume a large efficiency, the cloud will live one to two free fall
times.

In the most massive cloud complexes, the lifetime reaches 
as large as $\gtrsim30\,$Myr and is no longer short 
compared to stellar evolution timescales. Since massive stars are 
dying, instead of $L\propto M_{\ast}$, we have $L\propto \dot{M}_{\ast}$. 
In this limit, for a constant star formation rate 
$L/\dot{M}_{\ast}\approx5.7 \times10^{9}\,L_{\sun}/(M_{\sun}\,{\rm yr^{-1}}$). 
Inserting this into the derivation above requires an instantaneous 
SFR 
$\dot{M}_{\ast}\approx 0.8\,\msun\,{\rm yr}^{-1}\,(M_{\rm cloud}/10^{8}\,\msun)\,
\beta_{0.7}\,\langle \Sigma \rangle_{100}\,(1+\tau_{\rm IR})^{-1}$. 
This then requires a time $t_{\rm lifetime} \sim t_{\rm dyn}$ to dissociate the cloud 
(and must act in less than a few dynamical times so the momentum is not dissipated); 
if we require the total momentum injected at this rate equal the binding 
momentum of the cloud, we obtain $t_{\rm lifetime} \approx 1.4\,\beta^{\prime}\,t_{\rm ff}$ 
(where $\beta^{\prime}$ again depends very weakly on the mass profile and 
$\langle t_{\rm ff} \rangle \equiv (3\pi/32\,\langle\rho\rangle)^{1/2}$), giving
\begin{align}
\label{eqn:massive.cloud.efficiencies}
\frac{M_{\ast}}{M_{\rm cloud}} &\approx 
0.04\,\beta^{\prime}\,
\frac{\langle\Sigma\rangle_{100}}{1+\tau_{\rm IR}}\,
{\Bigl(} \frac{t_{\rm ff}}{4\,{\rm Myr}} {\Bigr)} \\ 
&\approx
0.04\,\beta^{\prime\prime}\,
\frac{\langle\Sigma\rangle_{100}^{1/2}}{1+\tau_{\rm IR}}\,
{\Bigl(} \frac{R_{e}}{10\,{\rm pc}} {\Bigr)}^{1/2} 
\end{align}

Note that for similar-$\langle\Sigma\rangle$ clouds, 
the characteristic dynamical time increases with cloud mass as 
$t_{\rm ff}\propto R_{e}^{0.5}\propto M_{\rm cloud}^{0.25}$. 
We therefore expect more massive clouds to have larger absolute lifetimes, 
a correlation clearly evident in Figure~\ref{fig:clump.lifetimes}. 
Moreover, according to Equation~\ref{eqn:massive.cloud.efficiencies}, 
the integrated star formation efficiency (in massive clouds, at least) should also 
be a weakly increasing function of the cloud mass $\propto M_{\rm cloud}^{0.25}$.
We also see this trend in our simulations, with efficiencies 
rising from $\approx1\%$ to $\approx10\%$ from $M_{\rm cl}\lesssim 10^{5}\,\msun$ 
to $M_{\rm cl}\gtrsim 10^{8}\,\msun$, 
or $M_{\ast,\,{\rm formed}} / M_{\rm cloud,\,max} \propto M_{\rm cloud,\,max}^{(0.2-0.3)}$ 
\citep[see also the discussion in][]{murray:molcloud.disrupt.by.rad.pressure}.

Figure~\ref{fig:clump.lifetimes} compares the cloud 
star formation efficiencies in runs without 
feedback. In {\em absolute} units, the lifetimes are not necessarily much longer -- many 
live $\sim 10\,$Myr; however the population with lifetimes $\sim1\,$Myr disappears 
and a large tail of clouds (often in global structures such as spiral arms) appears with 
lifetimes $\sim100\,$Myr. 
But, recall that in the feedback-free cases, the dense gas collapses rapidly. As such, the 
range of lifetimes in units of the free-fall time $\propto \rho^{-1/2}$ is much more broad, 
with clumps living as long as $\sim10-100\,t_{\rm ff}$. The distributions are very broad because 
it is often local regions within the clump running away to high densities, so the average 
$\langle \rho \rangle$ may or may not reflect this to varying degrees in a cloud. The more 
robust indicator is the integrated star formation efficiencies. As expected, from the runaway global 
SFR in these models, the star formation efficiency per-cloud is $\sim10-100\%$, with a 
mean of $\sim40\%$ (it is not $100\%$ because there is always some low-density 
material associated with clouds, which is left at even lower densities once dense regions have 
all collapsed). Feedback is necessary to reproduce the observed 
per-cloud cloud lifetimes and star formation efficiencies. 

Considering each feedback mechanism in turn, our results essentially mirror 
those for the cold gas and GMC density distributions. In all models, 
radiation pressure is important to resist collapse; removing it in the HiZ case 
gives results nearly identical to the feedback-free case. 
And runs in all models with {\em only} radiation pressure enabled give similar 
results. In the MW and SMC 
cases, gas heating on $\sim1\,$Myr timescales from ``fast'' stellar winds and 
photo-ionization can contribute comparably to cloud dissociation (removing 
radiation pressure does not result in total collapse if they 
remain, though the average star 
formation efficiency is slightly higher), and the longer-lived clouds in these 
models tend to be dissociated when SNe turn on (a few Myr).

\vspace{-0.7cm}

\section{Discussion}
\label{sec:discussion}


We have implemented a model for stellar feedback in SPH simulations of galaxies that incorporates physically motivated treatments of a number of the key feedback processes:   radiation pressure from UV, optical, and IR photons; 
the momentum injection, gas heating, and gas recycling from 
SNe (Types I \&\ II) and stellar winds; and HII photoionization-heating.   Our treatment of radiation pressure is  calibrated by comparison to more detailed radiative transfer calculations (Appendix \ref{sec:appendix:num.longrange}).
Our model of the ISM also includes molecular cooling, a simple treatment 
of HII destruction, and allows star formation only at very high densities within giant molecular clouds ($n\gg100 \, {\rm cm^{-3}}$). 

In this paper, we study the consequences of these feedback models
for the  global structure of the ISM in disk galaxies (velocity dispersions, 
Toomre $Q$, etc), the phase structure of the ISM, 
and the properties of GMCs.  We simulate four representative galaxy models to bracket the diversity of star-forming galaxies observed locally and at high redshift (Table~\ref{tbl:sims}; Figs.~\ref{fig:morph.hiz}-\ref{fig:morph.smc}).   In a companion paper, we study the galaxy-scale outflows driven by stellar feedback and show that galactic winds appear to remove the bulk of the gas from their host galaxies.

With physically plausible stellar feedback mechanisms enabled, we show that star-forming galaxies generically approach a quasi steady-state in which the properties of the ISM are self-regulated by the turbulence driven by stellar feedback.   GMCs form by self-gravity and survive for a few dynamical times before they are disrupted by stellar feedback; during this time, GMCs turn a few percent of their mass into stars (Fig.~\ref{fig:clump.lifetimes}).  The resulting galaxy-integrated star formation rates are in reasonable agreement with the Kennicutt-Schmidt law (Fig.~\ref{fig:sfh.vs.fb}; see also \paperone) and the global galaxy properties (dispersion, scale heights, Toomre $Q$ parameter; Fig.~\ref{fig:hr.vs.t}) and phase distribution (Fig.~\ref{fig:rho.dist}) of the ISM are consistent with observations.

The ISM phase distributions calculated in this paper are of course still limited in accuracy both by standard numerical restrictions and by the inability of any galaxy-scale calculation to capture small-scale mixing processes and the effects of (saturated) thermal conduction. Nonetheless, much of the ISM dynamics and phase structure we find is a consequence of  supersonic turbulence driven by stellar feedback.  Such turbulence is reasonably well modeled numerically by SPH (based on comparisons to grid-based methods; e.g., \citealt{kitsionas:2009.grid.sph.compare.turbulence,price:2010.grid.sph.compare.turbulence,bauer:2011.sph.vs.arepo.shocks}). Based on this and preliminary comparison with other codes, we expect that the inclusion or exclusion of different feedback mechanisms will result in much larger differences than the details of the numerical method.

\vspace{-0.3cm}
\subsection{The Problems of Feedback-Free Models}
\label{sec:discussion:nofb}

The  structure of the ISM in the presence of stellar feedback is markedly different than in  calculations that do not have any stellar feedback.  In the latter case, the gas rapidly radiates away its vertical thermal support and any gravitationally-induced turbulent fluctuations are damped on a crossing time.   This leads to runaway collapse of the gas and star formation well in excess of the Kennicutt-Schmidt relation ($\dot{M}_{\ast}\sim M_{\rm gas}/t_{\rm dyn}$; Fig.~\ref{fig:sfh.vs.fb}). Dense star-forming regions within GMCs collapse without 
limit, producing a large non-lognormal excess of high-density gas (Fig.~\ref{fig:rho.dist.all}).   These GMCs have very low 
virial parameters relative to observations (Fig.~\ref{fig:vir.param}) and turn $\sim 10-50 \%$ of their gas into stars into just a few cloud dynamical times (Fig.~\ref{fig:clump.lifetimes}). 

In other words, turbulence from global gravitational instabilities and the
collapse of GMCs  does {not} appear
sufficient to maintain low star formation efficiencies and prevent 
runaway collapse of the ISM (neither globally nor 
within GMCs). This result has been found in other simulations with
 sufficiently high resolution to follow individual GMCs 
\citep[see e.g.][]{bournaud:2010.grav.turbulence.lmc,
tasker:2011.photoion.heating.gmc.evol,
dobbs:2011.why.gmcs.unbound}.  
Although our calculations do not include inflow of gas from the IGM and thus we cannot explicitly 
assess whether such accretion can help maintain turbulence in galactic disks, the fact that the 
runaway is {\em local}, and occurs even while the disk maintains a large nominal gas velocity dispersion, suggests that gas inflow will not change these conclusions.

We caution that some properties of the ISM can be reasonable in feedback-free models, but for the 
wrong reasons. For example, without feedback disks still eventually reach $Q\sim1$ (Fig.~\ref{fig:hr.vs.t}); this is true even without star formation so long as cooling is included.
However, without feedback this occurs because of two processes that do 
not occur in real galaxies: first, all of the dense gas rapidly turns into stars, leaving only a small, tenuous gas mass behind that can more easily maintain $Q\sim1$; second, in doing so, the dense gas 
collapses without limit to $\gtrsim 10^{6}\,{\rm cm^{-3}}$ regions that  
become so small that they dynamically act like the collisionless stellar disks.

In our simulations without  stellar feedback, the {\em global} collapse of GMCs is eventually arrested by angular momentum.\footnote{This collapse might be very different in the presence of magnetic fields which can torque the GMCs and remove this angular momentum.}   The resulting GMCs
have roughly similar median surface densities and Larson's law scalings to 
models with feedback (Fig.~\ref{fig:clump.sigma}), but dynamically the GMCs are rotationally supported disks instead of  triaxial clouds. Moreover three-dimensional collapse 
still runs away, especially in dense sub-regions. Thus the median 
surface densities of GMCs are similar with and without feedback, but the  mass in dense 
molecular gas and the per-cloud SFRs are  an order-of-magnitude larger in the absence of feedback.
Finally, while clouds can in many cases be disrupted over $\sim10\,$Myr 
by shear in the disk, during this time the runaway collapse of the gas has turned an order-unity 
fraction of the GMC mass into stars (Fig.~\ref{fig:clump.lifetimes}).
As a result, although models without feedback can give rise to $Q\sim1$ disks with several observed disk and GMC scaling relations, gravity alone cannot (in our calculations) regulate the structure of the ISM in a way that is similar to models with explicit stellar feedback.

\subsection{How Does Feedback Regulate Star Formation?}
\label{sec:discussion:vsfbmech}

What maintains the low star formation efficiency in the presence of feedback?   One interpretation is that the {instantaneous} star formation efficiency of $\sim 1\%$ per dynamical time can be explained as a consequence of the statistics of the lognormal turbulent density field \citep{nordlund:1999.density.pdf.supersonic,vazquez-semadeni:2003.turb.reg.sfr,li:2004.turb.reg.sfr,krumholz.schmidt}. More recent -- and much higher resolution -- numerical work has highlighted, however, that star formation in such models proceeds more rapidly than previously appreciated ($\sim 30\%$ per dynamical time) if turbulence is {driven solely on large scales} \citep{padoan:2011.new.turb.collapse.sims}.    In these calculations, the rate of star formation is indeed set by the mass of gas above a critical density that can collapse; however, when self-gravity and rapid cooling are included in simulations with sufficient dynamic range, self-gravitating regions can ``detach'' from the large scale flow and collapse without limit. Even if their internal instantaneous SFR per dynamical time remains low, such self-gravitating regions quickly reach such high densities (short $t_{\rm dyn}$) that the global SFR per dynamical time becomes much larger than observed. 

This dynamical decoupling of high density regions is very similar to  what occurs in our simulations without feedback. Recall, these maintain a similar large-scale velocity dispersion as models with feedback.
\citet{bournaud:2010.grav.turbulence.lmc} also explicitly showed that high-resolution galaxy models with and without feedback maintain similar turbulent power spectra, but that the turbulent cascade from large scales is unable to dissociate clumps on small scales once they have become self-gravitating.  This leads to large deviations from a lognormal density distribution (Fig.~\ref{fig:rho.dist.all}), which leads to a star formation efficiency that is an order of magnitude larger than observed.

The difficulties of models in which turbulence is driven solely on large scales (via self-gravity or some other mechanism) points to the critical role of stellar feedback processes that can regulate star formation on {\em small} scales (e.g., by disrupting GMCs).   
Moreover, the high star formation efficiency found in high-resolution ``turbulent box'' simulations and  feedback-free galaxy simulations suggests that the feedback required to dissociate dense, self-gravitating regions (GMCs and cores within them) regulates the SFR rather differently than simple homogeneous turbulence driven on large scales. 

Which feedback mechanisms prevent the runaway collapse seen in the no-feedback models and in models in which turbulence is driven solely on large scales?   We show that this depends on the physical properties of the galaxy model, in particular the density of the ISM. 
In high-density regions (typical of the global ISM of starbursts and high-redshift galaxies and the interiors of GMCs), we find that radiation pressure is the dominant feedback mechanism (especially multiple-scatterings of IR photons).   Lower-mass clouds (more prevalent in the MW and SMC models), with lower mean densities, can be similarly disrupted by HII photoionization (e.g., \citealt{matzner02}).
In \paperone \, we showed that radiation pressure alone leads to galaxy models that lie roughly along the Kennicutt-Schmidt relation; including the additional feedback mechanisms in this paper typically results in smaller (factor $\sim2-3$) global SFRs by suppressing the collapse of gas into GMCs and 
by expelling more gas in galactic winds.   However, these additional feedback mechanisms have relatively little effect on the integrated lifetimes or star formation efficiencies of GMCs themselves.

In lower-density regions of the ISM (typical of  MW-like systems and dwarf galaxies), {\em global} self-regulation is much more strongly affected by heating processes (Fig.~\ref{fig:sfh.vs.fb}). 
Removing radiation pressure in these regimes can still lead to runaway collapse of the 
dense star-forming gas.   However, although 
dense regions run away, the feedback from the resulting star formation can act on the remaining lower-density gas to limit further collapse and help globally regulate the star formation.   Overall, SNe are the most important heating mechanism, since their integrated energetics dominate 
over stellar winds and photoionization-heating; however, within GMCs, 
stellar winds and photo-ionization are typically more important because they act on $\lesssim$\,Myr timescales while SNe do not explode until a few Myr later -- longer than the GMC lifetime.

\subsection{Generic Consequences of Strong Stellar Feedback}
\label{sec:discussion:observable}

As discussed in \S \ref{sec:discussion:nofb}, we find that  gaseous star-forming disks robustly self-regulate
at $Q\sim1$, which in turn sets the required velocity dispersion $\sigma$  
and the disk scale height (Fig.~\ref{fig:hr.vs.t}). This is independent of the details of feedback coupling mechanisms. In particular, energy/momentum input in 
excess of what is needed for $Q\sim1$ will be ``lost'' (in e.g. galactic winds) or 
eventually dissipated (since cooling times are short), and any gas with $Q\ll1$ will collapse, 
turning into stars and triggering more efficient feedback.


Likewise, we find that the properties of GMCs largely reflect the conditions for gravitational 
collapse starting from a $Q \sim 1$ disk. More precisely, this is true provided that feedback makes GMCs relatively short-lived, so that 
their instantaneous conditions are not that different from those ``at formation.''   In this context 
the $\sim M^{-2}$ mass function of GMCs is a 
consequence of the scale-free nature of gravity, with a cutoff near the global 
Jeans mass (Fig.~\ref{fig:clump.mf}). The observed size-mass and linewidth-size ($\delta v\propto R^{0.5}$) relations of GMCs reflect conditions 
for collapse in a supersonically turbulent $Q\sim1$ disk (Figs.~\ref{fig:clump.sigma}-\ref{fig:clump.linewidth}). 
This in turn requires a weak dependence 
of GMC densities on the mean density of the galactic disk, such that the linewidth-size 
relation normalization scales with the surface density of the disk as $\delta v/R^{0.5}\propto \Sigma^{0.5}$.

The fact that the GMC properties -- and some of the average disk properties   -- are largely determined by the conditions for gravitational collapse  
accounts for why other calculations with different physics  reproduce many of the same GMC/disk properties. The key criterion is that feedback is {sufficient} to  both disrupt GMCs and maintain the ISM in a marginally stable state against collapse.  If this can be satisfied, galaxy-averaged SFRs are set by the number of young stars needed to supply the necessary feedback, independent of {how} those stars form.    As a result, the global SFR -- and thus the global Schmidt-Kennicutt law -- emerges relatively independent of the assumed micro-scale (high density) star formation law, density threshold, or efficiency. In \paperone\ we varied these parameters widely and explicitly showed they have no effect on the SFR. Here, we similarly find that explicit 
models for molecular chemistry in the local SFR and cooling function have no effect on the SFR for the galaxies we model (Fig.~\ref{fig:molecules}). 

In low-resolution numerical simulations, an explicit treatment of the molecular fraction  can make large differences to the SFR, because such models are really implicit models for the fraction of gas at high densities. But in our calculations GMCs are resolved, and star formation occurs only at the highest densities within them ($n\gg 100\,{\rm cm^{-3}}$), which are already overwhelmingly molecular. At our resolution, in fact,  most molecular gas is non-star forming. As long as a cooling channel is available, gas will eventually collapse to the densities needed to self-shield and form sufficient young stars to regulate the galaxies' SFR via feedback.   An explicit treatment of the molecular chemistry is thus only dynamically important at low enough metallicities  that the cooling time by any channel becomes long relative to the dynamical time, roughly $Z \ll 0.01\,Z_{\sun}$ (see also 
\citealt{glover:2011.molecules.not.needed.for.sf}.)

These conclusions are reassuring in the sense that they suggest that our models are unlikely to be inaccurate on large scales, even if they are still missing some important physics. For example, magnetic fields and/or cosmic rays may contribute comparably to the ISM pressure in many galaxies. However, given that our results are robust to variations in the ``microphysics'' of how, precisely, feedback and cooling are implemented -- and self-regulation at $Q\sim1$ implies they will, in equipartition, contribute only $\sim10-30\%$ level corrections to the SFR needed to maintain stability -- they are unlikely to significantly alter our conclusions. 

The phase structure of the ISM is  much more sensitive to the microphysics of stellar feedback than global disk or GMC properties.   Without feedback, there is a runaway collapse of some of the gas at the highest resolvable densities $n\gg 10^{4}\,{\rm cm^{-3}}$, even though the amount of gas in GMCs and at modest densities $n\sim1-100\,{\rm cm^{-3}}$ is similar to models with feedback (Fig.~\ref{fig:rho.dist.all}).  Based on the calculations in this paper and in \paperone\ we conclude that the fraction of the ISM mass at high densities varies continuously with feedback strength, with more efficient feedback implying less gas at high densities  (e.g., Fig. 8 of \paperone).  This is  because the SFR itself adjusts depending on the efficiency of feedback and the SFR is set by the amount of mass in the high density ISM.   The dense gas is particularly sensitive to radiation pressure: removing this support allows GMCs to collapse to much smaller radii and higher densities, particularly in high surface density systems where photoionization feedback is less effective (Fig.~\ref{fig:sfhwind.morph.vs.fb.hiz} \& \ref{fig:clump.sigma}).   These results strongly suggest that observational probes of high density molecular gas (e.g., HCN) are particularly sensitive probes of the dominant stellar feedback mechanisms in galaxies.  We will pursue this connection quantitatively in future work.


 
\vspace{-0.15in}
\acknowledgments 
We thank Dusan Keres, Todd Thompson, and Lars Hernquist for helpful discussions.  Support for PFH was provided by the Miller Institute for Basic Research in Science, University of California Berkeley.  EQ is supported in part by the David and Lucile Packard Foundation.  NM is supported in part by NSERC
and by the Canada Research Chairs program.
\\

\bibliography{/Users/phopkins/Documents/lars_galaxies/papers/ms}


\begin{appendix}

\section{Numerical Tests of the Long-Range Radiation-Pressure Model}
\label{sec:appendix:num.longrange}

The model we introduce in \S~\ref{sec:stellar.fb:continuous.accel} for the treatment of 
long-range radiation pressure forces requires several simplifying assumptions, 
in lieu of the ability to perform full on-the-fly radiation hydrodynamics 
with dust. In this Appendix, we test these 
assumptions in detail to calibrate our simplified model.

\subsection{Calibration of the Model}
\label{sec:appendix:num.longrange.calibration}

First, we wish to check whether our calculation returns a reasonable 
approximation to the correct column densities and extinction of each star particle, and the corresponding rate at which the stellar momentum couples to the gas. 

\begin{figure}
    \centering
    \plotone{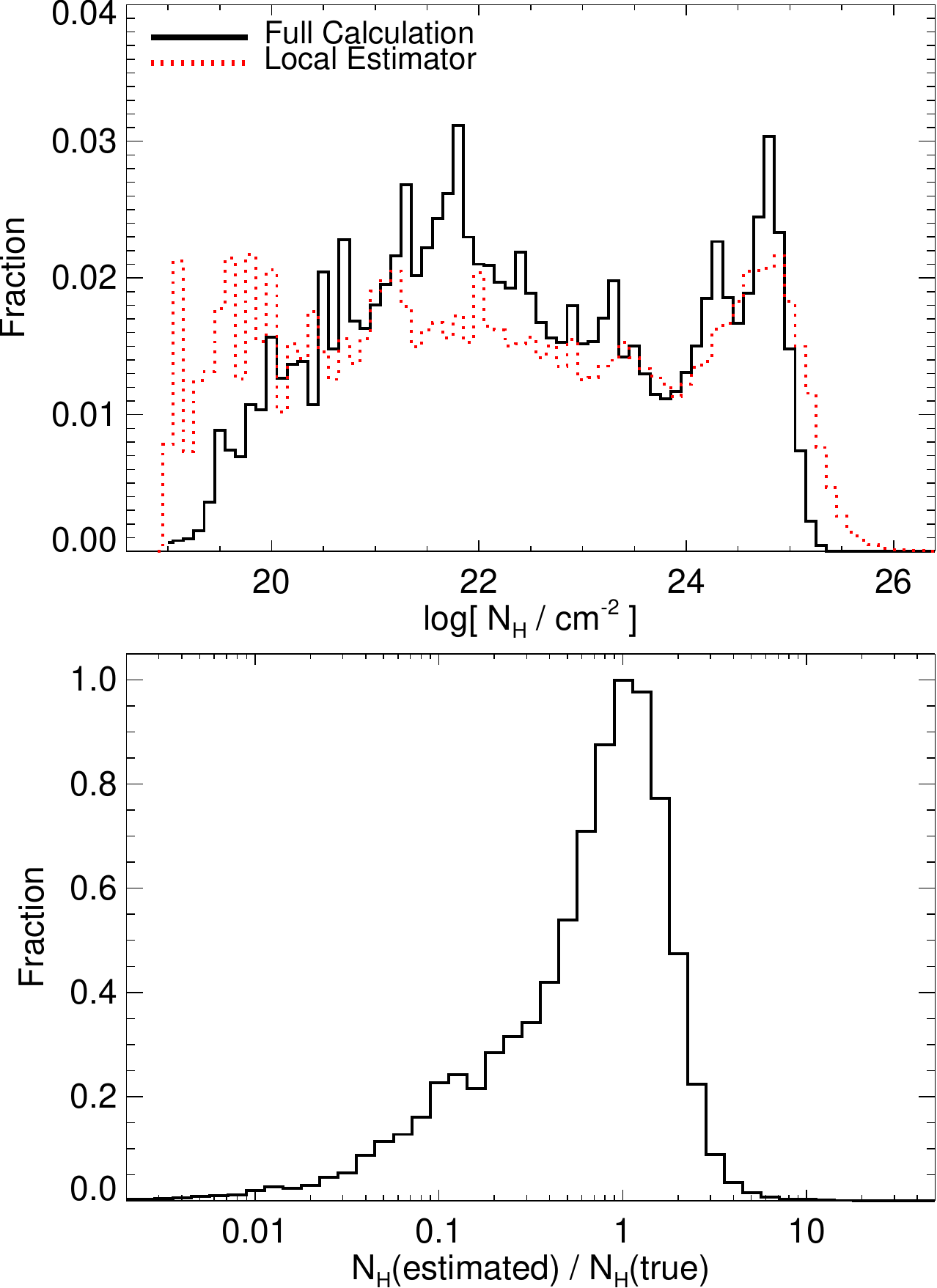}
    \caption{{\em Top:} Distribution of column densities between all sources (stars) and 
    absorbing gas particles in a standard (all feedback enabled) HiZ simulation.
    We compare the columns estimated using the local on-the-fly estimator used in 
    real time in the code (based on the local density and density gradient) and 
    that from a full (post-processing) integration over the gas distribution between all stars and gas. 
    The differences in time are minimal. The HiZ model is shown because it is most sensitive 
    to radiation pressure, but the other models are qualitatively similar. 
    {\em Bottom:} Distribution of the ratio $N_{H}({\rm estimated})/N_{H}({\rm true})$ 
    averaged around all sources (weighted by their contribution to the bolometric luminosity). 
    The local estimator we employ provides a good approximation to the true columns, with 
    an accurate mean and factor $\sim2$ scatter. This is because columns tend to be dominated 
    by gas near each star.
    The largest errors occur at the lowest columns, which tend to have comparable contributions 
    from gas at much larger radii (this gives the ``tail'' in the lower panel), but 
    these tend to be older stars that are no longer in gas-rich star-forming regions, and so 
    contribute only $\sim10\%$ of the luminosity. 
    \label{fig:nh.localvsfull}}
\end{figure}

We estimate the line-of-sight extinction of sources with our local column density estimator defined in Equation~\ref{eqn:columnproxy}, which uses the local density and density gradient. If the column is dominated by 
gas close to the sources, this should be a good approximation to the true column towards 
incident gas particles. Figure~\ref{fig:nh.localvsfull} compares this estimator to the exact 
line-of-sight column density integrated between each source and every gas particle in 
the simulation. We calculate the latter in post-processing at a given instant, integrating 
over the full simulation gas distribution. 
The top panel in Figure~\ref{fig:nh.localvsfull} compares the distribution of columns towards all sources, while the bottom panel shows the fraction of column density ratios (local estimate/true) averaged over all sources.   The results Figure~\ref{fig:nh.localvsfull}  are for our ``standard'' (all feedback enabled) HiZ model at a random time once the disk is in quasi steady-state, but the results are similar at all times and 
in the other simulations (modulo the absolute value of the typical columns). We focus on the HiZ 
case because it is the simulation where the radiation pressure has the largest effect, 
and therefore is most sensitive to any errors in the model. 

The two methods yield quite similar column density distributions. 
The local estimator yields a slightly broader and more uniform distribution, which is 
expected since it introduces some errors that broaden the distribution. 
This is evident in the distribution of $N_{H}({\rm estimated})/N_{H}({\rm true})$, 
which has a relatively narrow Gaussian core.
The dispersion in that distribution is reassuringly small -- a factor of $\approx2$ -- 
and most important the mean value is about unity. 
The biggest difference appears at very low columns:  below $N_{H}\sim 10^{20}-10^{21}\,{\rm cm^{-2}}$, there is a comparable column contributed from passing through almost any part of a typical galaxy disk, so the ``true'' columns are likely to have contributions from large radii that are not accounted for in our local estimator. 
This is manifest as the non-Gaussian ``tail'' towards lower $N_{H}({\rm estimated})/N_{H}({\rm true})$.
However only a relatively small fraction of the stellar luminosity ($\sim10-20\%$) samples this tail in the distribution.

If we consider all sources by 
age separately, the youngest stars tend to have the highest local column densities, 
and therefore be most accurately approximated by our local column estimator. 
This is because they tend to be located in the sites of star formation (near dense gas) and 
often have not cleared away all of the surrounding gas and dust with feedback. 
If we repeat the comparison in Figure~\ref{fig:nh.localvsfull} for ``old'' stars with 
age $>10^{7}$\,yr ($>10^{8}$\,yr), we find 
a median $N_{H}({\rm estimated})/N_{H}({\rm true})$ of $\sim0.7$ ($\sim0.5$), 
a scatter of a factor $\sim4$ ($\sim10$). This is because the older stars, having 
cleared away the dense gas from which they formed, can have 
larger contributions to the column from gas at very large distances. In principle, 
we could adopt an age-dependent ``fudge factor'' to account for this, but 
this introduces a number of additional uncertainties. In practice, 
since the older stars contribute 
fractionally less to the total luminosity especially in the UV, which is the band most 
sensitive to the column density estimator, this does not introduce much bias 
in the luminosity-weighted $N_{H}$ distribution. As a result, 
the effect on the actual momentum coupled is small.

\begin{figure}
    \centering
    \plotone{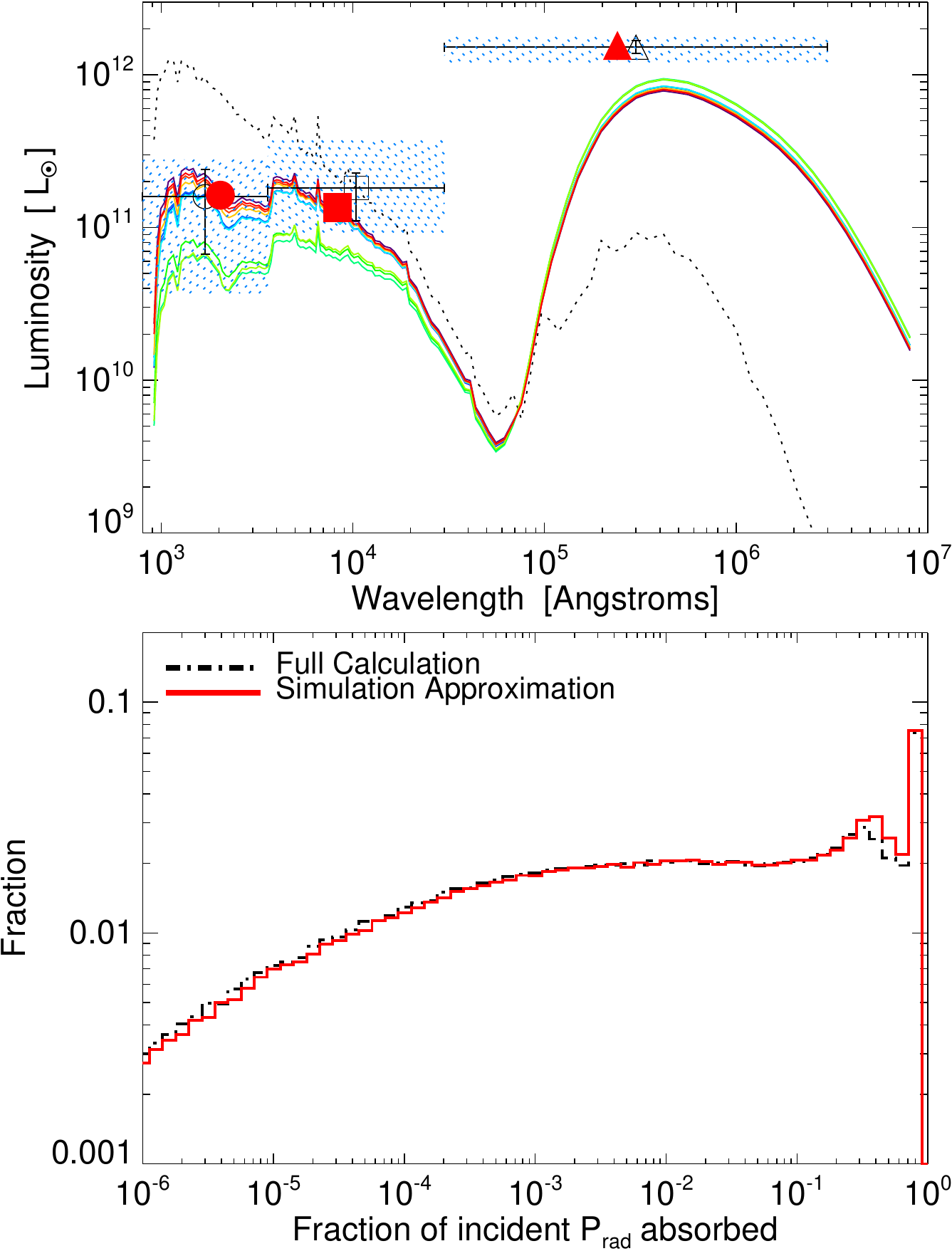}
    \caption{{\em Top:} Comparison of the emergent SED estimated with the 
    on-the-fly simulation approximations and a full radiative transfer calculation 
    using the SUNRISE code. 
    For the same HiZ snapshot as Figure~\ref{fig:nh.localvsfull}, we plot the 
    full emergent spectrum (plotted as $\lambda\,L_{\lambda}$) 
    calculated with RT (solid lines; we show 10 of 
    random sub-sampled sightlines). For comparison, the spectrum without any attenuation 
    is shown as the black dotted line. Integrating over our three wavelength intervals 
    (UV/Optical/IR), the resulting integrated luminosity is shown (open black points; 
    circle/square/triangle for UV/Optical/IR, respectively) with horizontal error bars 
    showing the wavelength range of integration and vertical error bars showing the 
    $\pm1\,\sigma$ range across all sightlines. 
    Red filled points show the average luminosity in each band estimated using the simulation 
    approximations. Shaded blue areas show the $\pm1\,\sigma$ observed range of luminosities 
    in each band, for observed galaxies with the same bolometric luminosity. 
    The local approximation predicts fluxes in each band within the $1\,\sigma$ range of 
    sightlines from a full RT calculation; both agree well with the observed ``average'' SEDs of 
    galaxies with similar SFR. 
    {\em Bottom:} Distribution of coupled momentum (specifically, distribution across all gas 
    particles of the radiation pressure momentum absorbed relative to the incident flux, 
    which depends on the SED shape and opacities). We compare the simulation approximation 
    and results again from a full RT calculation. The two distributions agree to within $\sim10\%$. 
    \label{fig:mom.and.sed.estimated}}
\end{figure}

Figure~\ref{fig:mom.and.sed.estimated}  shows the consequences 
of our local estimator for the emergent SED and the typical SED seen by each gas particle. 
First, consider the emergent spectrum. 
Using the SUNRISE code \citep{jonsson:sunrise.methods}, we can calculate the emergent spectrum 
from the galaxy along a large number of sightlines, uniformly sampling the sightlines to 
e.g.\ gas particles outside the disk. We do so using the exact same assumptions 
as in our simplified model about the scaling of dust-to-gas ratio with metallicity and emergent 
SED from all star particles as a function of age and metallicity from STARBURST99.
Figure~\ref{fig:mom.and.sed.estimated}  shows these results (as well as the un-processed spectrum) for a random subset of ten sightlines. 
Again, this is prohibitive to calculate in real-time, so we do so in post-processing for 
the same snapshot considered in Figure~\ref{fig:nh.localvsfull}. 
Next, recall that we simplify our calculation in the radiation pressure by integrating 
over three frequency ranges $L_{\rm UV}$ ($\lambda<3600\,$\AA), 
$L_{\rm Opt}$ ($3600$\,\AA$<\lambda<3\,\mu$) and $L_{\rm IR}$ ($\lambda > 3\,\mu$). 
We therefore integrate the emergent SUNRISE spectra over each range, 
and plot the resulting median luminosity in each range as well as its dispersion 
(here $14-86\%$ intervals). 

Figure~\ref{fig:mom.and.sed.estimated} also compares this 'exact' SED calculation to the approximate result given by our local estimator. Specifically, using the results from STARBURST99 to estimate the initial $L_{\rm UV}/L_{\rm Opt}$ for each star particle, we attenuate each 
according to the local column density estimator calculated for the particle 
(eqn.~\ref{eqn:attenuation})
and assume all of the absorbed energy in each of the UV and optical 
ranges is re-radiated as $L_{\rm IR}$. 
The resulting SED agrees well with the full radiation transfer calculation;
at all wavelengths it is within the $1\,\sigma$ sightline-to-sightline scatter of the 
full calculation, and generally within a factor of $<2$ of the mean in each 
waveband. In the IR, which contains most of the energy, the accuracy is much better, 
$\sim10\%$ (the largest uncertainties come when the net attenuation is very large). 
It is important to compare these theoretical fluxes to those measured in similar observed galaxies. 
In \S~\ref{sec:appendix:num.longrange.empirical} 
below, we discuss a variety of observational constraints on the fraction of the luminosity 
observed to escape in the UV/optical/IR, for galaxies of similar bolometric luminosity and/or 
SFR to the HiZ model considered here. From the references discussed there, we estimate 
an average empirical SED shape (shaded blue bands in Fig.~\ref{fig:mom.and.sed.estimated}), which is reasonably consistent with the model predictions.

Given the agreement in the emergent SED, we would expect the local model to 
reasonably capture the stellar momentum that couples to the gas.  Figure~\ref{fig:mom.and.sed.estimated} 
tests this explicitly. For the same simulation, we use the full radiative transfer results to 
calculate (in post-processing at a given instant) the net momentum deposition 
from radiation absorbed in each gas particle (using the full spectrum and dust opacity 
curve). We compare this to the actual coupled momentum using the simulation 
code approximations. Since we already know the total photon momentum is the same in either 
case, is is convenient to normalize the absorbed momentum by the incident momentum 
flux. This then tests whether the differences in emergent spectral shape 
(from our local column estimator), and the approximation 
of three wavebands instead of a full spectrum, as well as neglect of re-radiation changing 
the spatial distribution of emission on kpc scales, ultimately make a difference to the coupled 
momentum. 
Figure~\ref{fig:mom.and.sed.estimated} shows that the differences are small, 
at the level of tens of percent. This is smaller than the systematic uncertainties in our method (e.g., dust properties, unresolved clumping).

\subsection{The Effects of Photon ``Leakage''}
\label{sec:appendix:leakage}

\begin{figure}
    \centering
    \plotone{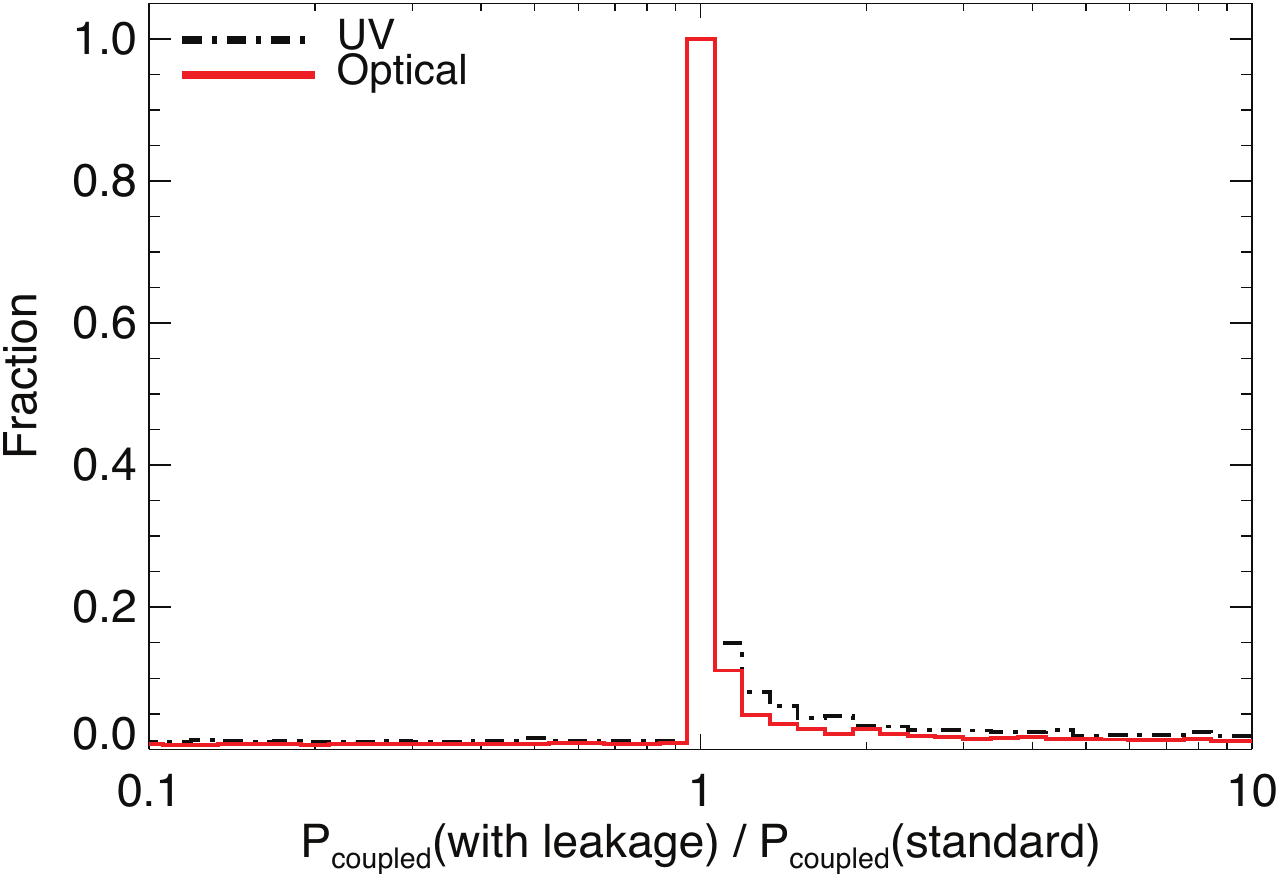}
    \caption{Effects of allowing for photon ``leakage'' -- i.e.\ assuming that the 
    ISM is clumpy and porous on sub-resolution scales. 
    We plot the ratio of coupled momentum from the UV and optical 
    wavebands - the distribution of this quantity is measured over 
    all gas particles for a given instant in the same HiZ 
    simulation. 
    The ``standard'' model takes the local column density 
    estimator as-is, and estimates the escape fraction from the spectrum of each star with 
    $f_{\rm esc}=\exp{(-\tau)}$ for each waveband. 
    The ``with leakage'' model assumes that each star is actually surrounded by 
    an (un-resolved) distribution of columns following Equations~\ref{eqn:fesc}-\ref{eqn:fesc.exact} 
    and uses this to calculate $f_{\rm esc}$. 
    The allowance for leakage only makes a difference when the average opacity is large, 
    at which point the contribution (in either case) to the global total luminosity in the relevant band 
    is small. As a result it has a negligible effect on our results.
    \label{fig:leakage}}
\end{figure}

One subtle complication is that the ISM is probably inhomogeneous 
on small scales well below what we model. 
In principle, that can allow photons to ``leak out'' 
from the region around their parent star at a rate much higher 
than the nominal $\exp(-\tauavg)$ expectation which we 
use above to estimate the attenuated spectrum. 
In \paperone\ we discuss how this leakage in the IR can affect 
the short-range radiation pressure term (and show the effects 
are generally small for the cases modeled here). 
Here, the question is how this might modify the initial attenuation of 
optical and UV photons used to treat the long-range radiation pressure forces. 

If a medium with mean opacity $\tauavg$ has a ``true'' column 
density distribution $P(\tau\,|\,\tauavg)$ across all sightlines, 
then the escape fraction is just 
\be
\label{eqn:fesc}
f_{\rm esc} = \int\,\exp{(-\tau)}\,dP(\tau\,|\, \tauavg)\ .
\ee
Fortunately, in \paperone\ we show that we can make a reasonable estimate of 
$dP(\tau\,|\, \tauavg)$. In ultra-high resolution simulations, we calculate
$dP(\tau\,|\, \tauavg)$ with $\sim1000$ lines-of-sight for each molecular cloud, 
and find the resulting column density distribution (per cloud) can be 
well-approximated with a near-universal function 
\be
\label{eqn:taudist.subgrid}
dP(\tau\,|\, \tauavg) \approx \frac{1}{2\,\sigma}\,
\exp{{\Bigl(}-\frac{|\ln{(\tau/\tauavg)}|}{\sigma}{\Bigr)}}\,
\frac{d\tau}{\tau}
\ee
with $\sigma=0.25-1.0$ (median $0.5$).  
This is also similar to the distribution estimated in even 
higher resolution simulations of individual GMCs and sub-cloud clumps 
\citep[see][]{ostriker:2001.gmc.column.dist}, 
and to observational estimates \citep{wong:2008.gmc.column.dist,goodman:2009.gmc.column.dist}. 
If the ISM is self-similar, then for any given $\tauavg$ the true 
$\tau$ distribution is given by the function above; 
we then solve for $f_{\rm esc}$:
\be
\label{eqn:fesc.exact}
f_{\rm esc} = \frac{\tauavg ^{-\frac{1}{\sigma}}}{2\,\sigma}\,
{\Bigl[}
\tauavg ^{\frac{1}{\sigma}}\,
{\rm E}_{1+1/\sigma}\,(\tauavg) 
+ \Gamma(\frac{1}{\sigma}) - 
\Gamma_{1/\sigma}(\tauavg)
{\Bigr]}
\ee
where the ${\rm E}$ and $\Gamma$ are incomplete 
exponential integral and Gamma functions. 
The exact functional form, however, is not important: the behavior is simple in 
two limits. For $\tauavg\ll 1$, this function is very close to 
$\exp(-\tauavg)$, since most sightlines are optically thin
(actually $f_{\rm esc}\rightarrow\exp{(-\tauavg/[1-\sigma^{2}])}$, 
a slightly {\em lower} value, owing to the presence of some dense sightlines). 
For $\tauavg\gg1 $, 
\be
f_{\rm esc}\rightarrow \frac{1}{2}\,\Gamma[1+\sigma^{-1}]\,\tauavg^{-1/\sigma}
\ee
i.e.\ the escape fraction declines as a power-law 
$\propto \tauavg^{-1/\sigma}$, rather than as an 
exponential $\exp{(-\tauavg)}$. This is straightforward to understand -- 
$f_{\rm esc}$ in this limit is essentially just the fraction of sightlines 
which are optically thin.

Figure~\ref{fig:leakage} shows the effects of replacing 
$f_{\rm esc} = \exp{(-\tauavg)}$, our usual estimator of attenuation, 
with this modified $f_{\rm esc}$, in the HiZ simulation. 
We specifically show the distribution of coupled momentum 
on all gas particles from the UV and optical wavebands
There is no significant difference. 
The reason is simple. For intermediate or low $\tauavg$, 
the escape fractions are large and very similar regardless of 
sub-structure. 
For high $\tauavg$, they can be very different, but they are both 
still very small. 
At $\tauavg\sim100$ (common for the optical opacities in dense 
star-forming regions), for example, the power-law $f_{\rm esc}$ is 
much larger than $\exp{(-\tauavg)}$, but is still 
$\sim10^{-7}-10^{-4}$ (for $\sigma=0.25-0.5$), so the amount of 
escaping optical light is still totally negligible compared to gravity. 
In short, the total UV/optical light seen by a given gas particle tends to be 
dominated by the stars that have low $\tauavg$ (even if they are rare), not by the 
small leakage from stars with high $\tauavg$. 
For the same reasons, repeating the experiment above but with a 
log-normal distribution assumed for $dP(\tau\,|\, \tauavg)$  (instead 
of the power-law distribution above) also makes no difference.

\subsection{Comparison to Purely Empirical SED Models}
\label{sec:appendix:num.longrange.empirical}

\begin{figure*}
    \centering
    \plotside{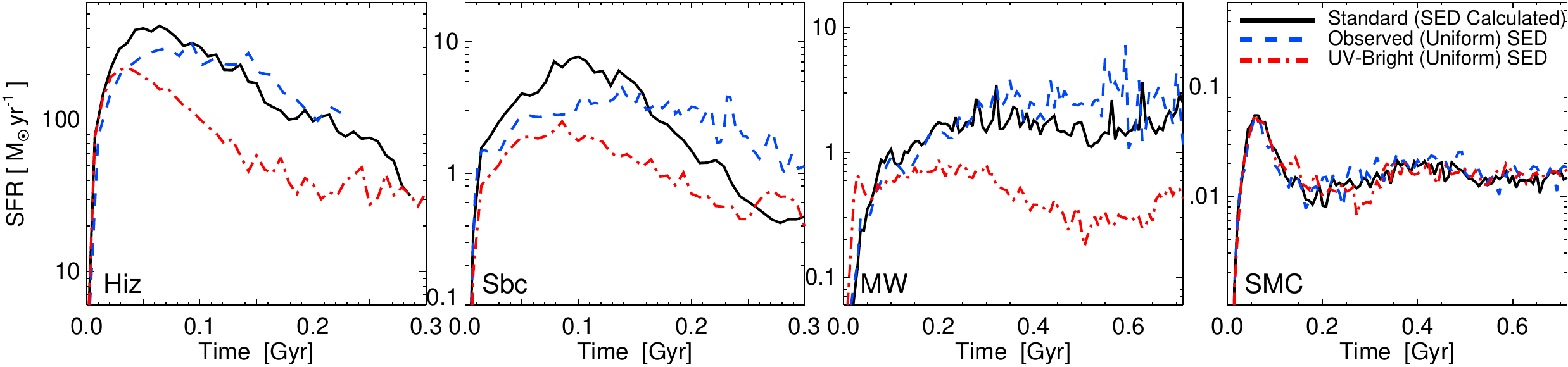}
    \caption{Comparison of our standard long-range radiation pressure 
    model (where the attenuation and SED shape of each star particle is calculated) 
    to one in which we simply force the SED shape to match a specific empirical 
    template. We show the resulting SFR versus time (as Figure~\ref{fig:sfh.vs.fb}) for each of the 
    HiZ/Sbc/MW/SMC models. 
    In each, we compare our standard model (black) in which the SED shapes 
    are self-consistently calculated on-the-fly, to a model where the SED shape 
    ($L_{\rm bol}$ is still determined self-consistently) 
    is forced to a constant fixed value chosen to match the observed mean 
    SED shape for similar observed galaxies to each model (blue). 
    These choices correspond to a relative proportion of $L_{\rm bol}$ in the 
    UV/optical/IR bands of 
    $(0.05,\,0.2,\,0.75),\ (0.07,\,0.23,\,0.7),\ (0.13,\,0.4,\,0.47),\ (0.3,\,0.3,\,0.4)$ 
    for the HiZ/Sbc/MW/SMC cases. 
    In each case, the self-consistent and empirically fixed models give SFRs 
    in reasonable agreement; this is because the emergent SEDs calculated 
    with the full models tend to agree well with the typical observed values 
    (Figure~\ref{fig:mom.and.sed.estimated}). However factor $\lesssim2$ differences 
    do arise, largely because the empirical model does not allow 
    spatial and/or time-dependent variations in the 
    emergent SED (for example, very young stars in the galaxy nucleus tend to be much more obscured 
    than older stars in the galaxy outskirts). 
    We also compare a model with fixed UV/optical/IR fractions, but much higher UV/optical fractions
    than actually observed (red): $(0.25,\,0.3,\,0.45)$ for the HiZ/Sbc/MW cases and $(0.4,\,0.4,\,0.2)$ 
    for the SMC. This artificially boosts the long-range radiation pressure, and significantly suppresses the 
    SFR in the HiZ/Sbc/MW models 
    (the effect is weak in the SMC case, because radiation pressure is relatively unimportant). 
    Accounting for attenuation changing the SED shapes and consequently radiation pressure is 
    critical; but small differences between empirical values and our simulation values 
    are not dominant uncertainties in the model.
    \label{fig:radiation.sim.vs.empirical}}
\end{figure*}

The average SED shape as a function of 
galaxy star formation rate or bolometric luminosity is 
well-studied \citep[see e.g.][]{cortese:2006.ir.vs.otherbands.vs.sfr,iglesias-paramo:2006.lir.luv.vs.sfr,
buat:2007.lir.luv.compiled,reddy:2008.z2.uv.ir.lfs,sargsyan:2009.luv.lir.compiled,
noll:2009.sed.vs.lbol,takeuchi:2010.extinction.vs.lbol}.
Specifically, using the relations in \citet{cortese:2006.ir.vs.otherbands.vs.sfr} between 
$L_{\rm UV}$, $L_{\rm IR}$, and $H$-band luminosity, with the same templates 
discussed therein to make the 
(small) correction between their wavebands and ours, we 
expect a typical luminosity fraction in each 
interval $(f_{\rm UV},\,f_{\rm Opt},\,f_{\rm IR})$ 
equal to $(0.05,\,0.2,\,0.75)$ for the HiZ, $(0.13,\,0.40,\,0.47)$ for the MW, 
$(0.3,\,0.3,\,0.4)$ for the SMC, 
and $(0.07,0.23,0.7)$ for the Sbc models. We stress that all of these values have factor $\gtrsim2$ scatter 
even in narrowly-selected galaxy samples.
This SED is shown for the HiZ model in Figure~\ref{fig:mom.and.sed.estimated}.  Although the  \citet{cortese:2006.ir.vs.otherbands.vs.sfr} sample is low-redshift, a similar result is obtained if we consider the 
high-redshift relation studied in \citet{reddy:2008.z2.uv.ir.lfs}. 
The shaded range in Figure~\ref{fig:mom.and.sed.estimated} corresponds to the empirical scatter in the luminosity at each 
band at fixed $L_{\rm bol}$ (estimated as the $14-86\%$ interval from the observed samples). 
The observed SED shape agrees well with the 
integrated fractions predicted by our models 
and the more detailed radiative transfer calculation. 
We find similar agreement for each of the other galaxy models as well. 

One  way we can test our 
long-range radiation pressure model is to calculate the bolometric luminosity and incident force 
on gas as in \S~\ref{sec:stellar.fb:continuous.accel} but to specify the
SED shape of each star particle (the emergent $L_{\rm UV}/L_{\rm Opt}/L_{\rm IR}$), 
to be a constant determined by the observed integral values for each galaxy model. 
This will of course gloss over spatial and temporal differences in the SED shape, 
but it has the advantage that we can directly take the SED motivated 
by observations for all ``typical'' particles.

In Figure~\ref{fig:radiation.sim.vs.empirical} we compare the 
star formation history for each of our standard galaxy models to
such a calculation with a fixed attenuated SED.  It is reassuring that the results are similar.    Of course, this is not that surprising given that the predicted  
SED $(f_{\rm UV},\,f_{\rm Opt},\,f_{\rm IR})$ from our simple attenuation model is
similar to those observed in each galaxy type.
Moreover, since the acceleration by the diffuse stellar radiation is primarily important only for gas that is out of the midplane, it turns out not to be a very bad 
approximation to assume that the typical particle sees much the same spectrum 
that the typical observer would see. 

For contrast, we also compare in Figure~\ref{fig:radiation.sim.vs.empirical} 
the results if we were to enforce a uniform SED shape with 
a higher UV/optical fraction than is actually observed -- 
which will substantially boost the coupled momentum. 
For the HiZ, Sbc, and MW models, this serves to dramatically 
suppress the SFR (especially at late times, because more of the gas 
is blown out of the disk in a wind). 
The SMC however is sufficiently low-luminosity 
that this has only a weak effect. 
This comparison highlights at least two important points.
First, on large scales, systems are generally not optically thick in the IR and 
so, unlike inside of GMCs, it is the UV/optical photons that tend to 
be absorbed and provide the radiation pressure force, so it is important 
that any model be able to reasonably approximate the average attenuation 
around stars in a self-consistent manner. 
Second, if there are times when dust obscuration is relatively weak, 
for example at very high redshifts (owing to low metallicities) or 
perhaps after phases of strong AGN feedback (either the AGN blowing away 
or dissociating the dust), then the 
long-range radiation pressure from stars can be dramatically 
boosted (even while the short-range effects are weakened), enhancing global 
outflows. Since this in turn removes more material, there is the possibility 
of unstable runaway that could substantially enhance quenching in 
such systems.

\section{Resolution \& Convergence Tests}
\label{sec:appendix:resolution}

\begin{figure}
    \centering
    \plotone{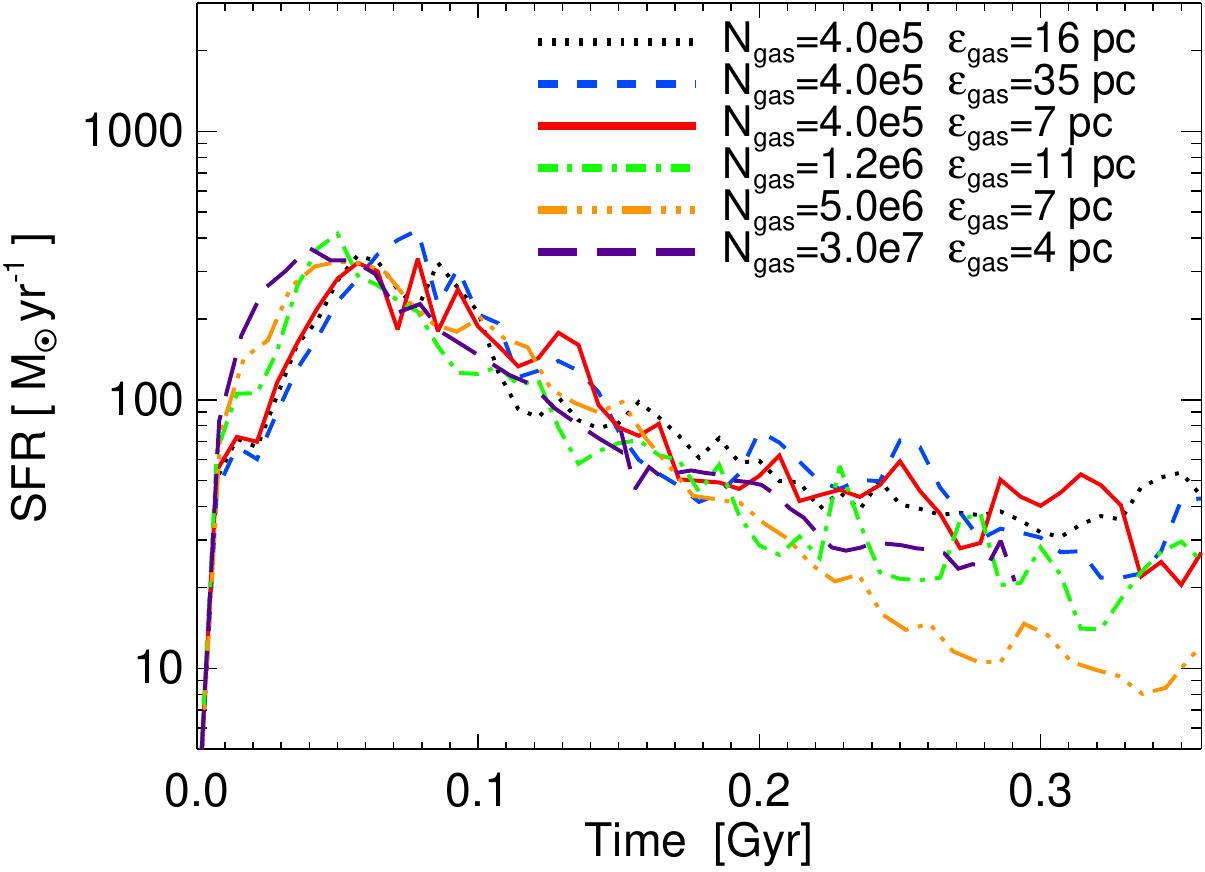}
    \caption{Resolution tests. We plot the SFR vs time 
    as Figure~\ref{fig:sfh.vs.fb}.
    We consider a ``standard'' (all feedback enabled) HiZ model at 
    a series in mass and force resolution ($N_{\rm gas}$ is the number of 
    gas particles in the star-forming disk, which 
    determines the mass resolution; $\epsilon_{\rm gas}$ is the 
    force softening). 
    The SFR converges quickly because it depends on the integral constraint of momentum 
    feedback ($\propto L \propto \dot{M}_{\ast}$) balancing gravity 
    (differences at late times owes to more efficient winds at higher resolution). 
    Recall, the Jeans length in these galaxies is several hundred pc 
    (Jeans mass $\gtrsim 10^{8}\,\msun$), so {\em all} of these models formally resolve the 
    Jeans scales. At lower resolution, our prescriptions do not have clear physical meaning. 
    \label{fig:resolution}}
\end{figure}

In Figure~\ref{fig:resolution}, we perform a spatial and mass resolution test for our HiZ simulation. 
We compare low, intermediate, standard, and high-resolution simulations 
of otherwise identical HiZ runs with all feedback enabled.\footnote{For 
technical reasons to ensure quantities such as random number generation 
were identical in these runs, this set was run on the same set of processors and so 
does not overlap with our standard HiZ run shown earlier; but the results 
from that run are completely consistent with the standard-resolution case here.}
We also compare, at low resolution, two additional runs with the same mass resolution but 
force resolution varied by a factor of $\approx5$. 
The star formation history converges reasonably quickly; the differences at 
the peak of the SFR are small even at low resolution. 
This is expected and discussed in detail in \paperone; since the equilibrium SFR is 
feedback-regulated, set by the balance of momentum injection with dissipation, it depends 
primarily on integral quantities (the total momentum injection $\propto L \propto \dot{M}_{\ast}$, 
approximately). As a result, it converges relatively quickly. 
In all cases, the higher-resolution runs begin to rise in SFR (and generally 
converge to their equilibrium structure) slightly earlier -- 
this simply owes to the fact that the most rapidly collapsing spatial scale resolved is on 
a smaller timescale as we go to higher resolution.
At late times, the higher-resolution 
simulations have a slightly lower SFR; this owes to their blowing slightly more 
efficient galaxy-scale outflows.



\end{appendix}

\end{document}